\documentclass[10pt,journal]{IEEEtran}

\ifCLASSOPTIONcompsoc
  \usepackage[caption=false,font=normalsize,labelfont=sf,textfont=sf]{subfig}
\else
  \usepackage[caption=false,font=footnotesize]{subfig}
\fi
\usepackage[cmex10]{amsmath}
\ifCLASSINFOpdf
   \usepackage[pdftex]{graphicx}
\else
   \usepackage[dvips]{graphicx}
\fi

\usepackage{xcolor,colortbl}

\definecolor{Gray}{gray}{0.85}
\definecolor{LightCyan}{rgb}{0.88,1,1}
\newcolumntype{a}{>{\columncolor{Gray}}c}
\newcolumntype{b}{>{\columncolor{white}}c}

\usepackage{amssymb}

\usepackage{epstopdf}

\usepackage{url}

\usepackage{graphicx}

\usepackage[caption=false]{subfig}

\usepackage{enumerate}
\usepackage{slashbox}
\usepackage{amsmath}

\usepackage{verbatim}

\usepackage{microtype}
\DisableLigatures{encoding = *, family = * }

\usepackage[nolist]{acronym}
\begin{acronym}
	\acro{AAA}{Authentication, Authorization and Accounting}
	\acro{ACL}{Access Control List}
	\acro{AKI}{Accountable Key Infrastructure}
	\acro{API}{Application Programming Interface}
    \acro{BSM}{Basic Safety Message}
    \acro{BYOD}{Bring Your Own Device}
	\acro{C2C-CC}{Car2Car Communication Consortium}
	\acro{C2C}{Car-to-Car}
	\acro{C2I}{Car-to-Infrastructure}
	\acro{CA}{Certification Authority}
	\acro{CN}{Common Name}
	\acro{CAM}{Cooperative Awareness Message}
	\acro{CAMP VSC3}{Crash Avoidance Metrics Partnership Vehicle Safety Consortium}
	\acro{CIA}{Confidentiality, Integrity and Availability}
	\acro{CRL}{Certificate Revocation List}
	\acro{CDN}{Content Delivery Network}
	\acro{COCA}{Cornell OnLine Certification Authority}
	\acro{CSR}{Certificate Signing Request}
	\acro{DAA}{Direct Anonymous Attestation}
	\acro{DDoS}{Distributed Denial of Service}
	\acro{DDH}{Decisional Diffie-Helman}
	\acro{DENM}{Decentralized Environmental Notification Message}
	\acro{DHT}{Distributed Hash Table}
	\acro{DL/ECIES}{Discrete Logarithm and Elliptic Curve Integrated Encryption Scheme}
	\acro{DoS}{Denial of Service}
	\acro{DoT}{Department of Transportation}
	\acro{DPA}{Data Protection Agency}
	\acro{DSRC}{Dedicated Short Range Communication}
	\acro{DSS}{Digital Signature Standard}
	\acro{ECU}{Electronic Control Unit}
	\acro{EDR}{Event Data Recorder}
	\acro{ETSI}{European Telecommunications Standards Institute}
	\acro{ECDSA}{Elliptic Curve Digital Signature Algorithm}
	\acro{ECC}{Elliptic Curve Cryptography}
	\acro{EVITA}{E-safety Vehicle Intrusion protected Applications}
	\acro{FOT}{Field Operational Testing}
	\acro{FPGA}{Field-Programmable Gate Array}
	\acro{GPA}{Global Passive Adversary}
	\acro{GN}{GeoNetworking}
	\acro{GS-VLR}{Group Signatures with Verifier Local Revocation}
	\acro{GS}{Group Signatures}
	\acro{GM}{Group Manager}
	\acro{GBA}{Generic Bootstrapping Architecture}
	\acro{GUI}{Graphic User Interface}
	\acro{HSM}{Hardware Security Module}
	\acro{HTTP}{Hypertext Transfer Protocol}
	\acro{IEEE}{Institute of Electrical and Electronics Engineers}
	\acro{IETF}{Internet Engineering Task Force}
	\acro{IoT}{Internet of Things}
	\acro{ITS}{Intelligent Transport System}
	\acro{IT}{Information Technologies}
	\acro{IMSI}{International Mobile Subscriber Identity}
	\acro{IMEI}{International Mobile Station Equipment Identity}
	\acro{IdP}{Identity Provider}
	\acro{IDS}{Intrusion Detection System}
	\acro{ISP}{Internet Service Provider}
	\acro{LEA}{Law Enforcement Agency}
	\acro{LCPP}{Lightweight Conditional Privacy Preservation}
	\acro{LTC}{Long Term Certificate}
	\acro{LTCA}{Long Term \acs{CA}}
	\acro{H-LTCA}{Home-LTCA}
	\acro{F-LTCA}{Foreign-LTCA}
	\acro{LDAP}{Lightweight Directory Access Protocol}
	\acro{LBS}{Location Based Service}
	\acro{LTE}{Long Term Evolution}
	\acro{LuST}{Luxembourg SUMO Traffic}
	\acro{MAC}{Message Authentication Code}
	\acro{MCA}{Message \ac{CA}}
	\acro{MEA}{Misbehavior Evaluation Authority}
	\acro{OBU}{On-Board Unit}
	\acro{OEM}{Original Equipment Manufacturer}
	\acro{OCSP}{Online Certificate Status Protocol}
	\acro{PCA}{Pseudonym \acs{CA}}
	\acro{PDP}{Policy Decision Point}
	\acro{PEP}{Policy Enforcement Point}
	\acro{PIR}{Private Information Retrieval}
	\acro{PKC}{Public Key Cryptography}
	\acro{PKCS}{Public Key Cryptosystem}
	\acro{PKI}{Public-Key Infrastructure}
	\acro{PRECIOSA}{Privacy Enabled Capability in Co-operative Systems and Safety Applications}
	\acro{PRESERVE}{Preparing Secure Vehicle-to-X Communication Systems}
	\acro{P2P}{peer-to-peer}
	\acro{PS}{Participatory Sensing}
	\acro{RA}{Resolution Authority}
	\acro{REST}{Representational State Transfer}
	\acro{RBAC}{Role Based Access Control}
	\acro{RCA}{Root \acs{CA}}
	\acro{RSU}{Roadside Unit}
	\acro{SAML}{Security Assertion Markup Language}
	\acro{SAS}{Sample Aggregation Service}
	\acro{SCMS}{Security Credential Management System}
	\acro{SCORE@F}{Système COopératif Routier Expérimental Français}
	\acro{SDSI}{Simple Distributed Security Infrastructure}
	\acro{SRAAC}{Secure Revocable Anonymous Authenticated Inter-Vehicle Communication}
	\acro{SeVeCom}{Secure Vehicle Communication}
	\acro{SIT}{Sichere Informationstechnologie}
	\acro{SLC}{Short-Lived Certificate}
	\acro{SoA}{Service-oriented-Approach}
	\acro{SIFS}{Short Inter Frame Space}
	\acro{SSO}{Single-Sign-On}
	\acro{SSL}{Secure Sockets Layer}
	\acro{SOAP}{Simple Object Access Protocol}
	\acro{TACK}{Temporary Anonymous Certified Key}
	\acro{TS}{Task Service}
	\acro{TLS}{Transport Layer Security}
	\acro{TPM}{Trusted Platform Module}
	\acro{TTP}{Trusted Third Party}
	\acro{TVR}{Ticket Validation Repository}
	\acro{URI}{Uniform Resource Identifier}
	\acro{UML}{Unified Modeling Language}
	\acro{VANET}{Vehicular Ad-hoc Network}
	\acro{V2I}{Vehicle-to-Infrastructure}
	\acro{V2V}{Vehicle-to-Vehicle}
	\acro{V2X}{\ac{V2V} and \ac{V2I}}
	\acro{VC}{Vehicular Communication}
	\acro{VM}{Virtual Machine}
	\acro{VSS}{\ac{VC} Security Subsystem}
	\acro{WAVE}{Wireless Access in Vehicular Environments}
	\acro{WSDL}{Web Services Discovery Language}
	\acro{W3C}{World Wide Web Consortium}
	\acro{V}{Vehicle}
	\acro{VANET}{Vehicular Ad-hoc Network}
	\acro{VLR}{Verifier-Local Revocation}
	\acro{VPKI}{Vehicular Public-Key Infrastructure}
	\acro{VM}{Virtual Machine}
	\acro{WS}{Web Service}
	\acro{WoT}{Web of Trust}
	\acro{WSACA}{\ac{WAVE} Service Advertisement \ac{CA}}
	\acro{XML}{Extensible Markup Language}
	\acro{XACML}{eXtensible Access Control Markup Language}
	\acro{ZKP}{Zero Knowledge Proof}
	\acro{3G}{3rd Generation}
\end{acronym}

\usepackage{makecell}

\usepackage{algorithm}

\usepackage{arydshln}

\usepackage{setspace}

\usepackage {xcolor}

\usepackage{footnote}

\usepackage{rotating}

\usepackage{soul}

\begin{document}
\title{SECMACE: Scalable and Robust Identity and Credential Management Infrastructure in Vehicular Communication Systems}

\author{\IEEEauthorblockN{Mohammad Khodaei, Hongyu Jin, and Panos Papadimitratos\\}
\IEEEauthorblockA{Networked Systems Security Group, KTH Royal Institute of Technology, Stockholm, Sweden \\
\emph{\{khodaei, hongyuj, papadim\}}@kth.se}}

\IEEEtitleabstractindextext{
\begin{abstract}
Several years of academic and industrial research efforts have converged to a common understanding on fundamental security building blocks for the upcoming \ac{VC} systems. There is a growing consensus towards deploying a special-purpose identity and credential management infrastructure, i.e., a \ac{VPKI}, enabling pseudonymous authentication, with standardization efforts towards that direction. In spite of the progress made by standardization bodies (IEEE 1609.2 and \acs{ETSI}) and harmonization efforts (\ac{C2C-CC}), significant questions remain unanswered towards deploying a \ac{VPKI}. Deep understanding of the \ac{VPKI}, a central building block of secure and privacy-preserving \ac{VC} systems, is still lacking. This paper contributes to the closing of this gap. We present SECMACE, a \ac{VPKI} system, which is compatible with the IEEE 1609.2 and \acs{ETSI} standards specifications. We provide a detailed description of our state-of-the-art \ac{VPKI} that improves upon existing proposals in terms of security and privacy protection, and efficiency. SECMACE facilitates multi-domain operations in the \ac{VC} systems and enhances user privacy, notably preventing linking \emph{pseudonyms} based on timing information and offering increased protection even against \emph{honest-but-curious} \ac{VPKI} entities. We propose multiple policies for the vehicle-\ac{VPKI} interactions, based on which and two large-scale mobility trace datasets, we evaluate the full-blown implementation of SECMACE. With very little attention on the \ac{VPKI} performance thus far, our results reveal that modest computing resources can support a large area of vehicles with very low delays and the most promising policy in terms of privacy protection can be supported with moderate overhead.
\end{abstract}

}

\maketitle

\IEEEdisplaynontitleabstractindextext

\section{Introduction}
\label{sec:secmace-introduction}

\acf{VC} systems can generally enhance transportation safety and efficiency with a gamut of applications, ranging from collision avoidance alerts to traffic conditions updates; moreover, they can integrate and enrich \acp{LBS}~\cite{ETSI-102-638, papadimitratos2009vehicular} and vehicular social networks~\cite{jin2016security}, and provide infotainment services. In \ac{VC} systems, vehicles are provided with \acp{OBU} to communicate with each other (\ac{V2V} communication), or with \acp{RSU} (\ac{V2I} communication). The deployment of such a large-scale system cannot materialize unless the security and privacy of the users are safeguarded~\cite{papadimitratos2006securing}. Standardization bodies (IEEE 1609.2 WG~\cite{1609-2016} and \acs{ETSI}~\cite{ETSI-102-638}) and harmonization efforts (\acf{C2C-CC}~\cite{c2c}) have reached a consensus to use \ac{PKC} to protect the \acl{V2X} communications~\cite{papadimitratos2008secure}: a set of Certification Authorities (CAs), constituting the \acf{VPKI}, provide credentials to legitimate vehicles; each vehicle is provided with a \ac{LTC} (and has the corresponding private key) to ensure accountable identification of the vehicle. To achieve unlinkability of messages originating the vehicle, a set of short-lived anonymized certificates, termed \emph{pseudonyms}, are used, along with the corresponding short-term private keys. The system maintains a mapping of these short-term identities to the vehicle long-term identity for accountability. Such ideas were elaborated by the \ac{SeVeCom} project~\cite{papadimitratos2008secure, leinmuller2006sevecom} and subsequent projects, e.g., \ac{CAMP VSC3}~\cite{whyte2013security} and \ac{PRESERVE}~\cite{preserve-url}, as well as technical standards, notably the IEEE 1609.2 WG~\cite{1609-2016}, \acs{ETSI}~\cite{ETSI-102-638}, and harmonization documents (\acs{C2C-CC}~\cite{c2c}).

In multi-domain \ac{VC} systems, each vehicle is registered with one \ac{LTCA}, responsible for issuing its \ac{LTC}, and it is able to obtain pseudonyms from any \ac{PCA}, a pseudonym provider. Vehicles digitally sign transmitted messages, e.g., \acp{CAM} or \acp{DENM}, with the private key, $k^i_v$, that corresponds to a currently valid pseudonym, $P^i_v$. The pseudonym is then attached to the signed messages to facilitate verification by any recipient. Upon reception, the pseudonym is verified (presuming a trust relationship with the pseudonym provider) before the message itself (signature validation). This ensures authenticity and integrity of the message and non-repudiation. Vehicles switch from one pseudonym (and the corresponding private key) to another one (ideally, non-previously used) to ensure message unlinkability (pseudonyms per se are inherently unlinkable if they are issued appropriately as it will become clear later). 

We propose SECMACE, a \ac{VPKI} system compatible with standards, which improves the state-of-the-art both in terms of security and privacy protection, as well as extensive evaluations of the system. In the following, we describe four technical aspects of our system that improves over the state-of-the-art. The \ac{VPKI} entities are, often implicitly, assumed to be fully trustworthy. Given the experience from recent mobile applications, we need to extend the adversarial model from fully trustworthy to \emph{honest-but-curious} \ac{VPKI} servers: they are \emph{honest}, i.e., thoroughly complying with the best practices, specified protocols, and system policies, but \emph{curious}, i.e., tempted to infer sensitive user information, thus harming user privacy. In the context of \ac{VC} systems, an \ac{LTCA} should not know which \ac{PCA} is targeted, and which pseudonyms are obtained by which vehicles, and for which period; the \ac{PCA} also should not be able to identify the real identity of the vehicles, or link successive pseudonym requests to a single vehicle. 

We propose a system that prevents misuse of the credentials, in particular towards a Sybil-based~\cite{douceur2002sybil} misbehavior: the acquisition of multiple simultaneously valid pseudonyms enables an attacker to inject multiple erroneous hazard notifications as if they were originated from multiple vehicles, thus misleading the system. In the light of a multi-domain \ac{VC} systems with a multiplicity of \acp{PCA}, each vehicle could obtain pseudonyms from any \ac{PCA}. This enables a compromised vehicle to obtain multiple sets of pseudonyms, valid simultaneously, from different \acp{PCA}, thus operating as a Sybil node. A general remedy to mitigate such a misbehavior~\cite{papadimitratos2007architecture} is to issue the pseudonyms with non-overlapping lifetimes and equip the vehicles with \acp{HSM}, using which all outgoing signatures are guaranteed to be signed under the private key of a single valid pseudonym at any time. However, our \ac{VPKI} design, per se, prevents Sybil-based misbehavior in a multi-domain \ac{VC} system without presuming trusted hardware. 

We further ensure that the pseudonyms themselves are not inherently linkable based on the timing information: a transcript of pseudonymously authenticated messages could be linked simply based on the pseudonym lifetime and issuance times~\cite{khodaei2014ScalableRobustVPKI}, and requests could act as user \emph{``fingerprints''}~\cite{khodaei2016evaluating}. Simply put, individually determined pseudonym lifetimes allow an observer to link pseudonyms of the same vehicle only by inspecting the successive pseudonym lifetimes (without even examining the content of the message). To mitigate this threat, we propose a privacy-preserving policy so that the timing information does not harm user privacy. 

We provide three generally applicable policies, including a privacy-preserving one, for vehicle-\ac{VPKI} interactions with a realistic evaluation of the workload that a \ac{VPKI} would face. The workload on the \ac{VPKI} servers mainly depends on the frequency of vehicle-\ac{VPKI} interactions and the duration for which vehicles request pseudonyms for. Towards dimensioning our \ac{VPKI}, we investigate the overall effects of these policies on the actual implementation of our \ac{VPKI} and we provide an extensive analysis of the suitability of different representative policies for vehicle-\ac{VPKI} interactions. We demonstrate that SECMACE, as the most promising \ac{VPKI} in terms of performance combined with the most promising policy in terms of privacy, introduces a very modest overhead. Eventually, any vehicle could obtain all required pseudonyms within a short delay, practically in real-time, for its participation in the system. 

SECMACE, a comprehensive security and privacy-preserving architecture for \ac{VC} systems, contributes a set of novel features: (i) multi-domain operation, (ii) increased user privacy protection, in the presence of honest-but-curious system entities even with limited collusion, and by eliminating pseudonym linking based on timing information, (iii) thwarting Sybil-based misbehavior, and (iv) multiple pseudonym acquisition policies. Beyond these features, we provide an extensive survey of the prior art and a detailed security and privacy analysis of our system. We further provide an extensive evaluation of the overall system performance including alternative pseudonym acquisition policies, and assessing its efficiency, scalability, and robustness based on an implementation of our \ac{VPKI} and two large-scale mobility traces. 

In the rest of the paper, we describe the related work (Sec.~\ref{sec:secmace-related-work}) and the problem statement (Sec.~\ref{sec:secmace-problem-statement}). We then explain our system entities and model with detailed security protocols (Sec.~\ref{sec:secmace-system-entities-design}). We describe the security and privacy analysis (Sec.~\ref{sec:secmace-security-and-privacy-analysis}), followed by the extensive experimental evaluation (Sec.~\ref{sec:secmace-vpki-servers-performance-evaluation}) before the conclusion (Sec.~\ref{sec:secmace-conclusions}). 

\section{Related Work}
\label{sec:secmace-related-work}

\setlength\dashlinedash{0.9pt}
\setlength\dashlinegap{2.5pt}
\setlength\arrayrulewidth{0.9pt}


	\begin{table*} [!thb]
		\caption{\acs{VPKI} Security Features and Properties Comparison (\checkmark: support, $\times$: no support)}
	    \centering
	    \resizebox{18.2cm}{!} {
	   	\begin{tabular}{|l||*{12}{c|}}\hline
	   		\rowcolor{gray!40}
	   		\backslashbox{\textbf{Schemes}}{\textbf{Properties}}
	   		& \makebox[4.5em]{\centering \shortstack{{\normalsize \textbf{IEEE 1609.2}} \\ {\normalsize \textbf{compliance}} \\ {}}}
	   		& \makebox[3.8em]{\centering \shortstack{{\normalsize \textbf{\acs{ETSI}}} \\ {\normalsize \textbf{compliance}} \\ {}}} 
	   		& \makebox[4.1em]{\centering \shortstack{{\normalsize \textbf{Long-term}} \\ {\normalsize \textbf{identifier}} \\ {}}}
	   		& \makebox[3em]{\centering \shortstack{{\normalsize \textbf{Multiple}} \\ {\normalsize \textbf{domain}} \\ {}}}
	   		& \makebox[3em]{\centering \shortstack{\\ {\normalsize \textbf{Sybil}} \\ {\normalsize \textbf{resilience}} \\ {}}}
	   		& \makebox[4.8em]{\centering \shortstack{\\{\normalsize \textbf{Cryptosystem}} \\ {\normalsize} \\ {}}} 
	   		& \makebox[2.2em]{\centering \shortstack{\\{\normalsize \textbf{Cryptographic algorithms}} {\scriptsize } \\ {} \\ {}}}
	   		& \makebox[4em]{\centering \shortstack{{\normalsize \textbf{Revocation}} \\ {} \\ {}\\ {} }}
	   		& \makebox[5.2em]{\centering \shortstack{{\normalsize \textbf{Accountability}} \\ {} \\ {}}}
	   		& \makebox[3em]{\centering \shortstack{\vspace{0.5em} \\ {} {\normalsize \textbf{Perfect}} \\ {\normalsize \textbf{forward}} \\ {\normalsize \textbf{privacy}}}}
	   		& \makebox[1.5em]{\centering \shortstack{{\normalsize \textbf{Refilling strategy}}\\ {}  \\ {}}}
	   		& \makebox[3.5em]{\centering \shortstack{{\normalsize \textbf{V$\leftrightarrow$\acs{VPKI} communication}} \\ {} \\ {} {\scriptsize}}} \\\hline\hline
	
	   		\rowcolor{gray!20}
	   		\shortstack{\textbf{Fischer et al. (\acs{SRAAC})~\cite{fischer2006secure}} \\ {} } & \shortstack{$\times$ \\ {}} & \shortstack{$\times$ \\ {}} & \shortstack{Certificate \\ {}} & \shortstack{$\times$ \\ {}} & \shortstack{$\times$ \\ {}} & \shortstack{\acs{PKC} \\ {}} & {\centering \shortstack{Magic-ink signature \\ with DSS}} & \shortstack{\checkmark \\ {}} & \shortstack{\checkmark \\ {}} & \shortstack{$\times$ \\ {}} & \shortstack{On-demand \\ {}} & {\centering \shortstack{Blind signature \\ without confidentiality}} \\ \hdashline 
	
			\shortstack{\textbf{Sha et al.~\cite{sha2006adaptive}} \\ {}} & \shortstack{$\times$ \\ {}} & \shortstack{$\times$ \\ {}} & \shortstack{Certificate \\ {}} & \shortstack{\checkmark \\ {}} & \shortstack{$\times$ \\ {}} & \shortstack{\acs{PKC} \\ {}} & \shortstack{\acs{ECC} public key \\ cryptography and RSA} & \shortstack{\checkmark \\ {}} & \shortstack{\checkmark \\ {} \\ {}} & \shortstack{\textbf{\textemdash} \\ {}} & \shortstack{On-demand \\ {}} & \shortstack{Symmetric-key cryptography \\ (session key)} \\\hdashline 

			\rowcolor{gray!20}
	  		\textbf{\acs{SeVeCom}~\cite{papadimitratos2008secure, kargl2008secure}} & $\times$ & $\times$ & \shortstack{Certificate} & \shortstack{$\times$} & \checkmark & \acs{PKC} & \acs{ECDSA} & \checkmark & \checkmark & $\times$ & Preloading & Secure wireline communication \\ \hdashline 
	   		
	   		\textbf{\acs{C2C-CC} pilot \acs{PKI}~\cite{c2c}} & \checkmark & \checkmark & \shortstack{Certificate} & \shortstack{\checkmark} & \shortstack{$\times$} & \acs{PKC} & \acs{ECDSA} &$\times$ & \checkmark & $\times$ & Preloading & \acs{DL/ECIES} over UDP \\ \hdashline 

			\rowcolor{gray!20}
	   		\shortstack{\textbf{Studer et al. (\acs{TACK})~\cite{studer2009tacking}} \\ {}} & \shortstack{$\times$ \\ {}} & \shortstack{$\times$ \\ {}} & \shortstack{\\ {} Group \\ user key} & \shortstack{\checkmark \\ {}} & \shortstack{$\checkmark$ \\ {}} & \shortstack{GS/\acs{PKC} \\ {}} & \shortstack{\acs{ECDSA} and \acs{VLR} GS \\ {}} & \shortstack{\checkmark \\ {}} & \shortstack{\checkmark \\ {}} & \shortstack{\checkmark \\ {}} & \shortstack{On-demand \\ {}} & {\centering \shortstack{GS without confidentiality \\ {}}} \\\hdashline 

	  		\shortstack{\textbf{Schaub et al. (V-tokens)~\cite{schaub2010v}} \\ {} \\ {} \\ {}} & \shortstack{$\times$ \\ {} \\ {} \\ {}} & \shortstack{$\times$ \\ {} \\ {} \\ {}} & \shortstack{Certificate \\ {} \\ {} \\ {}} & \shortstack{\checkmark \\ {} \\ {} \\ {}} & \shortstack{\checkmark \\ {} \\ {} \\ {}} & \shortstack{\acs{PKC} \\ {} \\ {} \\ {}} & \shortstack{\textemdash \\ {} \\ {} \\ {}} & \shortstack{$\times$ \\ {} \\ {} \\ {}} & \shortstack{\checkmark \\ {} \\ {} \\ {}} & \shortstack{\textbf{\textemdash} \\ {} \\ {} \\ {}} & \shortstack{On-demand \\ {} \\ {} \\ {}} & {\centering \shortstack{Blind signature scheme \& \\ anonymous communication \\ channel (e.g., onion routing)}} \\\hdashline 
	
			\rowcolor{gray!20}
	  		\textbf{Alexiou et al. (VeSPA)~\cite{khodaei2012secure, vespa2013}} & \checkmark & $\times$ & \shortstack{Certificate} & $\checkmark$ & $\times$ & \acs{PKC} & \acs{ECDSA} & \checkmark & \checkmark & $\times$ & On-demand & \acs{SSL}/\acs{TLS} \\\hdashline 
	
	   		\textbf{\acs{PRESERVE}~\cite{preserve-url}} & \checkmark & \checkmark & \shortstack{Certificate} & $\times$ & $\times$ & \acs{PKC} & \acs{ECDSA} & \checkmark & \checkmark & $\times$ & Preloading & \acs{DL/ECIES} over UDP \\ \hdashline 

			\rowcolor{gray!20}
	  		\textbf{Bi{\ss}meyer et al. (CoPRA)~\cite{bibmeyer2013copra}} & \checkmark & \checkmark & \shortstack{Certificate} & \checkmark & $\times$ & \acs{PKC} & \acs{ECDSA} \& ECIES & $\times$ & \checkmark & $\times$ & On-demand & \acs{DL/ECIES} over UDP \\\hdashline 
	  		
	   		\textbf{Gisdakis et al. (SEROSA)~\cite{gisdakis2013serosa}} & \checkmark & $\times$ & \shortstack{Certificate} & \checkmark & $\times$ & \acs{PKC} & \acs{ECDSA} & \checkmark & \checkmark & $\times$ & On-demand & \acs{SSL}/\acs{TLS} \\\hdashline 
	
			\rowcolor{gray!20}
	  		\shortstack{\textbf{Whyte et al. (\acs{CAMP VSC3})~\cite{whyte2013security, US-VPKI}} \\ {}} & \shortstack{\checkmark \\ {}} & \shortstack{$\times$ \\ {}} & \shortstack{Certificate \\ {}} & \shortstack{\checkmark \\ {}} & \shortstack{$\times$ \\ {}} & \shortstack{\acs{PKC} \\ {}} & {\centering \shortstack{\\ {} Butterfly key \\ expansion cryptography}} & \shortstack{\checkmark \\ {}} & \shortstack{\checkmark \\ {} } & \shortstack{\checkmark \\ {}} & \shortstack{Preloading \\ {}} & {\centering \shortstack{Asymmetric cryptography \\ over UDP }} \\\hdashline 
	   		
	   		\shortstack{\textbf{F{\"o}rster et al. (PUCA)~\cite{puca2014}} \\ {} \\ {} \\ {}} & \shortstack{$\times$ \\ {} \\ {} \\ {}} & \shortstack{$\times$ \\ {} \\ {} \\ {}} & \shortstack{Certificate \\ {} \\ {} \\ {}} & \shortstack{\checkmark \\ {} \\ {} \\ {}} & \shortstack{\checkmark \\ {} \\ {} \\ {}} & \shortstack{\acs{ZKP}/\acs{PKC} \\ {} \\ {} \\ {}} & \shortstack{\\ {} \acs{ECDSA}, Dynamic \\ accumulator and \\ CL signature} & \shortstack{$\times$ \\ {} \\ {} \\ {}} & \shortstack{$\times$ \\ {} \\ {} \\ {}} & \shortstack{$\times$ \\ {} \\ {} \\ {}} & \shortstack{On-demand \\ {} \\ {} \\ {}} & \shortstack{\acs{SSL}/\acs{TLS} over Tor \\ {} \\ {} \\ {}} \\\hdashline 
	   		
			\rowcolor{gray!20}
	   		\textbf{Khodaei et al. (SECMACE)} & \checkmark & $\times$ & \shortstack{Certificate} & \checkmark & \checkmark & \acs{PKC} & \acs{ECDSA} & \checkmark & \checkmark & \shortstack{\checkmark} & On-demand & \acs{SSL}/\acs{TLS} \\\hline 

			\shortstack{\textbf{Sun et al.~\cite{sun2007secure}} \\ {} \\ {} \\ {}} & \shortstack{$\times$ \\ {} \\ {}} & \shortstack{$\times$ \\ {} \\ {} \\ {}} & \shortstack{\\ {} Group \\ user key \\ {}} & \shortstack{$\times$ \\ {} \\ {} \\ {}} & \shortstack{$\times$ \\ {} \\ {} \\ {}} & \shortstack{GS \\ {} \\ {} \\ {}} & \shortstack{\\ {} Short GS with \\ provably-secure ID-based \\ signature scheme} & \shortstack{\checkmark \\ {} \\ {} \\ {}} & \shortstack{\checkmark \\ {} \\ {} \\ {}} & \shortstack{\checkmark \\ {} \\ {} \\ {}} & \shortstack{On-demand \\ {} \\ {} \\ {}} & \shortstack{\textemdash \\ {} \\ {} \\ {}} \\\hdashline 
	
			\rowcolor{gray!20}
			\shortstack{\textbf{Guo et al.~\cite{guo2007group}} \\ {} \\ {} \\ {}} & \shortstack{$\times$ \\ {} \\ {} \\ {}} & \shortstack{$\times$ \\ {} \\ {} \\ {}} & \shortstack{\\ {} Group \\ user key \\ {}} & \shortstack{\checkmark \\ {} \\ {} \\ {}} & \shortstack{$\times$ \\ {} \\ {} \\ {}} & \shortstack{GS \\ {} \\ {}} & \shortstack{\\ {} Short GS with optimization: \\ probabilistic verification \\ of signatures} & \shortstack{\checkmark \\ {} \\ {} \\ {}} & \shortstack{\checkmark \\ {} \\ {} \\ {}} & \shortstack{\checkmark \\ {} \\ {} \\ {}} & \shortstack{Annual preloading \\ {} \\ {} \\ {}} & \shortstack{Off-line annual renewal keys \\ {} \\ {} \\ {}} \\\hdashline 
	
			\shortstack{\textbf{Lin et al. (GSIS)~\cite{lin2007gsis}} \\ {}} & \shortstack{$\times$ \\ {}} & \shortstack{$\times$ \\ {}} & \shortstack{\\ {} Group \\ user key} & \shortstack{\checkmark \\ {}} & \shortstack{$\times$ \\ {}} & \shortstack{GS} & {\centering \shortstack{GS and \\ ID-based signature}} & \shortstack{\checkmark \\ {}} & \shortstack{\checkmark \\ {}} & \shortstack{\checkmark \\ {}} & \shortstack{On-demand \\ {}} & \shortstack{\textemdash \\ {}} \\\hdashline 

			\rowcolor{gray!20}
			\shortstack{\textbf{Wasef et al. (ECMV)~\cite{wasef2008ecmv}} \\ {}} & \shortstack{$\times$ \\ {}} & \shortstack{$\times$ \\ {}} & \shortstack{\\ {} Group \\ user key} & \shortstack{\checkmark \\ {}} & \shortstack{$\times$ \\ {}} & \shortstack{GS \\ {}} & \shortstack{Bilinear \& \\ ID-based cryptography} & \shortstack{\checkmark \\ {}} & \shortstack{\checkmark \\ {}} & \shortstack{\checkmark \\ {}} & \shortstack{On-demand \\ {}} & \shortstack{Encrypted tunnel leveraging \\ asymmetric cryptography} \\\hdashline 
	
			\shortstack{\textbf{Wasef et al. (PPGCV)~\cite{wasef2008ppgcv}} \\ {}} & \shortstack{$\times$ \\ {}} & \shortstack{$\times$ \\ {}} & \shortstack{\\ {} Group \\ user key} & \shortstack{$\times$ \\ {}} & \shortstack{$\times$ \\ {}} & \shortstack{GS} & \shortstack{Group key \\ cryptography} & \shortstack{\checkmark \\ {}} & \shortstack{\checkmark \\ {}} & \shortstack{\checkmark \\ {}} & \shortstack{On-demand \\ {}} & \shortstack{\textemdash \\ {}} \\\hdashline

			\rowcolor{gray!20}
			\shortstack{\textbf{Lu et al. (ECPP)~\cite{lu2008ecpp}} \\ {}} & \shortstack{$\times$ \\ {}} & \shortstack{$\times$ \\ {}} & \shortstack{\\ {} Group \\ user key} & \shortstack{\checkmark \\ {}} & \shortstack{$\times$ \\ {}} & \shortstack{GS \\ {}} & {\centering \shortstack{GS \& bilinear \\ pairing-based cryptography}} & \shortstack{\checkmark \\ {}} & \shortstack{\checkmark \\ {}} & \shortstack{\checkmark \\ {}} & \shortstack{On-demand \\ {}} & \shortstack{Non-secure 5.9 GHz \acs{DSRC} \\ {}} \\\hline 
	
			\shortstack{\textbf{Calandriello et al.~\cite{calandriello2011performance}} \\ {}} & \shortstack{$\times$ \\ {}} & \shortstack{$\times$ \\ {}} & \shortstack{\\ {} Group \\ user key} & \shortstack{$\times$ \\ {}} & \shortstack{\checkmark \\ {}} & \shortstack{Hybrid \\ {}} & \shortstack{\acs{ECDSA} and \acs{VLR} GS \\ {}} & \shortstack{\checkmark \\ {}} & \shortstack{\checkmark \\ {}} & \shortstack{\checkmark \\ {}} & \shortstack{On-demand \\ {}} & \shortstack{\acs{SSL}/\acs{TLS} \\ {}} \\\hline  

	   	\end{tabular}
	    \label{table:secmace-VPKI-properties-comparison}
	    }
	  \vspace{-0.5em}
	\end{table*}

Pseudonymous authentication was elaborated by \acs{SeVeCom}~\cite{papadimitratos2008secure, leinmuller2006sevecom}, \acs{PRESERVE}~\cite{preserve-url}, \acs{CAMP VSC3}~\cite{whyte2013security, US-VPKI}, standardization bodies (IEEE 1609.2 and \acs{ETSI}), and harmonization efforts (\acs{C2C-CC}~\cite{c2c}). Several proposals follow the \acs{C2C-CC} architecture, e.g., \acs{PRESERVE}~\cite{preserve-url, bibmeyer2013copra}, entailing direct \ac{LTCA}-\ac{PCA} communication during the pseudonym acquisition process. This implies that the \ac{LTCA} can learn the targeted \ac{PCA}. As a consequence, the \ac{LTCA} can link the real identity of the vehicle with its corresponding pseudonyms based on the timing information~\cite{khodaei2014ScalableRobustVPKI}: the exact time of request could be unique, or one of few, and thus linkable by the \ac{LTCA}, as it might be unlikely in a specific region to have multiple requests at a specific instance.

A ticket based approach was proposed in~\cite{khodaei2012secure, vespa2013, gisdakis2013serosa}: the \ac{LTCA} issues authenticated, yet anonymized, tickets for the vehicles to obtain pseudonyms from a \ac{PCA}. There is no direct \ac{LTCA}-\ac{PCA} communication and the \ac{PCA} does not learn any user-related information through the pseudonym acquisition process. However, the \ac{LTCA} can still learn when and from which \ac{PCA} the vehicle shall obtain pseudonyms during the ticket issuance phase, because this information will be presented (by the vehicle) and will be included in the authenticated ticket (by the \ac{LTCA}). The pseudonym acquisition period can be used to infer the active vehicle operation period, and the targeting \ac{PCA} could be used to infer a rough location (assuming the vehicle chooses the nearest \ac{PCA}) or the affiliation (assuming the vehicle can only obtain pseudonyms from the \ac{PCA} it is affiliated to, or operating in) of the vehicle. 

A common issue for all schemes proposed in the literature is that the \ac{PCA} can trivially link the pseudonyms issued for a vehicle as a response to a single pseudonym request~\cite{studer2009tacking, schaub2010v, khodaei2012secure, vespa2013, bibmeyer2013copra, gisdakis2013serosa, puca2014}. \acs{CAMP VSC3}~\cite{whyte2013security, US-VPKI} proposes a proxy-based scheme that the registration authority (a proxy to validate, process, and forward pseudonym requests to the \ac{PCA}) aggregates and shuffles all requests within a large period of time before forwarding them to the \ac{PCA}, so that the \ac{PCA} cannot identify which pseudonyms belong to which vehicles. Our system can also be configured to prevent an honest-but-curious \ac{PCA} from linking a set of pseudonyms issued for a vehicle (as discussed in Sec.~\ref{sec:secmace-security-and-privacy-analysis}).
	
The idea of enforcing non-overlapping pseudonym lifetimes was first proposed in~\cite{papadimitratos2007architecture}. The motivation is to prevent an adversary from equipping itself with multiple valid identities, and thus affecting protocols of collection of multiple inputs, e.g., based on voting, by sending out redundant false, yet authenticated, information, e.g., fake traffic congestion alerts, or fake misbehavior detection votes~\cite{raya2007eviction}. Though this idea has been accepted, a number of proposals~\cite{khodaei2012secure, vespa2013, bibmeyer2013copra, gisdakis2013serosa} do not prevent a vehicle from obtaining simultaneously valid pseudonyms via multiple pseudonym requests. The existence of multiple \acp{PCA} deteriorate the situation: a vehicle could request pseudonyms from multiple \acp{PCA}, e.g., by requesting multiple tickets from the \ac{LTCA} or reusing a ticket, while each \ac{PCA} is not aware whether pseudonyms for the same period were issued by any other \ac{PCA}.~\cite{schaub2010v} prevents a vehicle from obtaining multiple simultaneously valid pseudonyms by enabling the \acp{PCA} communicating with each other, e.g., a distributed hash table. SECMACE (including its predecessor work~\cite{khodaei2014ScalableRobustVPKI, khodaei2016evaluating}) prevents Sybil-based misbehavior on the infrastructure side without the need for an additional entity, i.e., extra interactions or intra-\ac{VPKI} communications. More specifically, it ensures that each vehicle can only have one valid pseudonym at any time in a multi-domain environment: the \ac{LTCA} maintains a record of ticket acquisitions for each vehicle, thus preventing a vehicle from obtaining multiple simultaneously valid tickets. 

Beyond the standards specifications (classic \ac{PKC}), there have been proposals to use anonymous authentication by leveraging \ac{GS}~\cite{boneh2004group, boneh2004short}. Each group member is equipped with a group public key, common to all the group members, and a distinct group signing key. Group signing keys can be used to sign messages and these signatures can be verified with the group public key. The signer is kept anonymous as its signatures (even the signatures of two identical messages) cannot be linked. A group signing key itself can be used to sign outgoing messages~\cite{lin2007gsis}. However, \ac{GS} themselves exhibit high computational delay to sign \ac{VC} messages~\cite{calandriello2011performance}. For example, the signing delay with \ac{GS-VLR}~\cite{boneh2004group} is around 67 times higher than that with the \ac{ECDSA}-256, and the verification delay with the former one is around 11 times higher than that of the latter (for the same security level, i.e., 128 bits)~\cite{calandriello2011performance}.~\cite{puca2014} proposes a fully anonymous scheme using \acfp{ZKP} for the vehicle-\ac{PCA} authentication with the consequence that compromised \acp{OBU} can be revoked only \emph{``manually''} with involvement of the owners. 

This naturally leads us to the protocol of a hybrid approach~\cite{studer2009tacking, calandriello2011performance, PapadiCLH:C:08}. In~\cite{calandriello2011performance}, a vehicle generates public/private key pairs and \emph{``self-certifies''} the public keys on-the-fly with its own group signing key to be used as the pseudonyms. Such schemes eliminate the need to request pseudonyms from the \ac{VPKI} repeatedly. Upon reception of messages signed under a new pseudonym, the \ac{GS} and the pseudonym are verified before the message itself (signature validation); if the pseudonym is cached, only the signatures of the messages need to be verified. Performance improvement relies on the lifetime of each pseudonym: the longer the pseudonym lifetime is, the less frequent pseudonym verifications are needed. However, this trades off the linkability of the messages that are signed under the same pseudonym. Moreover, allowing a vehicle to generate its own pseudonyms also makes Sybil-based misbehavior possible. In~\cite{studer2009tacking}, a vehicle requests a new pseudonym every time it enters a new region. A pseudonym request is signed with the group signing key of the vehicle, thus it is kept anonymous. Sybil-based misbehavior is prevented by fixing the random number (which is changed periodically) that used for \ac{GS} generations. Therefore, a vehicle cannot request two pseudonyms through two pseudonym requests with the same random number. However, it presumes that the random number should be negotiated and bound to the vehicle before requesting pseudonyms, while it is unclear how this can be done without disclosing the vehicle identity.

Table~\ref{table:secmace-VPKI-properties-comparison} shows a comparison of all, to the best of our knowledge, existing proposals for a \ac{VPKI} or its main building blocks with respect to their security properties. Only a few of the works evaluated the performance of their implementation while in the light of the \ac{VC} large-scale multi-domain environment, the efficiency and scalability of a \ac{VPKI} should be extensively evaluated. In this paper, we extensively evaluate the performance of the full-blown implementation of our \ac{VPKI}. Beyond the scope of this paper, SECMACE can be highly beneficial in other application domains, e.g., secure and privacy-preserving \ac{LBS} provision~\cite{jin2017resilient}.

\section{Problem Statement}
\label{sec:secmace-problem-statement}

\subsection{System Model and Assumptions}
\label{subsec:secmace-system-model-and-assumptions}

We assume that a \ac{VPKI} consists of a set of authorities with distinct roles: the \ac{RCA}, the highest-level authority, certifies other lower-level authorities; the \ac{LTCA} is responsible for the vehicle registration and the \ac{LTC} issuance; the \ac{PCA} issues pseudonyms for the registered vehicles; and the \ac{RA} is able to initiate a process to resolve a pseudonym, thus identifying the long-term identity of a (misbehaving, malfunctioning, or outdated~\cite{Papadi:C:08}) vehicle, i.e., the pseudonym owner. We further assume that each vehicle is only registered to its \emph{\ac{H-LTCA}}, the \emph{policy decision and enforcement point}, and is reachable by the registered vehicles in its \emph{domain}. A \emph{domain} is defined as a set of vehicles, registered with their \ac{H-LTCA}, subject to the same administrative regulations and policies~\cite{khodaei2015VTMagazine}. Each domain is governed by only one \ac{H-LTCA}, while there are several \acp{PCA} active in one or multiple domains. Each vehicle, depending on the policies and rules, can cross to \emph{foreign} domains and communicate with the \emph{\ac{F-LTCA}} towards obtaining pseudonyms. Trust between two domains can be established with the help of an \ac{RCA}, or through cross certification between them. All vehicles registered in the system are provided with \acp{HSM}, ensuring that private keys never leave the \ac{HSM}. We assume that the certificates of higher-level authorities are installed on the \acp{OBU}, which are loosely synchronized with the \ac{VPKI} servers. 

\subsection{Adversarial Model}
\label{subsec:secmace-adversarial-model}

We adhere to the assumed adversarial behavior defined in the literature~\cite{papadimitratos2006securing} and in this paper, we are primarily concerned with adversaries that seek to abuse the \ac{VPKI}. Nonetheless, we consider a stronger adversarial model: rather than assuming fully trustworthy \ac{VPKI} entities, we consider them to be \emph{honest-but-curious}. Such servers correctly execute the security protocols, but the servers function towards collecting or inferring user sensitive information based on the execution of the protocols. Such honest-but-curious \ac{VPKI} servers could link pseudonym sets provided to the users, through the \ac{VPKI} operations, e.g., pseudonyms issuance, thus, tracing vehicle activities. Our adversarial model considers multiple \ac{VPKI} servers collude, i.e., share information that each of them individually infers with the others, to harm user privacy. The nature of collusion can vary, e.g., depending on who is the owner or administrator of any two or more colluding servers. We analyze the effects of collusion by different \ac{VPKI} entities in Sec.~\ref{subsec:honest-but-curious-vpki-servers}.

In a multi-\ac{PCA} environment, \emph{internal adversaries}, i.e., malicious (compromised) clients, raise two challenges. First, they could obtain multiple simultaneously valid pseudonyms, thus misbehaving each as multiple registered legitimate-looking vehicles. Second, they can degrade the operations of the system by mounting a clogging \ac{DoS} attack against the \ac{VPKI} servers. \emph{External adversaries}, i.e. unauthorized entities, can try to harm the system operations by launching a \ac{DoS} (or a \ac{DDoS}), thus degrading the availability of the system. But they are unable to successfully forge messages or `crack' the employed cryptosystems and cryptographic primitives.

\subsection{Security and Privacy Requirements}
\label{subsec:secmace-security-and-privacy-requirements}

The security and privacy requirements for the \ac{V2X} communications are described in the literature~\cite{papadimitratos2006securing}. Here, we only focus on the security and privacy requirements on vehicle-\ac{VPKI} interactions, intra-\ac{VPKI} actions, and relevant requirements in the face of honest-but-curious \ac{VPKI} entities. 

\begin{itemize}
	\item \emph{R1. Authentication and communication integrity, and confidentiality:} All vehicle-\ac{VPKI} interactions should be authenticated, i.e., both interacting entities should corroborate the sender of a message and the liveness of the sender. We further need to ensure the communication integrity, i.e., exchanged messages should be protected from any alternation. To provide confidentiality, the content of sensitive information, e.g., exchanged messages between a vehicle and a \ac{VPKI} entity to obtain pseudonyms, should be kept secret from other entities. 

	\item \emph{R2. Authorization and access control:} Only legitimate, i.e., registered, and authenticated vehicles should be able to be serviced by the \ac{VPKI}, notably obtain pseudonyms. Moreover, vehicles should interact with the \ac{VPKI} entities according to the system protocols and policies, and domain regulations. 
	
	\item \emph{R3. Non-repudiation, accountability and eviction (revocation):} All relevant operation and interactions with the \ac{VPKI} entities should be non-repudiable, i.e., no entity should be able to deny having sent a message. Moreover, all legitimate system entities, i.e., registered vehicles and \ac{VPKI} entities, should be accountable for their actions that could interrupt the operation of the \ac{VPKI} or harm the vehicles. In case of any deviation from system policies, the misbehaving entities should be evicted from the system. 
	
	\item \emph{R4. Anonymity (conditional):} Vehicles should participate in the \ac{VC} system \emph{anonymously}, i.e., vehicles should communicate with others without revealing their long-term identifiers and credentials. Anonymity is conditional in the sense that the corresponding long-term identity can be retrieved by the \ac{VPKI} entities, and accordingly revoked, if a vehicle deviates from system policies, e.g., submitting faulty information.  
	
	\item \emph{R5. Unlinkability:} In order to achieve \emph{unlinkability}, the real identity of a vehicle should not be linked to its corresponding pseudonyms; in other words, the \ac{LTCA}, should know neither the targeted \ac{PCA} nor the actual pseudonym acquisition periods, nor the credentials themselves. Moreover, successive pseudonym requests should not be linked to the same requester and to each other. The \ac{PCA} should not be able to retrieve the long-term identity of any requester, or link multiple pseudonym requests (of the same requester). Furthermore, an external observer should not be able to link pseudonyms of a specific vehicle based on information they carry, notably their timing information\footnote{This does not relate to location information that vehicular communication messages, time- and geo-stamped signed under specific pseudonyms, carry.}. In order to achieve \emph{full unlinkability}, which results in perfect forward privacy, no single entity (even the \ac{PCA}) should be able to link a set of pseudonyms issued for a vehicle as a response to a single request. 
	
	The level of anonymity and unlinkability is highly dependent on the \emph{anonymity set}, i.e., the number of active participants and the resultant number of requests to obtain pseudonyms, e.g., all vehicles serviced by one \ac{PCA}; because pseudonyms carry the issuer information, the \ac{VPKI} should enhance user privacy by rendering any inference (towards linking, thus tracking, vehicles) hard. 

	\item \emph{R6. Thwarting Sybil-based attacks:} The \ac{VPKI} should not issue multiple simultaneously valid pseudonyms for any vehicle.

	\item \emph{R7. Availability:} The \ac{VPKI} should remain operational in the presence of benign failures (system faults or crashes) and be resilient to resource depletion attacks, e.g., \ac{DDoS} attack. 

\end{itemize}

\section{Security System Entities and Design}
\label{sec:secmace-system-entities-design}

\begin{figure}[t!]
    \centering
	\includegraphics[width=0.4\textwidth,height=0.4\textheight,keepaspectratio] {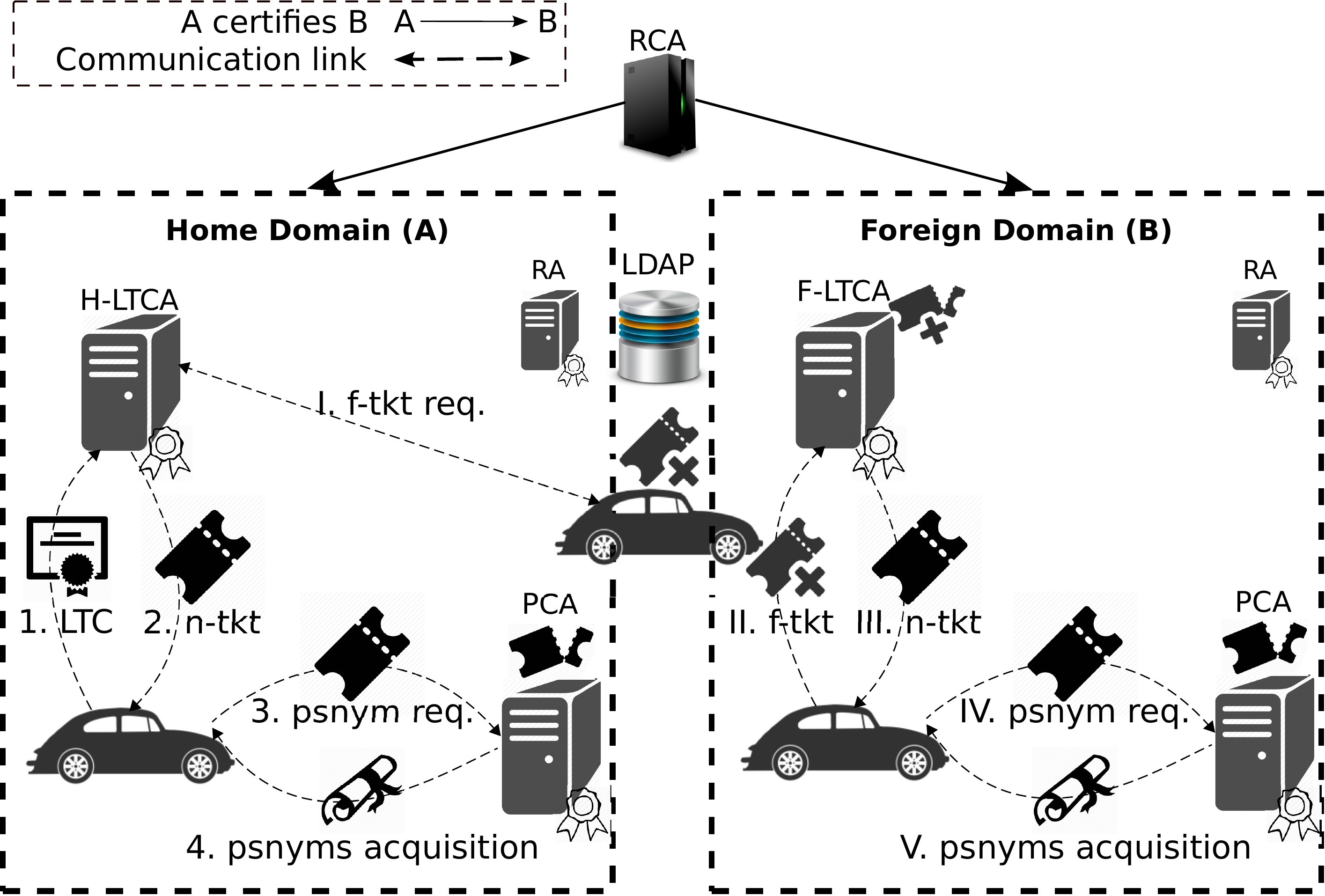}
	\caption{Pseudonym acquisition overview in the home and foreign domains.}
	\label{fig:secmace-system-overview-with-two-domains}
	\vspace{-0.5em}
\end{figure}

\subsection{System Overview}
\label{subsec:secmace-system-overview}

Fig.~\ref{fig:secmace-system-overview-with-two-domains} illustrates our \ac{VPKI} assuming two distinct domains: the home domain ($A$) and a foreign domain ($B$). In the registration phase, each \ac{H-LTCA} registers vehicles within its domain and maintains their long-term identities. At the bootstrapping phase, each vehicle needs to discover the \ac{VPKI}-related information, e.g., the available \acp{PCA} in its home domain, or the desired \ac{F-LTCA} and \acp{PCA} in a foreign domain, along with their corresponding certificates. To facilitate the overall intra-domain and multi-domain operations, a vehicle first finds such information from a \ac{LDAP}~\cite{sermersheim2006lightweight} server. This is carried out without disclosing the real identity of the vehicle. The vehicle, i.e., the \ac{OBU}, \emph{``decides''} when to trigger the pseudonym acquisition process based on different parameters, e.g., the number of remaining valid pseudonyms, the residual trip duration, and the networking connectivity. We presume connectivity to the \ac{VPKI}, e.g., via \acp{RSU}; should the connectivity be intermittent, the \ac{OBU} could initiate pseudonym provisioning proactively when there is connectivity. 

The \ac{H-LTCA} authenticates and authorizes vehicles, which authenticate the \ac{H-LTCA} over a mutually authenticated \ac{TLS}~\cite{dierks2008transport} tunnel. This way the vehicle obtains a \emph{native ticket} ($n\textnormal{-}tkt$) from its \ac{H-LTCA} while the targeted \ac{PCA} or the actual pseudonym acquisition period is hidden from the \ac{H-LTCA}; the ticket is anonymized and it does not reveal its owner's identity (Protocol~\ref{protocol:secmace-ticket-provision-in-home-domain}). The ticket is then presented to the intended \ac{PCA}, over a unidirectional (server-only) authenticated \ac{TLS}, for the vehicle to obtain pseudonyms (Protocol~\ref{protocol:secmace-pseudonym-provision-in-home-domain}). 

When the vehicle travels in a foreign domain, it should obtain new pseudonyms from a \ac{PCA} operating in that domain; otherwise, the vehicle would stand out with pseudonyms from another \ac{PCA}. The vehicle first requests a \emph{foreign ticket} ($f\textnormal{-}tkt$) from its \ac{H-LTCA} (without revealing its targeted \ac{F-LTCA}) so that the vehicle can be authenticated and authorized by the \ac{F-LTCA}. In turn, the \ac{F-LTCA} provides the vehicle with a new ticket ($n\textnormal{-}tkt$), which is native within the domain of the \ac{F-LTCA} to be used for pseudonym acquisition in that (foreign) domain. The vehicle then interacts with its desired \ac{PCA} to obtain pseudonyms. Obtaining an $f\textnormal{-}tkt$ is transparent to the \ac{H-LTCA}: the \ac{H-LTCA} cannot distinguish between native and foreign ticket requests. This way, the \ac{PCA} in the foreign domain cannot distinguish native requesters from the foreign ones. For liability attribution, our scheme enables the \ac{RA}, with the help of the \ac{PCA} and the \ac{LTCA}, to initiate a resolution process, i.e., to resolve a pseudonym to its long-term identity. Each vehicle can interact with any \ac{PCA}, within its home or a foreign domain, to fetch the \ac{CRL}~\cite{solo2002internet} and perform \ac{OCSP}~\cite{santesson2013x} operations, authenticated with a current valid pseudonym. 

\subsection{Pseudonym Acquisition Policies}
\label{subsec:secmace-pseudonym-acquisition-policies}

The choice of policy for obtaining pseudonyms has diverse ramifications: on the \ac{VPKI} performance as well as the user privacy. The policy determines the volume of the workload (pseudonym requests and related computation and communication latencies) imposed to the \ac{VPKI}. The timing of requests can reveal information that could allow linking pseudonyms. To systematically investigate the effect of diverse on-demand pseudonym acquisition methods, here we define three specific representatives, first proposed in~\cite{khodaei2016evaluating}. 

\textbf{User-controlled (User-defined) Policy (P1):} A vehicle requests pseudonyms for its residual (ideally entire) trip duration at the start of trip. We presume each vehicle \emph{precisely estimates} the trip duration in advance, e.g., based on automotive navigation systems, previous trips, or user input. The \ac{PCA} determines the pseudonym lifetime, either fixed for all vehicles or flexible for each requester. Additional pseudonyms should be requested if the actual trip duration exceeds the estimated one, to ensure that the vehicle is always equipped with enough valid pseudonyms throughout the entire trip.

\textbf{Oblivious Policy (P2):} The vehicle interacts with the \ac{VPKI} every $\Gamma_{P2}$ seconds (determined by the \ac{PCA} and fixed for all users) and it requests pseudonyms for the entire $\Gamma_{P2}$ time interval; this continues until the vehicle reaches its destination. This results in over-provisioning of pseudonyms only during the last iteration. The difference, in comparison to P1, is that either the vehicle does not know the exact trip duration, or, it does not attempt to estimate, or possibly, overestimate it; thus, P2 is oblivious to the trip duration. 

\begin{table} [t!]
	\caption{Notation used in the protocols}
	\centering
	\resizebox{0.5\textwidth}{!}
		{
		\renewcommand{\arraystretch}{1.1}
		\begin{tabular}{l | *{1}{c} r}
			\hline \hline
			$(P^{i}_{v})_{pca}$, $P^{i}_{v}$ & a pseudonym signed by the \acs{PCA} \\\hline
			$(LK_v, Lk_v)$ & long-term public/private key pairs \\\hline
			$(K^i_v, k^i_v)$ & \shortstack{pseudonymous public/private key pairs, \\ corresponding to current valid pseudonym} \\\hline
			$Id_{req}, Id_{res}$ & request/response identifiers \\ \hline
			$Id_{ca}$  & \acl{CA} unique identifier \\\hline
			$LTC$ & \acl{LTC} \\\hline
			$(msg)_{\sigma_{v}}$ & a signed message with the vehicle's private key \\\hline
			$N, Rnd$ & nonce, a random number \\\hline
			$t_{now}, t_s, t_e$ & fresh/current, starting, and ending timestamps \\\hline
			$t_{date}$ & timestamp of a specific day \\\hline
			$n\textnormal{-}tkt$, $(n\textnormal{-}tkt)_{ltca}$ & native ticket \\\hline
			$f\textnormal{-}tkt$, $(f\textnormal{-}tkt)_{ltca}$ & foreign ticket \\\hline
			$H()$ & hash function \\\hline
			$Sign(Lk_{}, msg)$ & signing a message with the private key ($Lk$) \\\hline 
			$Verify(LTC_{}, msg)$ & verifying a message with the public key (in the \ac{LTC}) \\ \hline 
			$\tau_{P}$ & pseudonym lifetime \\\hline
			$\Gamma_{Px}$ & interacting period/interval with the \ac{VPKI} for policy x \\\hline
			$IK$ & identifiable key \\\hline
			$V$ & vehicle \\\hline
			$\zeta, \chi, \xi$ & temporary variables \\\hline
			\hline
		\end{tabular}
		\renewcommand{\arraystretch}{1}
		\label{table:secmace-protocols-notation}
		}
	\vspace{-0.5em}
\end{table}

\textbf{Universally Fixed Policy (P3):} The \ac{H-LTCA}, as the policy decision point in its domain, has predetermined universally fixed interval, $\Gamma_{P3}$, and pseudonym lifetime, $\tau_{P}$. At the start of its trip, a vehicle requests pseudonyms for the \emph{``current''} $\Gamma_{P3}$, out of which useful (non-expired) ones are actually obtained for the residual trip duration within $\Gamma_{P3}$. For the remainder of the trip, the vehicle requests pseudonyms for the entire $\Gamma_{P3}$ at each time. This policy issues time-aligned pseudonyms for all vehicles; thus, timing information does not harm user privacy. With P3, if the vehicle can \emph{estimate} the trip duration, it can obtain all required pseudonyms at the start of its trip by interacting with the \ac{VPKI} multiple times. However, if the vehicle does not attempt to estimate the trip duration, it should interact with the \ac{VPKI} servers every $\Gamma_{P3}$ seconds to obtain pseudonyms. A strict limitation in using this policy is that partial pseudonym acquisition in $\Gamma^{i}_{P3}$ is not allowed, i.e., the vehicle must request pseudonyms for the entire $\Gamma^{i}_{P3}$. The reasons are twofold: (i) the \ac{PCA} should not distinguish among different requests, and (ii) if the vehicle needs more pseudonyms during the same $\Gamma^{i}_{P3}$, it cannot request yet another ticket because the \ac{H-LTCA} only issues one ticket for a single vehicle for each $\Gamma^{i}_{P3}$. 

\begin{algorithm}[t]
	\floatname{algorithm}{Protocol}
	\renewcommand{\thealgorithm}{1}
	\caption{Ticket Provisioning from the \ac{H-LTCA}}
	\label{protocol:secmace-ticket-provision-in-home-domain}
	{\scriptsize
		\begin{flalign}
			V &: {\textbf{P1: }(t_s, t_e) \gets (t_s, t_e)} \\ 
			  &\hphantom{:} \; \; {\textbf{P2: } (t_s, t_e) \gets (t_s, \Gamma_{P2})} \\ 
			  &\hphantom{:} \; \; {\textbf{P3: } (t_s, t_e) \gets (t_{date}+\Gamma^{i}_{P3}, t_{date}+\Gamma^{i+1}_{P3})} \\
			V &: \zeta \leftarrow (Id_{req}, H(Id_{pca}\|Rnd_{n\textnormal{-}tkt}), t_s, t_e) \\
			V &: (\zeta)_{\sigma_{v}} \leftarrow Sign(Lk_v, \zeta) \\ 
			V\rightarrow{H\textnormal{-}\ac{LTCA}} &: ((\zeta)_{\sigma_{v}}, \ac{LTC}_v, N, t_{now}) \\
			H\textnormal{-}\ac{LTCA} &: \text{Verify}(\ac{LTC}_v, (\zeta)_{\sigma_{v}}) \\
			H\textnormal{-}\ac{LTCA} &: {IK_{n\textnormal{-}tkt} \leftarrow H(\ac{LTC}_v || t_s || t_e || Rnd_{IK_{n\textnormal{-}tkt}})} \\ 
			H\textnormal{-}\ac{LTCA} &: \chi \leftarrow (H(Id_{pca} \| Rnd_{n\textnormal{-}tkt}), IK_{n\textnormal{-}tkt}, t_s, t_e) \\
			H\textnormal{-}\ac{LTCA} &: (n\textnormal{-}tkt)_{\sigma_{h\textnormal{-}ltca}} \leftarrow Sign(Lk_{h\textnormal{-}ltca}, \chi) \\ 
			V\leftarrow{H\textnormal{-}\ac{LTCA}}  &: (Id_{res}, (n\textnormal{-}tkt)_{\sigma_{h\textnormal{-}ltca}}, Rnd_{IK_{n\textnormal{-}tkt}}, N\textnormal{+}1, t_{now}) \\
			V &: \text{Verify}(LTC_{h\textnormal{-}ltca}, (n\textnormal{-}tkt)_{\sigma_{h\textnormal{-}ltca}}) \\
			V &: {H(\ac{LTC}_v || t_s || t_e || Rnd_{IK_{n\textnormal{-}tkt}}) \stackrel{?}{=} IK_{n\textnormal{-}tkt}}
		\end{flalign}
		\vspace{-1em}
	}
	\setcounter{equation}{0}
\end{algorithm} 

\subsection{\ac{VPKI} Services and Security Protocols}

In this section, we provide the detailed description of the protocols using the notation in Table~\ref{table:secmace-protocols-notation}. For \ac{UML} diagrams of the security and privacy protocols, we refer the reader to our prior work~\cite{khodaei2014ScalableRobustVPKI}. 

\textbf{Ticket Acquisition (Protocol~\ref{protocol:secmace-ticket-provision-in-home-domain}):} Assume the \ac{OBU} decides to obtain pseudonyms from a specific \ac{PCA}. If the relevant policy is P1, each vehicle \emph{estimates} the trip duration $[t_{s}, \: t_{e}]$ (step~\ref{protocol:secmace-ticket-provision-in-home-domain}.1, i.e., step 1 in Protocol~\ref{protocol:secmace-ticket-provision-in-home-domain}). While with P2, each vehicle requests pseudonyms for $[t_{s}, \: t_{s}+\Gamma_{P2}]$ (step~\ref{protocol:secmace-ticket-provision-in-home-domain}.2). If the relevant policy is P3, the vehicle calculates the trip duration based on the date of travel, $t_{date}$, and the actual time of travel corresponding to the universally fixed interval $\Gamma_{P3}$ of that specific \ac{PCA} (step~\ref{protocol:secmace-ticket-provision-in-home-domain}.3). Then, the vehicle prepares a request and calculates the hash value of the concatenation of its desired \ac{PCA} identity and a random number, i.e., $H(Id_{pca} \| Rnd_{n\textnormal{-}tkt})$ (step~\ref{protocol:secmace-ticket-provision-in-home-domain}.4). This conceals the targeted \ac{PCA}, the actual pseudonym acquisition periods, and the choice of policy from the \ac{LTCA}. In case of cross-domain operation, the vehicle interacts with the \ac{H-LTCA} to obtain an $f\textnormal{-}tkt$ and it concatenates its targeted F-\ac{LTCA} (instead of the desired \ac{PCA}) and a random number. The vehicle then signs the request (step~\ref{protocol:secmace-ticket-provision-in-home-domain}.5) and sends it to its \ac{H-LTCA} to obtain an $n\textnormal{-}tkt$ (step~\ref{protocol:secmace-ticket-provision-in-home-domain}.6). Upon a successful validation of the \ac{LTC} and verification of the request (step~\ref{protocol:secmace-ticket-provision-in-home-domain}.7), the \ac{H-LTCA} generates the \emph{``ticket identifiable key''} ($IK_{n\textnormal{-}tkt}$) to bind the ticket to the \ac{LTC}: $H(\ac{LTC}_{v} || t_s || t_e || Rnd_{IK_{n\textnormal{-}tkt}})$ (steps~\ref{protocol:secmace-ticket-provision-in-home-domain}.8); this prevents the \ac{H-LTCA} from mapping the ticket to a different \ac{LTC} during resolution process. The \ac{H-LTCA} then issues an \emph{anonymous} ticket, $(n\textnormal{-}tkt)_{\sigma_{h\textnormal{-}ltca}}$ (step~\ref{protocol:secmace-ticket-provision-in-home-domain}.9-\ref{protocol:secmace-ticket-provision-in-home-domain}.10). The ticket is anonymous in the sense it does not reveal the actual identity of its owner, i.e., the \ac{H-LTCA} issues tickets without the provided ticket revealing the actual identity of the requester. Thus, the \ac{PCA} cannot infer the actual identity of the ticket owner, or distinguish between two tickets, even if the two tickets come from the same vehicle. Next, the \ac{H-LTCA} delivers the ticket to the vehicle (step~\ref{protocol:secmace-ticket-provision-in-home-domain}.11). Finally, the vehicle verifies the ticket and $IK_{n\textnormal{-}tkt}$ (steps~\ref{protocol:secmace-ticket-provision-in-home-domain}.12-\ref{protocol:secmace-ticket-provision-in-home-domain}.13). In case of cross-domain operation, the vehicle interacts with the F-\ac{LTCA} and presents the $f\textnormal{-}tkt$ to obtain an $n\textnormal{-}tkt$ in the foreign domain. Thus, it can interact with the \acp{PCA} within the foreign domain as a \emph{``local''} vehicle.

\textbf{Pseudonym Acquisition (Protocol~\ref{protocol:secmace-pseudonym-provision-in-home-domain}):} With an $n\textnormal{-}tkt$ at hand, the vehicle interacts with the targeted \ac{PCA} to obtain pseudonyms. The vehicle initiates a protocol to generate the required \ac{ECDSA} public/private key pairs (which could be generated off-line) and sends a request to the \ac{PCA} (steps~\ref{protocol:secmace-pseudonym-provision-in-home-domain}.1-\ref{protocol:secmace-pseudonym-provision-in-home-domain}.2). Upon reception and successful ticket verification (step~\ref{protocol:secmace-pseudonym-provision-in-home-domain}.3), the \ac{PCA} verifies the targeted \ac{PCA} (step~\ref{protocol:secmace-pseudonym-provision-in-home-domain}.4), and whether or not the actual period of requested pseudonyms (i.e., $[t'_{s}, \: t'_{e}]$) falls within the period specified in the ticket (i.e., $[t_s, t_e]$): ${[t'_{s}, \: t'_{e}] \subseteq ([t_s, t_e])_{n\textnormal{-}tkt}}$ for P1 or P2, or $[t'_{s}, \: t'_{e}]=([t_s, t_e])_{n\textnormal{-}tkt}$ for P3 (steps~\ref{protocol:secmace-pseudonym-provision-in-home-domain}.5-\ref{protocol:secmace-pseudonym-provision-in-home-domain}.6). Then, the \ac{PCA} initiates a proof-of-possession protocol to verify the ownership of the corresponding private keys, $k^i_v$.\footnote{As an optimization, the \ac{PCA} can probabilistically verify $(K^i_v)_{\sigma_{k^i_v}}$.} The \ac{PCA} generates the \emph{``pseudonym identifiable key''} ($IK_{P_v^i}$) to bind the pseudonyms to the ticket; this prevents the compromised (malicious) \ac{PCA} from mapping the pseudonyms to a different ticket during the resolution process. It then issues the pseudonyms (steps~\ref{protocol:secmace-pseudonym-provision-in-home-domain}.7-\ref{protocol:secmace-pseudonym-provision-in-home-domain}.12), and delivers the response (step~\ref{protocol:secmace-pseudonym-provision-in-home-domain}.13). Finally, the vehicle verifies the pseudonyms and $IK_{P_{v}^{i}}$ (steps~\ref{protocol:secmace-pseudonym-provision-in-home-domain}.14-\ref{protocol:secmace-pseudonym-provision-in-home-domain}.17).

\begin{algorithm}[!t]
	\floatname{algorithm}{Protocol}
	\renewcommand{\thealgorithm}{2}
	\caption{Pseudonym Provisioning from the \ac{PCA}}
	\label{protocol:secmace-pseudonym-provision-in-home-domain}
	{\scriptsize
		\begin{flalign}
			V &: \zeta \leftarrow {(Id_{req}, Rnd_{n\textnormal{-}tkt}, t'_{s}, t'_{e}, (n\textnormal{-}tkt)_{\sigma_{h\textnormal{-}ltca}},} \notag \\
			  & \; \; \; \; \; \; \; \; \; \; \; \; \; \; \; \; \; \; \; \; {\{(K^1_v)_{\sigma_{k^1_v}}, \dots, (K^n_v)_{\sigma_{k^n_v}}\}, N, t_{now})} \\
		 	V\rightarrow{\ac{PCA}} &: (\zeta) \\
			\ac{PCA} &: \text{Verify}(LTC_{ltca}, (n\textnormal{-}tkt)_{\sigma_{ltca}}) \\
			\ac{PCA} &: H(Id_{this\textnormal{-}pca}\|Rnd_{n\textnormal{-}tkt}) \stackrel{?}{=} H(Id_{pca}\|Rnd_{n\textnormal{-}tkt}) \\
			\ac{PCA} &: \textbf{P1 or P2: } {[t'_{s}, \: t'_{e}] \overset{?}{\subseteq} ([t_s, t_e])_{n\textnormal{-}tkt}}\\ 
			 &\hphantom{:} \; \; \textbf{P3: } {[t'_{s}, \: t'_{e}] \overset{?}{=} ([t_s, t_e])_{n\textnormal{-}tkt}}\\
			\ac{PCA} &: \textbf{for } \text{i} \leftarrow 1, n \textbf{ do} \\ 
				\ac{PCA} &: \; \; \; \; \text{Verify}(K^{i}_{v}, (K^i_v)_{\sigma_{k^i_v}}) \\ 
				\ac{PCA} &: \; \; \; \; {IK_{P_v^i} \gets H(IK_{n\textnormal{-}tkt} || K^i_v || t_{s}^i || t_{e}^i || Rnd_{IK_{P^{i}_{v}}})} \\ 
				\ac{PCA} &: \; \; \; \; {\xi \leftarrow (K^i_v, IK_{P_v^i}, t_{s}^i, t_{e}^i)} \\
				\ac{PCA} &: \; \; \; \; (P^i_v)_{\sigma_{pca}} \leftarrow Sign(Lk_{pca}, \xi) \\ 
			\ac{PCA} &: \textbf{end for} \\ 
			V\leftarrow{\ac{PCA}} &: (Id_{res}, \{(P^1_v)_{\sigma_{pca}}, \dots, (P^n_v)_{\sigma_{pca}}\}, \\\notag
			 & \: \: \: \; \; \; \; \; \; \; \; \; \; \; \; \{Rnd_{IK_{P^{1}_{v}}}, \dots, Rnd_{IK_{P^{n}_{v}}}\}, N\textnormal{+}1, t_{now}) \\
			V &: \textbf{for } \text{i} \leftarrow 1, n \textbf{ do} \\ 
			V &: \; \; \; \; \text{Verify}(LTC_{pca}, P^i_v) \\
			V &: \; \; \; \; {H(IK_{n\textnormal{-}tkt} || K^i_v || t_{s}^i || t_{e}^i || Rnd_{IK_{P^{i}_{v}}})} \stackrel{?}{=} IK_{P_v^i} \\
			V &: \textbf{end for} 
		\end{flalign}
		\vspace{-1em}
	}
	\setcounter{equation}{0}
\end{algorithm} 


\textbf{Pseudonym Resolution and Revocation (Protocol~\ref{protocol:secmace-single-domain-ticket-resolution-and-revocation}):} The \ac{RA} requests the \ac{PCA} to map the pseudonym to the corresponding ticket, i.e., $n\textnormal{-}tkt$, which the \ac{PCA} has stored (steps~\ref{protocol:secmace-single-domain-ticket-resolution-and-revocation}.1-\ref{protocol:secmace-single-domain-ticket-resolution-and-revocation}.3). The \ac{PCA} verifies the request and maps the pseudonym to the corresponding $n\textnormal{-}tkt$ (steps~\ref{protocol:secmace-single-domain-ticket-resolution-and-revocation}.4-\ref{protocol:secmace-single-domain-ticket-resolution-and-revocation}.5). If needed, the \ac{PCA} includes all the valid (non-expired) pseudonyms, issued for this ticket, to the \ac{CRL}, thus evicting the misbehaving vehicle from the system (step~\ref{protocol:secmace-single-domain-ticket-resolution-and-revocation}.6). Then, it sends the $n\textnormal{-}tkt$ to the \ac{RA} (steps~\ref{protocol:secmace-single-domain-ticket-resolution-and-revocation}.7-\ref{protocol:secmace-single-domain-ticket-resolution-and-revocation}.9). The \ac{RA} verifies the response and calculates the $IK_{P^{i}_{v}}$ to confirm that the \ac{PCA} has correctly resolved the pseudonym to the corresponding $n\textnormal{-}tkt$ (steps~\ref{protocol:secmace-single-domain-ticket-resolution-and-revocation}.10-\ref{protocol:secmace-single-domain-ticket-resolution-and-revocation}.11). The output of this process is the $n\textnormal{-}tkt$; in case of a cross-domain resolution, one additional interaction is needed to resolve the foreign ticket, the $f\textnormal{-}tkt$. As a continuation, the \ac{RA} resolves the ticket with the help of the corresponding \ac{H-LTCA} (step~\ref{protocol:secmace-single-domain-ticket-resolution-and-revocation}.12, detailed in Protocol~\ref{protocol:secmace-real-identity-resolution-and-revocation}).

\begin{algorithm}[!t]
	\floatname{algorithm}{Protocol}
	\renewcommand{\thealgorithm}{3}
	\caption{Pseudonym Resolution and Revocation}
	\label{protocol:secmace-single-domain-ticket-resolution-and-revocation}
	{\scriptsize
		\begin{align}
		\ac{RA} &: \zeta \leftarrow (Id_{req}, P^{i}_{v}) \\
		\ac{RA} &: (\zeta)_{\sigma_{ra}} \leftarrow Sign({Lk_{ra}}, \zeta) \\ 
		\ac{RA}\rightarrow{\ac{PCA}} &: ((\zeta)_{\sigma_{ra}}, \ac{LTC}_{ra}, N, t_{now}) \\
		\ac{PCA} &: \text{Verify}(LTC_{ra}, (\zeta)_{\sigma_{ra}}) \\
		\ac{PCA} &: \{n\textnormal{-}tkt, Rnd_{IK_{P^{i}_{v}}}\} \leftarrow \text{Resolve}(P^{i}_{v}) \\
		\ac{PCA} &: Id_{req} \stackrel{?}{=} \text{\emph{`revoke':}} \text{ Add}(P^{i}_{v},\text{ CRL}) \\
		\ac{PCA} &: \chi \leftarrow (Id_{res}, n\textnormal{-}tkt, Rnd_{IK_{P^{i}_{v}}}) \\
		\ac{PCA} &:  (\chi)_{\sigma_{pca}} \leftarrow Sign({Lk_{pca}}, \chi) \\ 
		\ac{RA} \leftarrow{\ac{PCA}} &: ((\chi)_{\sigma_{pca}}, N\textnormal{+}1, t_{now}) \\ 
		\ac{RA} &: \text{Verify}(LTC_{pca}, \chi)  \\ 
		\ac{RA} &: {H(IK_{n\textnormal{-}tkt} || K^i_v || t^{i}_{s} || t^{i}_{e} || Rnd_{IK_{P^{i}_{v}}}) \stackrel{?}{=} IK_{P_v^i} } \\ 
		\ac{RA} &: \text{ResolveLTC}(n\textnormal{-}tkt) 
		\end{align}
		\vspace{-1em}
	}
\end{algorithm}

\begin{algorithm}[!t]
	\floatname{algorithm}{Protocol}
	\renewcommand{\thealgorithm}{4}
	\caption{\ac{LTC} Resolution and Revocation}
	\label{protocol:secmace-real-identity-resolution-and-revocation}
	{\scriptsize
		\begin{align}
		\ac{RA} &: \zeta \leftarrow (Id_{req}, n/f\textnormal{-}tkt, N, t_{now}) \\ 
		\ac{RA} &: (\zeta)_{\sigma_{ra}} \leftarrow Sign({Lk_{ra}}, \zeta) \\
		\ac{RA}\rightarrow H\textnormal{-}{\ac{LTCA}} &: ((\zeta)_{\sigma_{ra}}, \ac{LTC}_{ra}) \\ 
		H\textnormal{-}\ac{LTCA} &: \text{Verify}(LTC_{ra}, (\zeta)_{\sigma_{ra}}) \\ 
		H\textnormal{-}\ac{LTCA} &: \{LTC_v, Rnd_{IK_{n\textnormal{-}tkt}}\} \leftarrow \text{Resolve}(n/f\textnormal{-}tkt) \\ 
		H\textnormal{-}\ac{LTCA} &: Id_{req} \stackrel{?}{=} \text{\emph{`revoke':}} \text{ Add}(LTC_v,\text{ CRL}) \\ 
		H\textnormal{-}\ac{LTCA} &: \chi \leftarrow (Id_{res}, LTC_v, Rnd_{IK_{n\textnormal{-}tkt}}, N\textnormal{+}1, t_{now}) \\ 
		H\textnormal{-}\ac{LTCA} &: (\chi)_{\sigma_{h\textnormal{-}ltca}} \leftarrow Sign({Lk_{h\textnormal{-}ltca}}, \chi) \\ 
		\ac{RA}\leftarrow H\textnormal{-}{\ac{LTCA}} &: (\chi)_{\sigma_{h\textnormal{-}ltca}} \\
		\ac{RA} &: \text{Verify}(LTC_{h\textnormal{-}ltca}, \chi) \\
		\ac{RA} &: H(\ac{LTC}_v || t_s || t_e || Rnd_{IK_{n\textnormal{-}tkt}}) \stackrel{?}{=} IK_{n/f\textnormal{-}tkt}
		\end{align}
		\vspace{-1em}
	}
	\setcounter{equation}{0}
\end{algorithm}

\begin{table*} [t!]
	\caption{Information held by honest-but-curious Entities}
	\centering
	\Large
	\resizebox{1\textwidth}{!}
	{
		\renewcommand{\arraystretch}{1.8}
		\begin{tabular}{ | c | *{1}{c} | *{1}{c} | }
			\hline
			\Huge \textbf{Honest-but-curious (colluding) Entities} & \Huge \textbf{Information Leaked} & \Huge \textbf{Security and Privacy Implications} \\\hline
			
			\Huge {$H\textnormal{-}LTCA$} & \Huge $id_H, t_s, t_e$ & \Huge An \ac{H-LTCA} knows during when the registered vehicles wish to obtain pseudonyms. \\\hline
			
			\Huge {$PCA_{H_i}$} & \Huge $t_s, t_e, P_{H_i}$ & \Huge A \ac{PCA} in the home domain can link the pseudonyms it issued for a same request, but it cannot link those for different requests. \\\hline
			
			\Huge {$H\textnormal{-}LTCA$, $F\textnormal{-}LTCA$} & \Huge $id_H, id_F, t_s, t_e$ & \Huge Collusion among \acp{LTCA} from different domains does not reveal additional information. \\\hline
			
			\Huge {$PCA_{H}$, $PCA_{F}$} & \Huge $t_s, t_e, P_{H}, P_{F}$ & \Huge Collusion among \acp{PCA} from different domains does not reveal additional information. \\\hline
			
			\Huge {$H\textnormal{-}LTCA$, $PCA_{H}$} & \Huge $id_H, t_s, t_e, P_{H}$ & \Huge The pseudonyms they issued can be linked and the vehicle identities within the same domain can be derived. \\\hline
			
			\Huge {$F\textnormal{-}LTCA$, $PCA_{F}$} & \Huge $id_F, t_s, t_e, P_{F}$ & \Huge The pseudonyms they issued can be linked but the real identities of the vehicles cannot be derived. \\\hline
			
			\Huge {$H\textnormal{-}LTCA$, $F\textnormal{-}LTCA$, $PCA_{F}$} & \Huge $id_H, id_F, t_s, t_e, P_{F}$ & \Huge 
			\makecell{Colluding H-\acp{LTCA}, F-\ac{LTCA} and $\ac{PCA}_{F}$ can link the pseudonyms issued in the foreign domain with the \\ real identities of the vehicles if the $PCA_{F}$ is the issuer of the pseudonyms.} \\\hline
			
			\Huge {$H\textnormal{-}LTCA$, $F\textnormal{-}LTCA$, $PCA_{F}$, $PCA_{F}$} & \Huge $id_H, id_F, t_s, t_e, P_{H}, P_{F}$ & \Huge 
			\makecell{Colluding H-\acp{LTCA}, F-\ac{LTCA}, $\ac{PCA}_{H}$ and $\ac{PCA}_{F}$ can link the pseudonyms issued in the home and foreign domains \\ with the real identities of the vehicles if the $PCA_{H}$ and $PCA_{F}$ are the issuers of the pseudonyms.} \\\hline
			
			\Huge {$V, H\textnormal{-}LTCA$, $F\textnormal{-}LTCA$, $PCA_{F}$, $PCA_{F}$} & \Huge $id_H, id_F, t_s, t_e, P_{H}, P_{F}$ & \Huge 
			\makecell{Colluding vehicle, H-\ac{LTCA}, F-\ac{LTCA}, $\ac{PCA}_{H}$, and $\ac{PCA}_{F}$ could result in generating invalid $IK_{n\textnormal{/}f\textnormal{-}tkt}$, $IK_{n\textnormal{-}tkt}$, or $IK_{P_{v}}$, respectively.} \\\hline
		\end{tabular}
		\renewcommand{\arraystretch}{1}
		\label{table:secmace-VPKI-entity-knowledge}
	}
\end{table*}

\textbf{\ac{LTC} Resolution and Revocation (Protocol~\ref{protocol:secmace-real-identity-resolution-and-revocation}):} The \ac{RA} queries the corresponding \ac{H-LTCA} to have the vehicle identified, i.e., resolving the \ac{LTC} of the vehicle. The \ac{RA} prepares a request and sends the ticket serial number to \ac{H-LTCA} (steps~\ref{protocol:secmace-real-identity-resolution-and-revocation}.13-\ref{protocol:secmace-real-identity-resolution-and-revocation}.15). Upon the request verification, the \ac{H-LTCA} resolves, and possibly revokes, the \ac{LTC} corresponding to the $n\textnormal{-}tkt$ (steps~\ref{protocol:secmace-real-identity-resolution-and-revocation}.16-\ref{protocol:secmace-real-identity-resolution-and-revocation}.18). The H-\ac{LTCA} delivers the response back to the \ac{RA} (steps~\ref{protocol:secmace-real-identity-resolution-and-revocation}.19-\ref{protocol:secmace-real-identity-resolution-and-revocation}.21). Upon reception of the response, the \ac{RA} verifies the signature and confirms if the \ac{H-LTCA} has mapped the correct $\ac{LTC}_v$ by validating the $IK$ (steps~\ref{protocol:secmace-real-identity-resolution-and-revocation}.22-\ref{protocol:secmace-real-identity-resolution-and-revocation}.23).

\section{Security and Privacy Analysis}
\label{sec:secmace-security-and-privacy-analysis}

We analyze the achieved security and privacy of our \ac{VPKI} with respect to the requirements presented in Sec.~\ref{subsec:secmace-security-and-privacy-requirements}. All the communication runs over secure channels, i.e., \ac{TLS} with uni- or bidirectional authentication, thus we achieve \emph{authentication, communication integrity} and \emph{confidentiality} (R1). The \ac{H-LTCA} authenticates and authorizes the vehicles based on the registration and their revocation status, and makes appropriate decisions. It grants a \emph{service-granting ticket}, thus enabling the vehicles to request pseudonyms from any \ac{PCA} by presenting its anonymous ticket. The \ac{PCA} then grants the service, based on prior established trust, by validating the ticket (R2). Given the ticket acquisition request is signed with the private key corresponding to the vehicle's \ac{LTC} and pseudonym acquisition entails a valid ticket, the system provides \emph{non-repudiation and accountability} (R3). Moreover, the \ac{LTCA} and the \ac{PCA} calculate ticket and pseudonym identifiable keys ($IK_{tkt}$ and $IK_{P}$) to bind them to the corresponding \ac{LTC} and ticket respectively (R3). 

According to the protocol design, the vehicle conceals the identity of its targeted \ac{PCA} with $H(Id_{pca} || Rnd_{n\textnormal{-}tkt})$, and the targeted F-\ac{LTCA} when operating in a foreign domain. With P1 and P2, the vehicle hides the actual pseudonym acquisition periods, i.e. $[t'_{s}, \: t'_{e}]$, while only $[t_s, \: t_e]$ is revealed to the \ac{LTCA}. With P3, requesting intervals fall within the \emph{``universally''} fixed $\Gamma_{P3}$ (along with aligned pseudonyms lifetimes); thus timing information cannot be used to link two successive pseudonyms as they are time-aligned with those of all other active vehicles that obtain pseudonyms by the same \ac{PCA} (R4, R5). This is further discussed in Sec.~\ref{subsec:secmace-ticket-and-pseudonym-lifetime-policies}. Moreover, the separation of duties between the \ac{LTCA} and the \ac{PCA} provides \emph{conditional anonymity}, but revoked under special circumstances, e.g., misbehavior (R3).

The \ac{H-LTCA} enforces a policy that each vehicle cannot obtain tickets with overlapping lifetime: upon receiving a request, the \ac{H-LTCA} checks if a ticket was issued for the requester during that period. This ensures that no vehicle can obtain more than a single valid ticket to request multiple simultaneously valid pseudonyms. Moreover, a ticket is implicitly bound to a specific \ac{PCA}; thus, it cannot be used more than once or be reused for other \acp{PCA}. The \ac{PCA} also issues the pseudonyms with non-overlapping lifetimes; all in all, no vehicle can be provided with more than one valid pseudonym at any time; thus, Sybil-based misbehavior is thoroughly thwarted within a multi-domain \ac{VC} environment (R6). We achieve availability in the face of a crash failure by mandating load-balancers and server redundancy~\cite{khodaei2014ScalableRobustVPKI}; in case of a \ac{DDoS} attack, we use a puzzle technique as a mitigation approach (R7), further discussed in Sec.~\ref{subsubsec:secmace-ddos-attack}. 

The \ac{OBU} could request pseudonyms for a period depending on the policy, determined by the user (P1) or fixed by the \ac{VPKI} (P2 and P3). Based on the policy and the pseudonym lifetime, the \ac{OBU} automatically calculates the number of pseudonyms to obtain. Clearly, there is a trade off: the longer the pseudonym refill interval is, i.e., the higher the number of pseudonyms in a single request is, the less frequent vehicle-\ac{VPKI} interactions are. But the higher the chance for a \ac{PCA} to trivially link the issued pseudonyms for the same vehicle as a response to a single request. With our scheme, we can configure the system to reduce the number of pseudonyms per request to one to achieve \emph{full unlinkability}, thus enhancing user privacy. To do so, we configure the system with P3 so that $\Gamma_{P3}$ is equal to $\tau_{P}$, and have each vehicle requesting pseudonyms for a duration of $[t_s, t_s+\tau_{P}]$, i.e., obtaining a single pseudonym with a different ticket. This implies that a \ac{PCA} cannot link any two pseudonyms issued for a single vehicle. But this configuration increases the frequency of vehicle-\ac{VPKI} interactions. The performance of our \ac{VPKI} to issue fully-unlinkable pseudonyms is evaluated in Sec.~\ref{subsubsec:secmace-unlinkable-pseudonyms-evaluation}.

\begin{figure*} [!htb]
	\centering
	\subfloat[P1: User-controlled policy]{
	\includegraphics[trim=1cm 0.1cm 0.7cm 0cm, clip=true, totalheight=0.338\textheight, width=0.338\textwidth,angle=0,keepaspectratio] {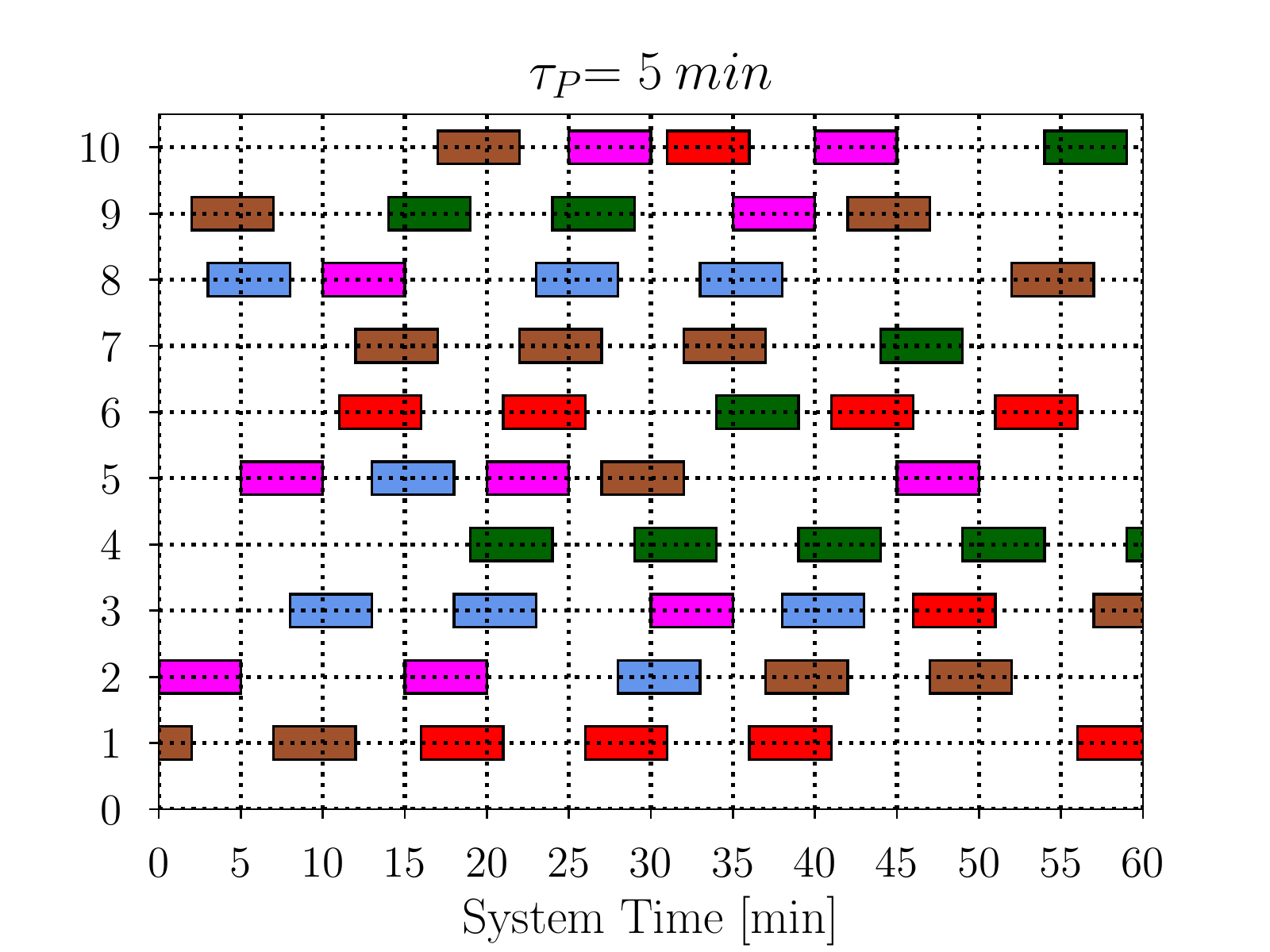}}
	\subfloat[P2: Oblivious policy] 
	{\includegraphics[trim=1cm 0.1cm 0.7cm 0cm, clip=true, totalheight=0.338\textheight, width=0.338\textwidth,angle=0,keepaspectratio] {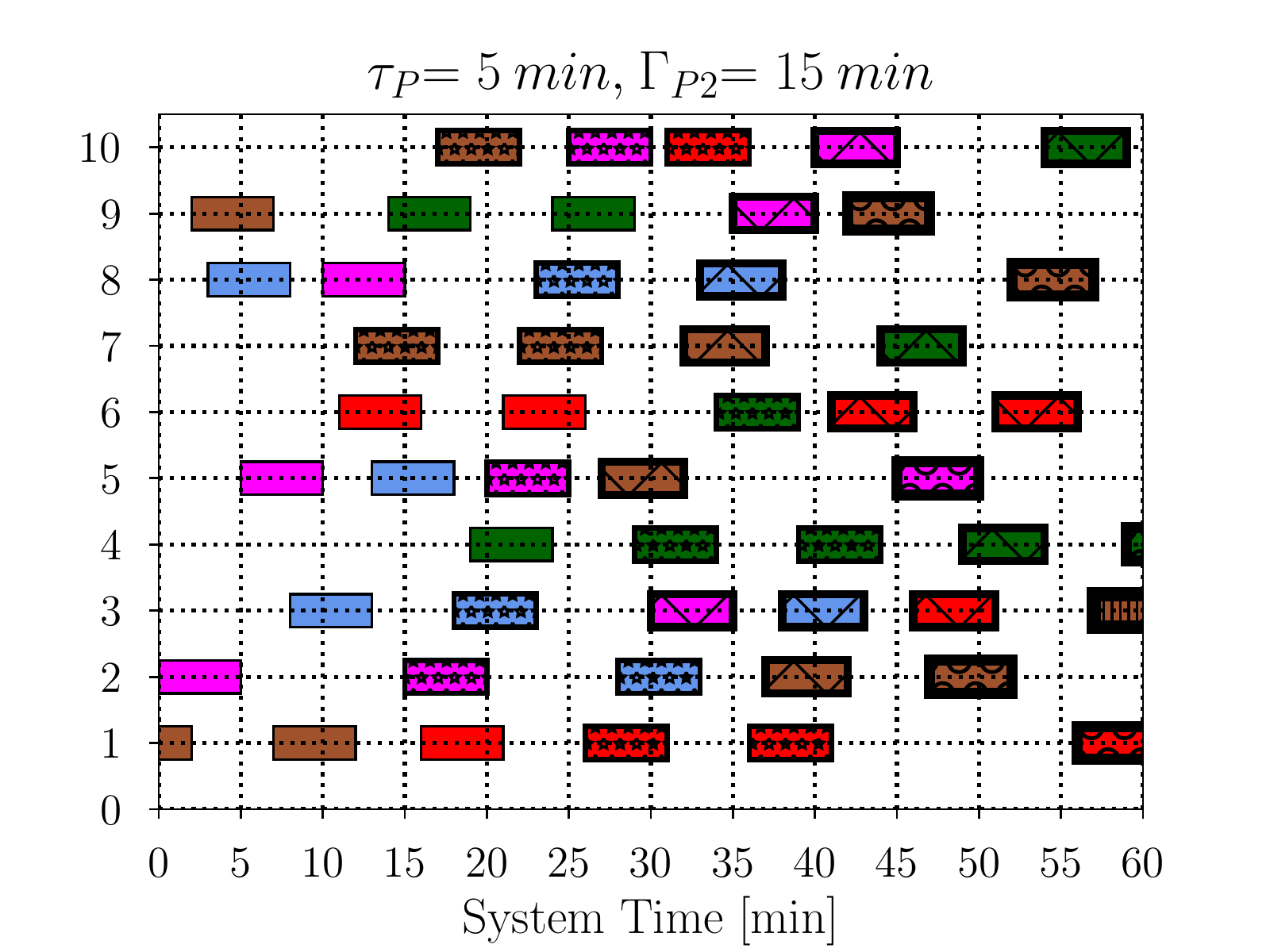}}
	\subfloat[P3: Universally fixed policy] 
	{\includegraphics[trim=1cm 0.1cm 0.7cm 0cm, clip=true, totalheight=0.338\textheight, width=0.338\textwidth,angle=0,keepaspectratio] {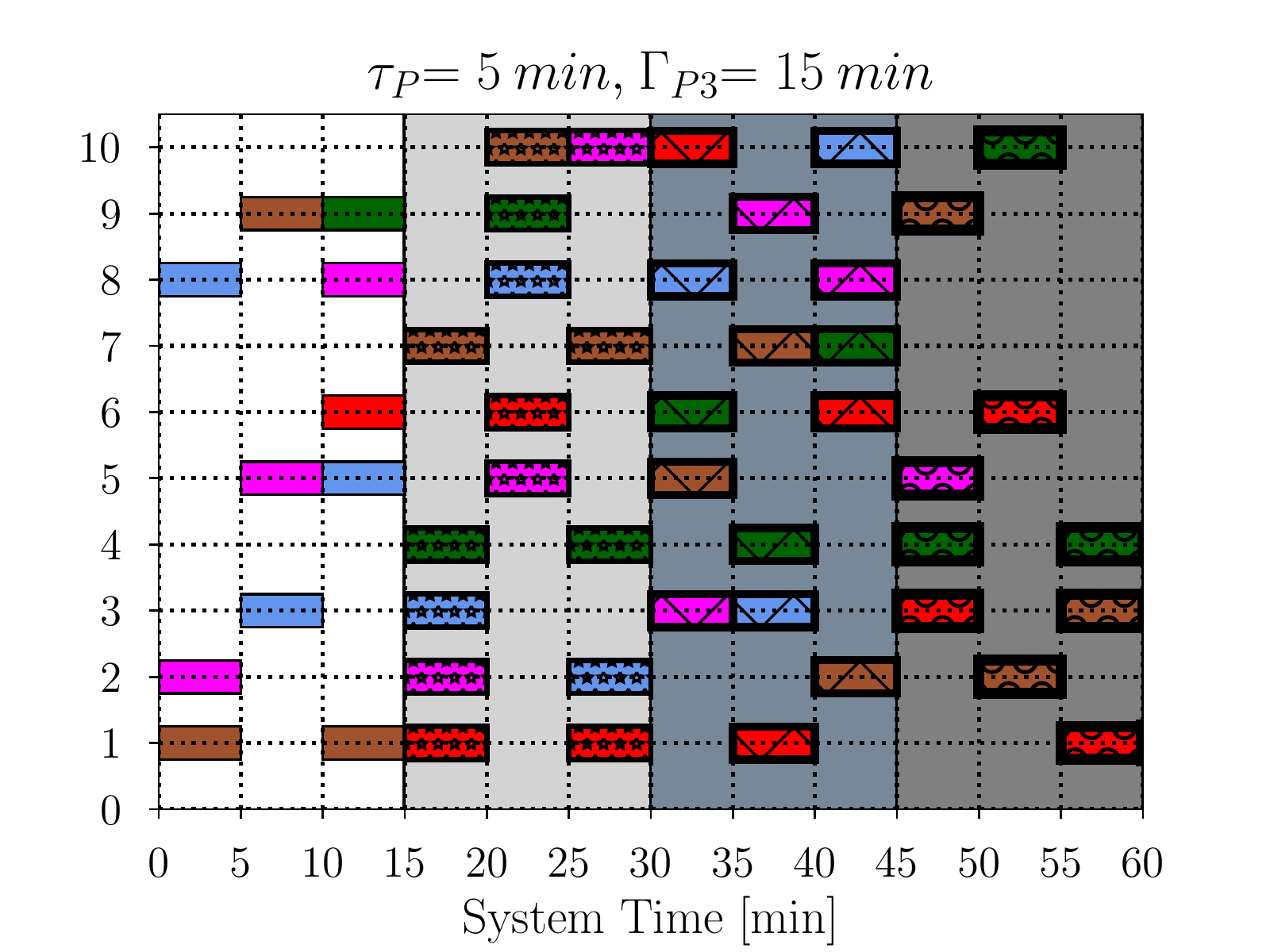}}
	\caption{Pseudonym acquisition policies (Each color shows the non-overlapping pseudonym lifetime).}
	\label{fig:secmace-pseudonym-lifetime-policies}
	\vspace{-0.5em}
\end{figure*}

\begin{table}[!t]
	\caption{Notation used in security \& privacy analysis}
	\centering
	\resizebox{0.4\textwidth}{!}
	{
		\begin{tabular}{l | *{1}{c} r}
			\hline \hline
			$PCA_{H_i}$ & $PCA_i$ in the home domain \\\hline
			$PCA_{H}$ & a set of \acp{PCA} in the home domain \\\hline
			$PCA_{F}$ & a set of \acp{PCA} in the foreign domain \\\hline
			$id_{H}$ & identities of the vehicles in the home domain \\\hline
			$id_{F}$ & identities of the vehicles in the foreign domain \\\hline
			$P_{H}$ & pseudonyms issued by the \acp{PCA} in the home domain \\\hline
			\hline
		\end{tabular}
		\label{table:secmace-sec-analysis-notations}
	}
	\vspace{-0.5em}
\end{table}

\subsection{Honest-but-curious \ac{VPKI} Servers}
\label{subsec:honest-but-curious-vpki-servers}

We further consider the privacy sensitive information that can be inferred by colluding \ac{VPKI} servers within the home or across domains. Table~\ref{table:secmace-VPKI-entity-knowledge} presents this information and the privacy implications when different honest-but-curious \ac{VPKI} entities collude, based on notation summarized in Table~\ref{table:secmace-sec-analysis-notations}. A single entity cannot fully de-anonymize a user due to the separation of duties in our design. Collusion by \ac{H-LTCA} and F-\acp{LTCA}, or $\ac{PCA}_{H}$ and $\ac{PCA}_{F}$ from the same or different domains, do not reveal any useful information to link the user identities with their pseudonyms. However, collusion by \ac{H-LTCA} and the $\ac{PCA}_{H}$ enables them to link the vehicle identities with their pseudonyms. Collusion by the F-\ac{LTCA} and the $\ac{PCA}_{F}$ does not reveal the real identities of the vehicles, but their pseudonyms and the foreign domain identifier. Collusion by \ac{H-LTCA}, F-\ac{LTCA} and $\ac{PCA}_{F}$ enables them to link the issued pseudonyms with their long term identifiers. Additionally, collusion by \ac{H-LTCA}, F-\ac{LTCA}, $\ac{PCA}_{H}$, and $PCA_{F}$ results in linking the vehicle identities and their corresponding pseudonyms in the home and foreign domains. Finally, collusion of a vehicle and the \ac{H-LTCA}, the \ac{F-LTCA}, or the \ac{PCA}, could yield invalid $IK_{n\textnormal{/}f\textnormal{-}tkt}$, $IK_{n\textnormal{-}tkt}$, or $IK_{P^{i}_{v}}$, respectively.

\subsection{Ticket and Pseudonym Lifetime Policies}
\label{subsec:secmace-ticket-and-pseudonym-lifetime-policies}

Fig.~\ref{fig:secmace-pseudonym-lifetime-policies} displays a transcript of eavesdropped pseudonyms for different pseudonym acquisition policies ($\Gamma_{}$=15 min, $\tau_{P}$=5 min). We assume an \ac{LTCA}, a \ac{PCA}, or an external observer attempts to link pseudonyms of the same vehicle based on the timing information of the credentials. With P1 and P2 (Fig.~\ref{fig:secmace-pseudonym-lifetime-policies}.a and~\ref{fig:secmace-pseudonym-lifetime-policies}.b), requests could act as user \emph{``fingerprints''}: the exact time of requests and all subsequent requests until the end of trip could be unique, or one of few, and thus linkable even by an external observer as it might be unlikely in a specific region to have multiple requests at a unique instance. In Fig.~\ref{fig:secmace-pseudonym-lifetime-policies}.a, the pseudonym (colored in magenta) in row 2 expires at system time 5 while the only pseudonym valid at time 5 is located in row 5. Thus, an external observer could simply link these two pseudonyms based on the pseudonym lifetimes. With P2, not only an external observer could link two successive pseudonyms of the same vehicle, but also a \ac{PCA} could link two sets of pseudonyms for the same requester based on the timing information: in Fig.~\ref{fig:secmace-pseudonym-lifetime-policies}.b, the second pseudonym in row 8 (colored in magenta) is the last pseudonym issued for the first iteration of $\Gamma_{P2}$, which expires at system time 15. The only pseudonym starting from 15 in the second iteration of $\Gamma_{P2}$ is the second pseudonym in row 2 (with a repeated asterisk pattern colored in magenta).

This vulnerability is thwarted by P3 (Fig.~\ref{fig:secmace-pseudonym-lifetime-policies}.c): the requesting intervals fall within \emph{``universally''} fixed interval ($\Gamma_{P3}$) and the issued pseudonyms are aligned with global system time (\ac{PCA} clock); thus, at any point in time all vehicles in a given domain will be transmitting under pseudonyms which are indistinguishable based on timing information alone. This results in eliminating any distinction among pseudonym sets, i.e., an anonymity set equal to the number of active requests. Hence, not only an external observer, but also the \ac{PCA} could not distinguish among pseudonyms sets, thus, protecting user privacy. The same policy should be applied for the ticket acquisition, during which the \ac{H-LTCA} fixes the timing interval to be the same for all requesters; thus preventing an \ac{LTCA} from linking successive requests from a vehicle.

\section{Performance Evaluation}
\label{sec:secmace-vpki-servers-performance-evaluation}

Without a large-scale deployment of \ac{VC} systems, we resort to realistic large-scale mobility traces. Based on these, we determine the period the vehicles need pseudonyms. We extract two features of interest from the mobility traces, i.e., the departure time and the trip duration for each vehicle, and we apply policies described in Sec.~\ref{subsec:secmace-pseudonym-acquisition-policies} to create the workload for the \ac{VPKI} to assess the performance, i.e., scalability, efficiency, and robustness, of the full-blown implementation of our \ac{VPKI} for a large-scale deployment. We evaluate performance with two mobility traces and we only plot the results for both traces and policies if they are significantly different than each other. The main functionality of interest are: ticket and pseudonym acquisition, \ac{CRL} update, pseudonyms validation with \ac{OCSP}, and pseudonym resolution. The main metric is the \emph{end-to-end pseudonym acquisition latency}, i.e., the delay from the initialization of protocol~\ref{protocol:secmace-ticket-provision-in-home-domain} till the successful completion of protocol~\ref{protocol:secmace-pseudonym-provision-in-home-domain}, measured at the vehicle.

\subsection{Experimental Setup}
\label{subsec:secmace-experimental-setup}

\textbf{\ac{VPKI} Testbed and Detailed Implementation:} We allocate \acfp{VM} for distinct \ac{VPKI} servers. Table~\ref{table:secmace-vm-SIS-clients-specifications} details the specification of the distinct servers and the clients. Our full-blown implementation is in C++ and we use FastCGI~\cite{heinlein1998fastcgi} to interface Apache web-server. We use XML-RPC~\cite{xmlrpc-c} to execute a remote procedure call on the servers. Our \ac{VPKI} interface is language-neutral and platform-neutral as we use Protocol Buffers~\cite{protocol-buffer} for serializing and de-serializing structured data. For the cryptographic protocols and primitives (\ac{ECDSA} and \ac{TLS}), we use OpenSSL with \ac{ECDSA}-256 public/private key pairs according to the standards~\cite{ETSI-102-638, 1609-2016}; other algorithms and key sizes are compatible for our implementation. We run our experiments in a testbed with both servers and clients (emulating \acp{OBU}) running on the \acp{VM}: this essentially eliminates the network propagation delays of the vehicle-\ac{VPKI} connectivity. As such a connectivity would vary greatly based on the actual vehicle-\ac{VPKI} connectivity, we do not consider it here. 

\begin{table}[!t]
	\centering
		\caption{Servers and Clients Specifications} 
		\resizebox{0.42\textwidth}{!}
		{
			\begin{tabular}{l | *{4}{c} r}
				& \ac{LTCA} & \ac{PCA} & \ac{RA} & Client \\ \hline 
				Number of entities & 1 & 1 & 1 & 1 \\ 
				Dual-core CPU (Ghz) & 2.0 & 2.0 & 2.0 & 2.0 \\ 
				BogoMips          & 4000 & 4000 & 4000 & 4000 \\ 
				Memory            & 2GB & 2GB & 1GB & 1GB \\ 
				Database          & MySQL & MySQL & MySQL & MySQL \\ 
			\end{tabular}
			\label{table:secmace-vm-SIS-clients-specifications}
		}
\end{table}

\begin{table}[!t]
	\centering
	\caption{Mobility Traces Information}
	\resizebox{0.42\textwidth}{!}
    {
		\begin{tabular}{l | *{4}{c} r}
			& Tapas-Cologne & \acs{LuST} \\ \hline
			Number of vehicles & 75,576 & 138,259 \\
			Number of trips  & 75,576 & 287,939 \\ 
			Duration of snapshot (hour) & 24 & 24 \\
			Available duration of snapshot (hour) & 2 (6-8 AM) & 24 \\
			Average trip duration (sec) & 590.49 & 692.81 \\ 
		\end{tabular}
		\label{table:secmace-mobility-traces-information}
	}
	\vspace{-0.8em}
\end{table}

\textbf{Mobility Traces:} We use two microscopic vehicle mobility datasets: Tapas-Cologne~\cite{uppoor2014generation} and \ac{LuST}~\cite{codeca2015lust}. The former one represents the traffic demand information across the Cologne urban area (available only for 2 hours, 6-8 AM), while the latter presents a full-day realistic mobility pattern in the city of Luxembourg. Table~\ref{table:secmace-mobility-traces-information} shows the mobility traces information for the two datasets. 

\begin{figure}[!t]
	\centering
	\subfloat[Issuing a ticket]{
		\includegraphics[trim=0.1cm 0.1cm 0.5cm 0.6cm, clip=true, totalheight=0.135\textheight,angle=0,keepaspectratio] {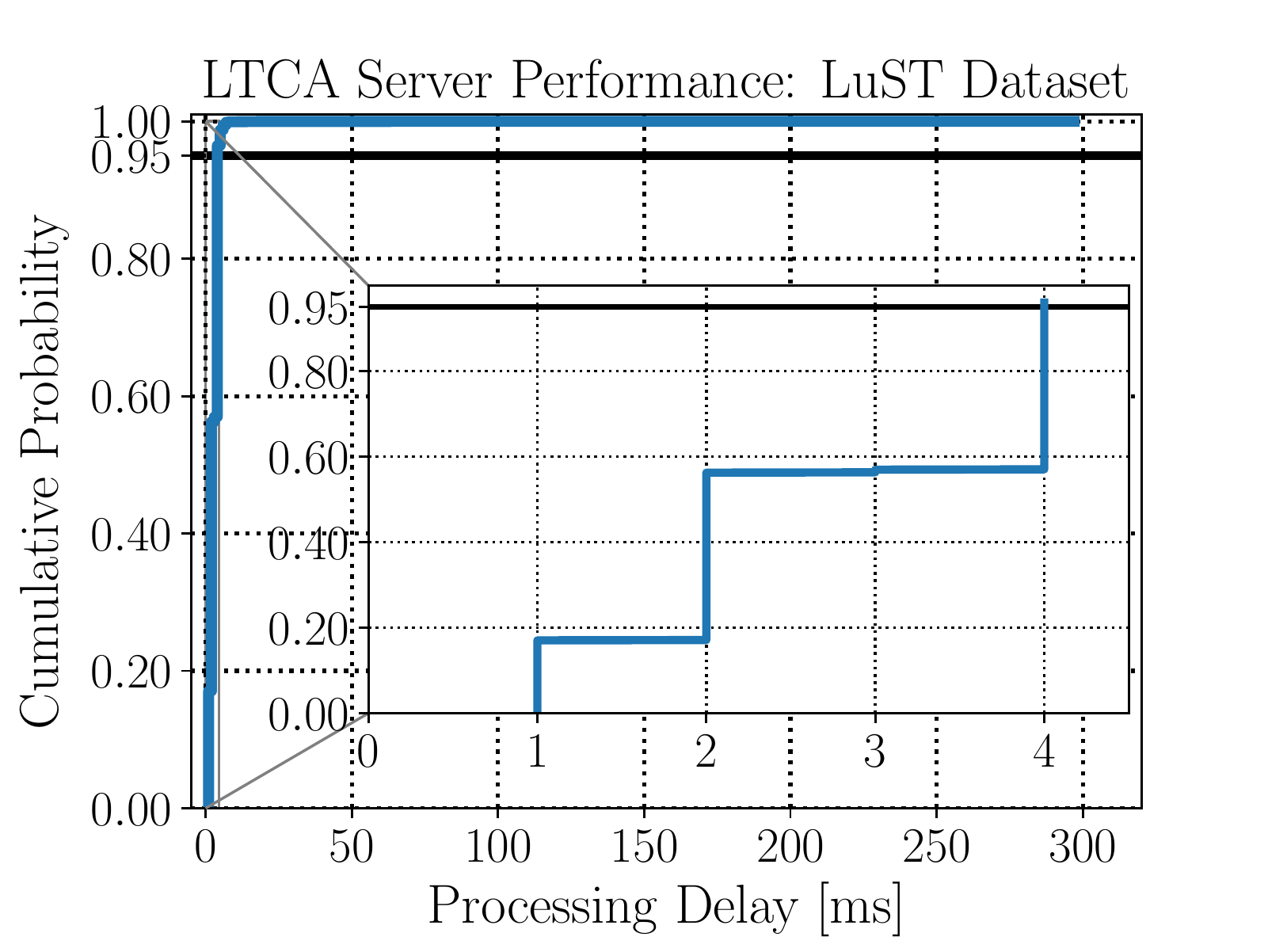}
		}
		\subfloat[Issuing pseudonyms]{
		\includegraphics[trim=0cm 0.1cm 0.5cm 0.6cm, clip=true, totalheight=0.135\textheight,angle=0,keepaspectratio] {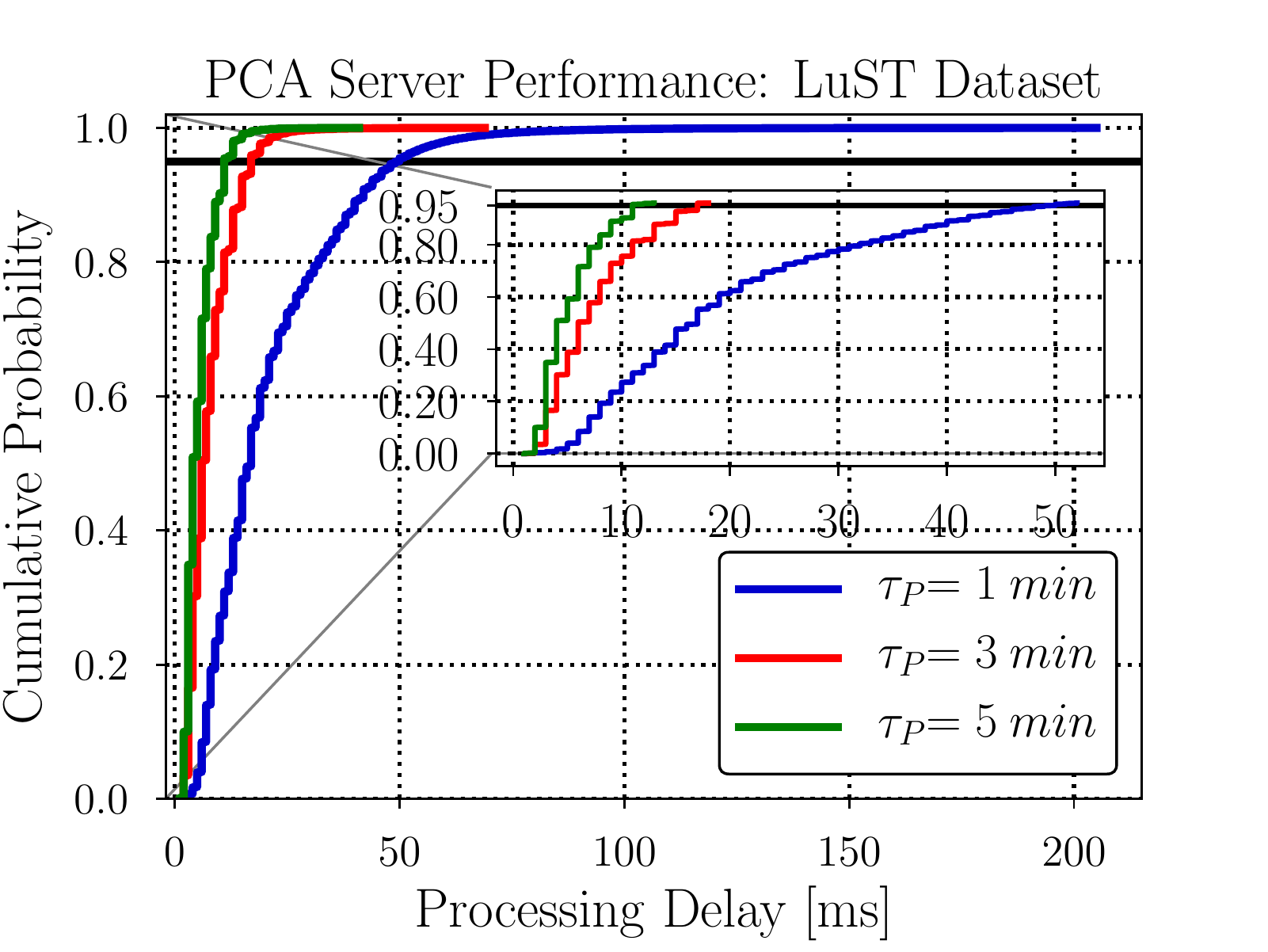}
		}
	\caption{CDF of server processing delay for P1, \ac{LuST} dataset.}
	\label{fig:secmace-vpki-processing-delay}
	\vspace{-0.3em}
\end{figure}

\textbf{Choice of Parameter:} The choice of parameter for $\Gamma_{P2/P3}$ and $\tau_{P}$ mainly determines the frequency of interaction with the \ac{VPKI} and the volume of workload imposed to the \ac{PCA}: the shorter the pseudonym lifetimes are, the greater number of pseudonyms will be requested, thus a higher workload is imposed on the \ac{PCA}. We evaluate the overall performance of the \ac{VPKI} servers to issue pseudonyms with short lifetimes, e.g., 60 sec, to investigate the behavior of the servers under a high-workload condition. 

\subsection{\ac{VPKI} Server Performance}
\label{subsec:secmace-vpki-server-performance}

\subsubsection{Ticket and Pseudonym Provisioning} 
\label{subsubsec:ticket-pseudonym-provisioning}

Fig.~\ref{fig:secmace-vpki-processing-delay}.a illustrates the CDF of a single ticket issuance processing delay for the Tapas dataset. For example, $F_x(t=4 \: ms)=0.95$, or $Pr\{t\leq4 \: ms\}=0.95$. Fig.~\ref{fig:secmace-vpki-processing-delay}.b shows the processing delay for issuing pseudonyms with different lifetimes for Tapas dataset. As illustrated, with $\tau_{P}$=1 min, around 95\% of requesters are served less than 52 ms: $F_x(t=52 \: ms)=0.95$, i.e., $Pr\{t\leq52 \: ms\}=0.95$. The results confirm the efficiency and scalability of our system.

\subsubsection{End-to-end Latency}
\label{subsubsec:end-to-end-latency}

\begin{figure} [!t] 
	\centering
	\subfloat[Tapas dataset]{ 
		\includegraphics[trim=0cm 0cm 0cm 0.5cm, clip=true, totalheight=0.25\textheight, width=0.25\textwidth, angle=0, keepaspectratio] {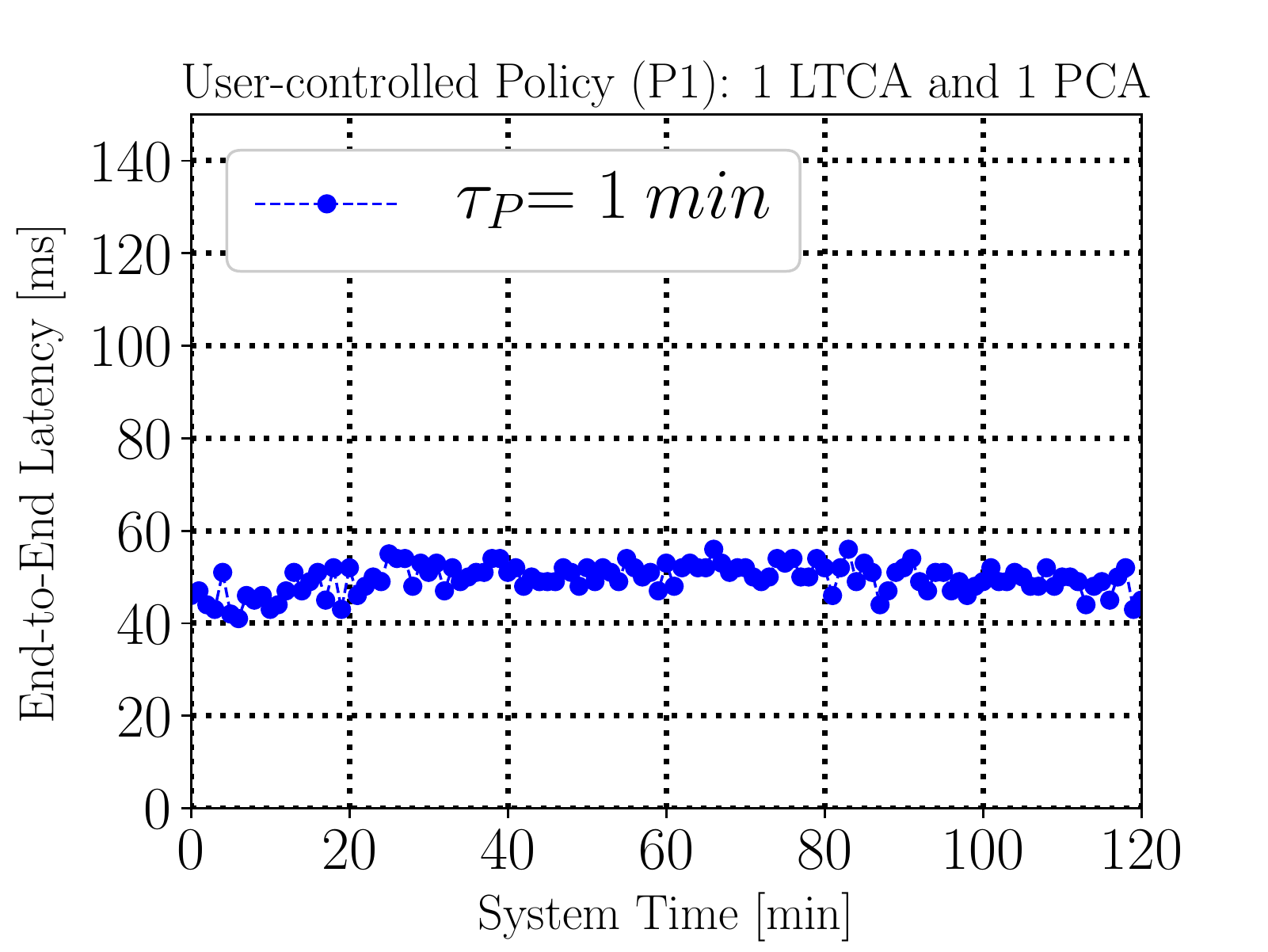}}
	\subfloat[\ac{LuST} dataset]{ 
		\includegraphics[trim=0cm 0cm 0cm 0.5cm, clip=true, totalheight=0.25\textheight, width=0.25\textwidth, angle=0, keepaspectratio] {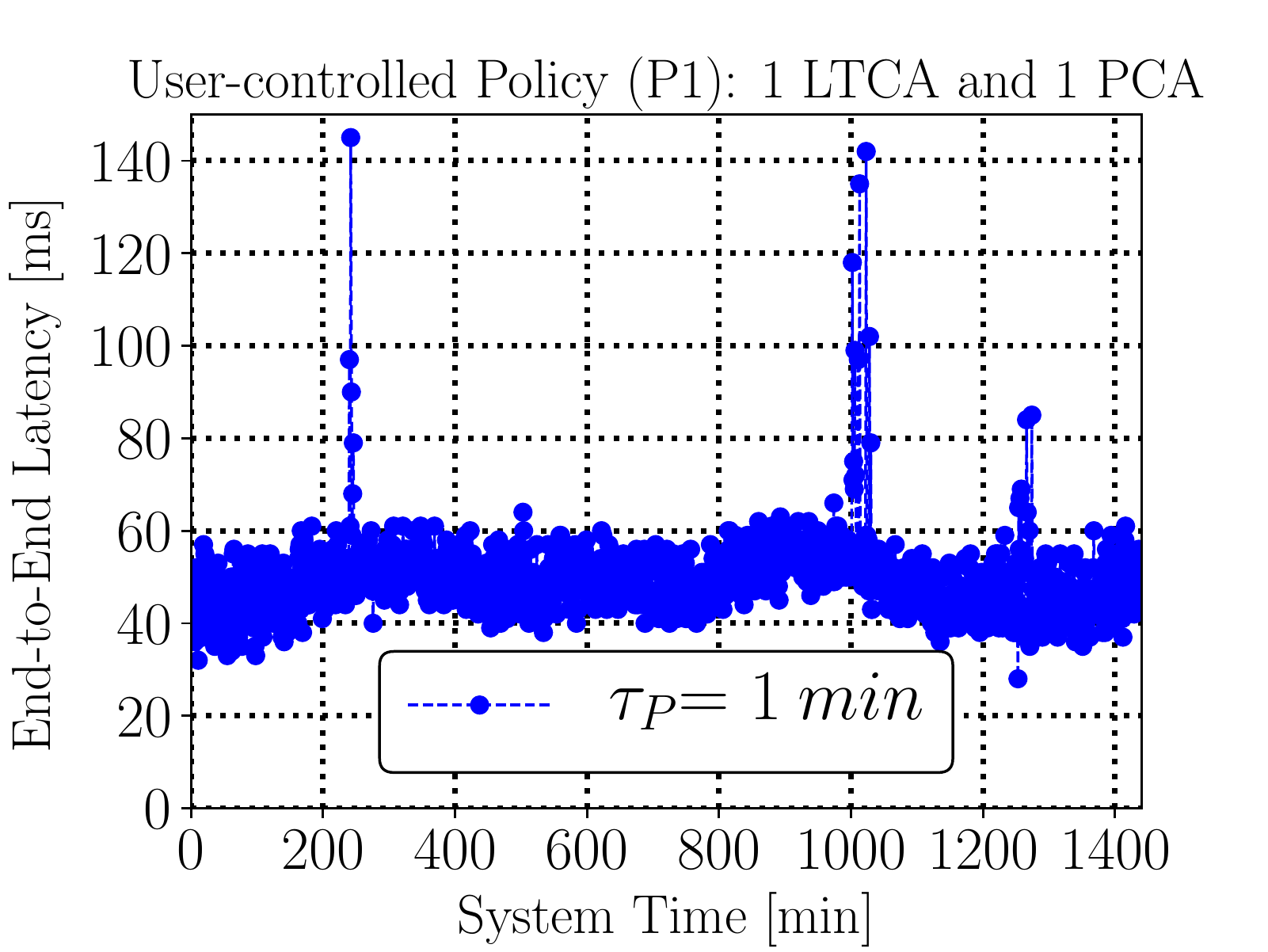}}
	\\
	\subfloat[Tapas dataset]{
		\includegraphics[trim=0cm 0cm 0cm 0.5cm, clip=true, totalheight=0.25\textheight, width=0.25\textwidth, angle=0, keepaspectratio] {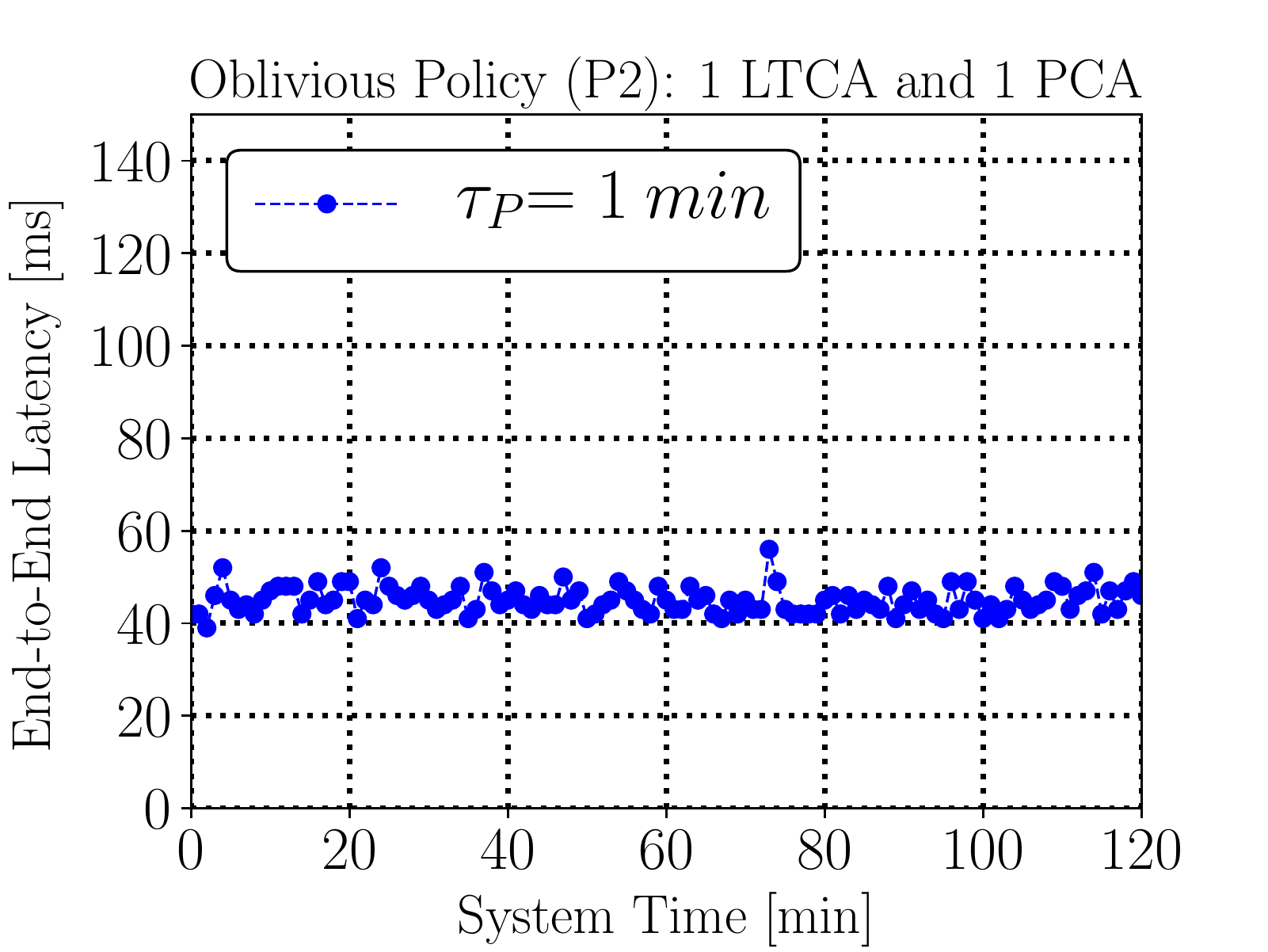}}
	\subfloat[\ac{LuST} dataset]{ 
		\includegraphics[trim=0cm 0cm 0cm 0.5cm, clip=true, totalheight=0.25\textheight, width=0.25\textwidth, angle=0, keepaspectratio] {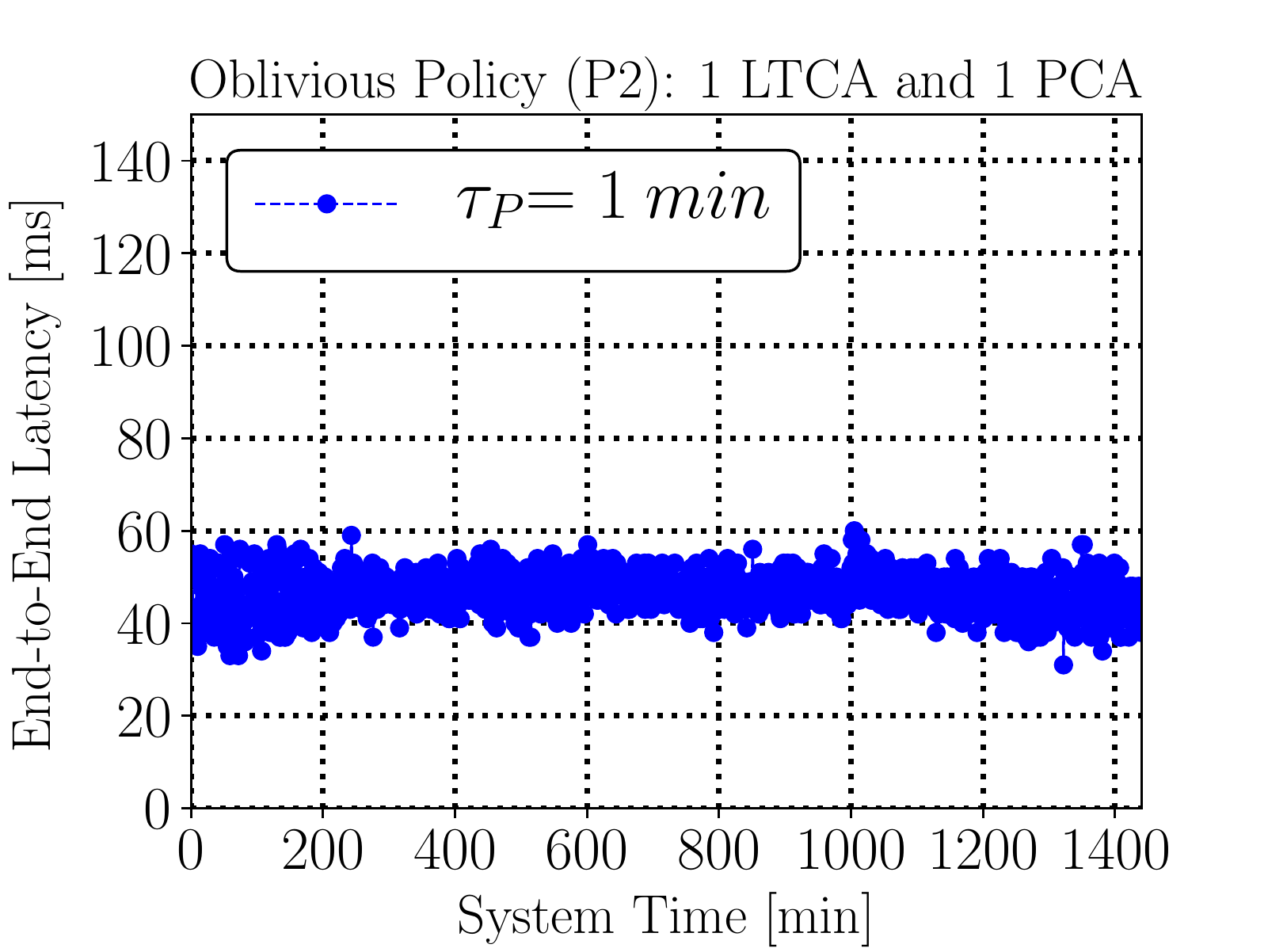}}
	\\
	\subfloat[Tapas dataset]{
		\includegraphics[trim=0cm 0cm 0cm 0.5cm, clip=true, totalheight=0.25\textheight, width=0.25\textwidth, angle=0, keepaspectratio] {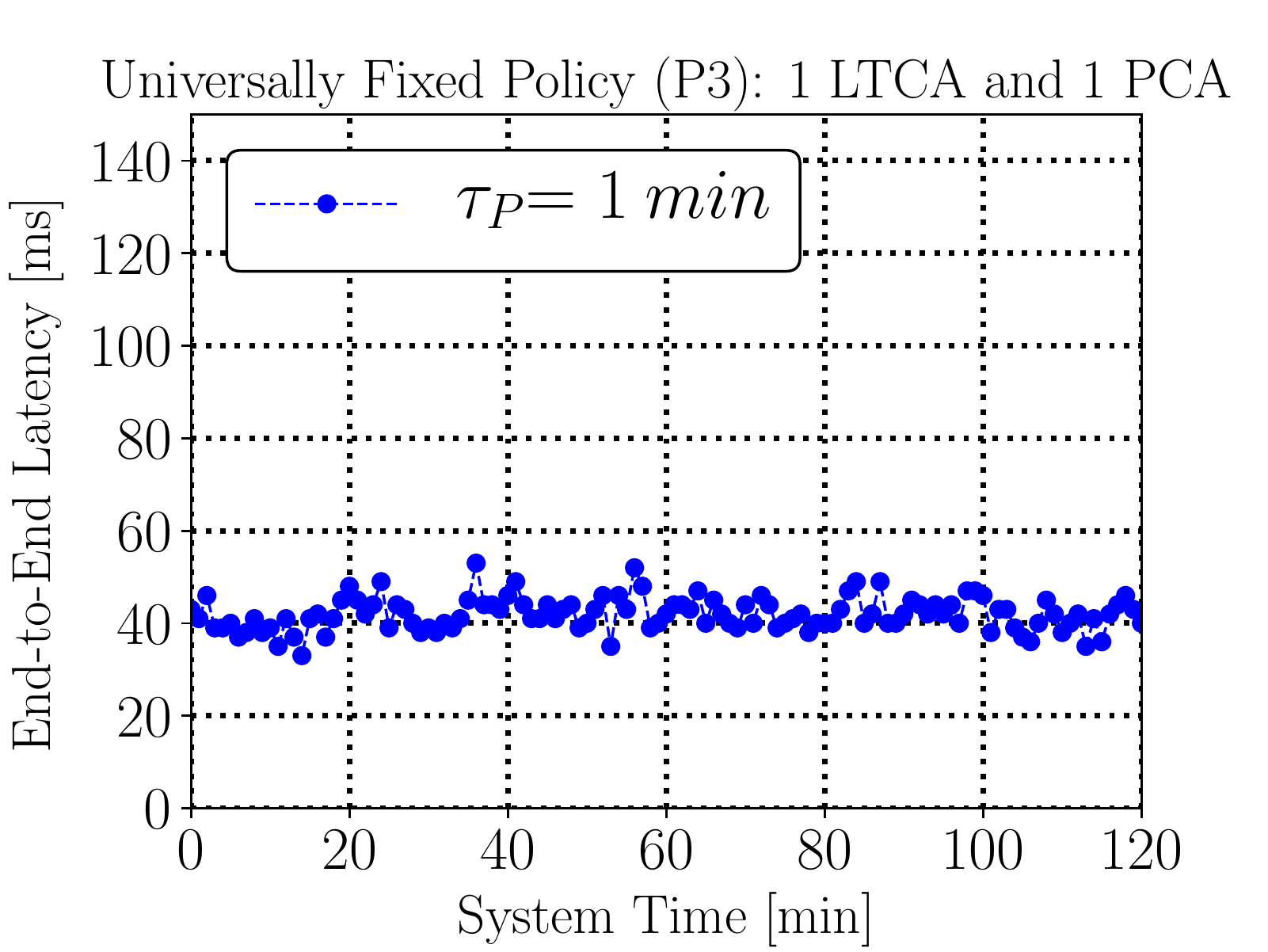}}
	\subfloat[\ac{LuST} dataset]{ 
		\includegraphics[trim=0cm 0cm 0cm 0.5cm, clip=true, totalheight=0.25\textheight, width=0.25\textwidth, angle=0, keepaspectratio] {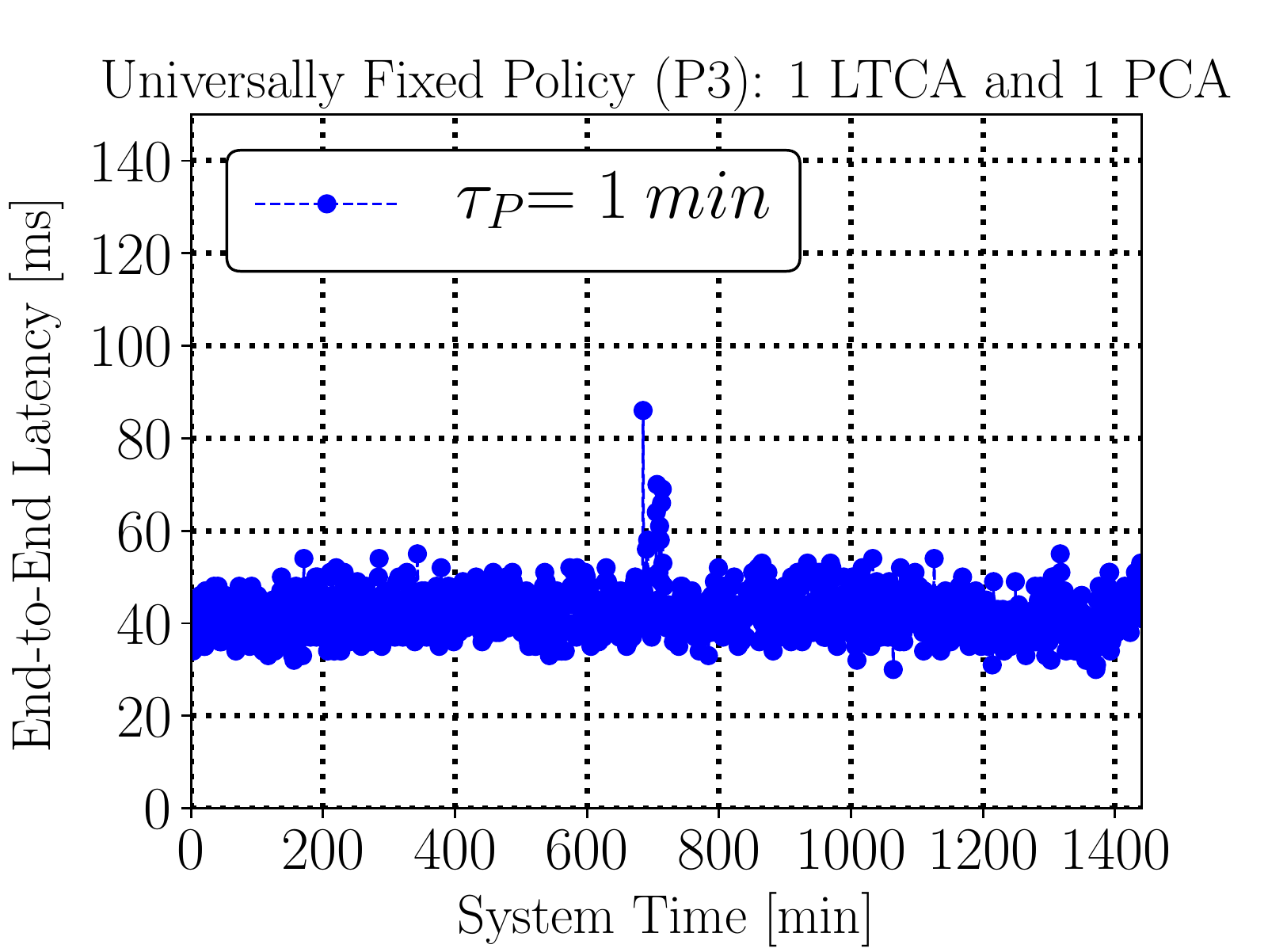}}
	\caption{Average end-to-end latency for different policies ($\Gamma_{}$=10 min, $\tau_{P}$=1 min). The first row (a, b), the second row (c, d), and the third row (e, f) show the average end-to-end latency for user-controlled policy (P1), oblivious policy (P2), and universally fixed policy (P3), respectively. The first column represents the delays for Tapas dataset while the second one shows the delay for \ac{LuST} dataset.} 
	\label{fig:secmace-tapas_lust_average_latencies}
	\vspace{-1em}
\end{figure}

\begin{table}[!t]
	\caption{End-to-end Latency Statistics \protect\linebreak ($\Gamma_{}$=10 min, $\tau_{P}$=1 min).}
	\centering
	\resizebox{0.5\textwidth}{!}
	{
		\begin{tabular}{| l || *{3}{c} || *{3}{c} |}
			\hline
			& \textbf{Tapas-P1} & \textbf{Tapas-P2} & \textbf{Tapas-P3} & \textbf{\acs{LuST}-P1} & \textbf{\acs{LuST}-P2} & \textbf{\acs{LuST}-P3} \\\hline 
			\textbf{Maximum (ms)}               & 296 & 320 & 5,168 & 3,062 & 248 & 5,545  \\\hline 
			\textbf{Minimum (ms)}               & 19  & 26 & 18 & 18 & 26 & 18 \\\hline 
			\textbf{Average (ms)}               & 50.29 & 45.56 & 42.76 & 51.78 & 47.58 & 43.10 \\\hline
			\textbf{Std. Deviation}        & 16.26 &  12.10 & 26.19 & 35.40 & 11.72 & 23.53 \\\hline 
			\textbf{Variance} & 264.35 & 146.5 & 685.83 & 1253.22 & 137.25 & 553.84 \\\hline 
			$\mathbf{Pr\{t\leq x \: (ms)\}=0.99}$ & 102 & 83 & 69 & 110 & 80 & 70 \\\hline
		\end{tabular}
		\label{table:secmace-latency-statistics-for-each-policy}
	}
	\vspace{-1em}
\end{table}

\begin{table*}[!t]
	\caption{Latency Statistics for different policies, Tapas dataset ($\Gamma_{}$=10 min)} 
	\centering
	\resizebox{0.95\textwidth}{!}
	{
		\begin{tabular}{| l ||| *{3}{c} || *{3}{c}  || *{3}{c} |}
			\hline
			\shortstack{\textbf{Policy}} & \textbf{} & \textbf{P1} & \textbf{} & \textbf{} & \textbf{P2} & \textbf{} & \textbf{} & \textbf{P3} & \textbf{} \\\hline 
			\shortstack{\textbf{Metrics} \\ {} \\ {}} & \shortstack{\\ {} \textbf{Avg. E2E} \\ \textbf{latency (ms)}} & \shortstack{\textbf{Avg. no.} \\ \textbf{of psnyms}} & \shortstack{\textbf{Total no.} \\ \textbf{of psnyms}} & \shortstack{\textbf{Avg. E2E} \\ \textbf{latency (ms)}} & \shortstack{\textbf{Avg. no.} \\ \textbf{of psnyms}} & \shortstack{\textbf{Total no.} \\ \textbf{of psnyms}} & \shortstack{\textbf{Avg. E2E} \\ \textbf{latency (ms)}} & \shortstack{\textbf{Avg. no.} \\ \textbf{of psnyms}} & \shortstack{\textbf{Total no.} \\ \textbf{of psnyms}} \\\hline\hline 
			\textbf{$\tau_{P}$=30 (sec)}  & 75.65 & 20.17 & 1,524,227 & 74.17 & 20 & 2,226,560 & 60.51 & 14.63 & 2,196,277 \\\hline 
			\textbf{$\tau_{P}$=60 (sec)}  & 50.29 & 10.33 & 781,060 & 45.56 & 10 & 1,113,280 & 42.76 & 7.47 & 995,291 \\\hline 
			\textbf{$\tau_{P}$=120 (sec)} & 44.26 & 5.42 & 409,355 & 40.70 & 5 & 556,640 & 35.07 & 3.85 & 578,099 \\\hline 
			\textbf{$\tau_{P}$=180 (sec)} & 41.56 & 3.77 & 285,359 & \textemdash\textemdash & \textemdash\textemdash & \textemdash\textemdash & \textemdash\textemdash & \textemdash\textemdash & \textemdash\textemdash \\\hline 
			\textbf{$\tau_{P}$=240 (sec)} & 35.20 & 2.96 & 223,578 & \textemdash\textemdash & \textemdash\textemdash & \textemdash\textemdash & \textemdash\textemdash & \textemdash\textemdash & \textemdash\textemdash \\\hline 
			\textbf{$\tau_{P}$=300 (sec)} & 35.21 & 2.46 & 186,116 & 33.86 & 2 & 222,656 & 32.19 & 1.70 & 255,384 \\\hline 
			\textbf{$\tau_{P}$=360 (sec)} & 34.62 & 2.13 & 161,211 & \textemdash\textemdash & \textemdash\textemdash & \textemdash\textemdash & \textemdash\textemdash & \textemdash\textemdash & \textemdash\textemdash \\\hline 
			\textbf{$\tau_{P}$=420 (sec)} & 33.40 & 1.90 & 143,498 & \textemdash\textemdash & \textemdash\textemdash & \textemdash\textemdash & \textemdash\textemdash & \textemdash\textemdash & \textemdash\textemdash \\\hline 
			\textbf{$\tau_{P}$=480 (sec)} & 32.17 & 1.72 & 125,074 & \textemdash\textemdash & \textemdash\textemdash & \textemdash\textemdash & \textemdash\textemdash & \textemdash\textemdash & \textemdash\textemdash \\\hline 
			\textbf{$\tau_{P}$=540 (sec)} & 31.74 & 1.58 & 119,481 & \textemdash\textemdash & \textemdash\textemdash & \textemdash\textemdash & \textemdash\textemdash & \textemdash\textemdash & \textemdash\textemdash \\\hline 
			\textbf{$\tau_{P}$=600 (sec)} & 31.63 & 1.47 & 111,237 & 32.23 & 1 & 111,328 & 31.63 & 1 & 150,071 \\\hline 
		\end{tabular}
		\label{table:secmace-tapas-latency-statistics-for-each-policy}
	}
\end{table*}

We are primarily concerned with the \emph{end-to-end latency}, i.e., the delay for pseudonym acquisition, measured at the vehicle, calculated from the initialization of Protocol~\ref{protocol:secmace-ticket-provision-in-home-domain} till the successful completion of Protocol~\ref{protocol:secmace-pseudonym-provision-in-home-domain}.\footnote{The processing time to generate the key pairs is not considered here as the \ac{OBU} can generate them off-line.} Table~\ref{table:secmace-latency-statistics-for-each-policy} details the latency statistics to obtain pseudonyms with different policies for the two datasets. Figs.~\ref{fig:secmace-tapas_lust_average_latencies} show the average latency for the vehicles with different pseudonym acquisition policies. With P1 (Figs.~\ref{fig:secmace-tapas_lust_average_latencies}.a and~\ref{fig:secmace-tapas_lust_average_latencies}.b), each vehicle requests all required pseudonyms at once. With $\tau$=1 min, 99\% of the requesters for the Tapas and \ac{LuST} datasets are served within less than 102 ms and 110 ms respectively. As we see, there are some sudden jumps in Fig.~\ref{fig:secmace-tapas_lust_average_latencies}.b: the principal reason is that P1 allows vehicles to request pseudonyms for any trip duration; thus, long trip durations result in requesting more pseudonyms at once. 

\begin{figure} [!t] 
	\begin{center}
		\centering
		\includegraphics[trim=0.5cm 0.1cm 1cm 0.7cm, clip=true,totalheight=0.4\textheight, width=0.4\textwidth, angle=0, keepaspectratio] {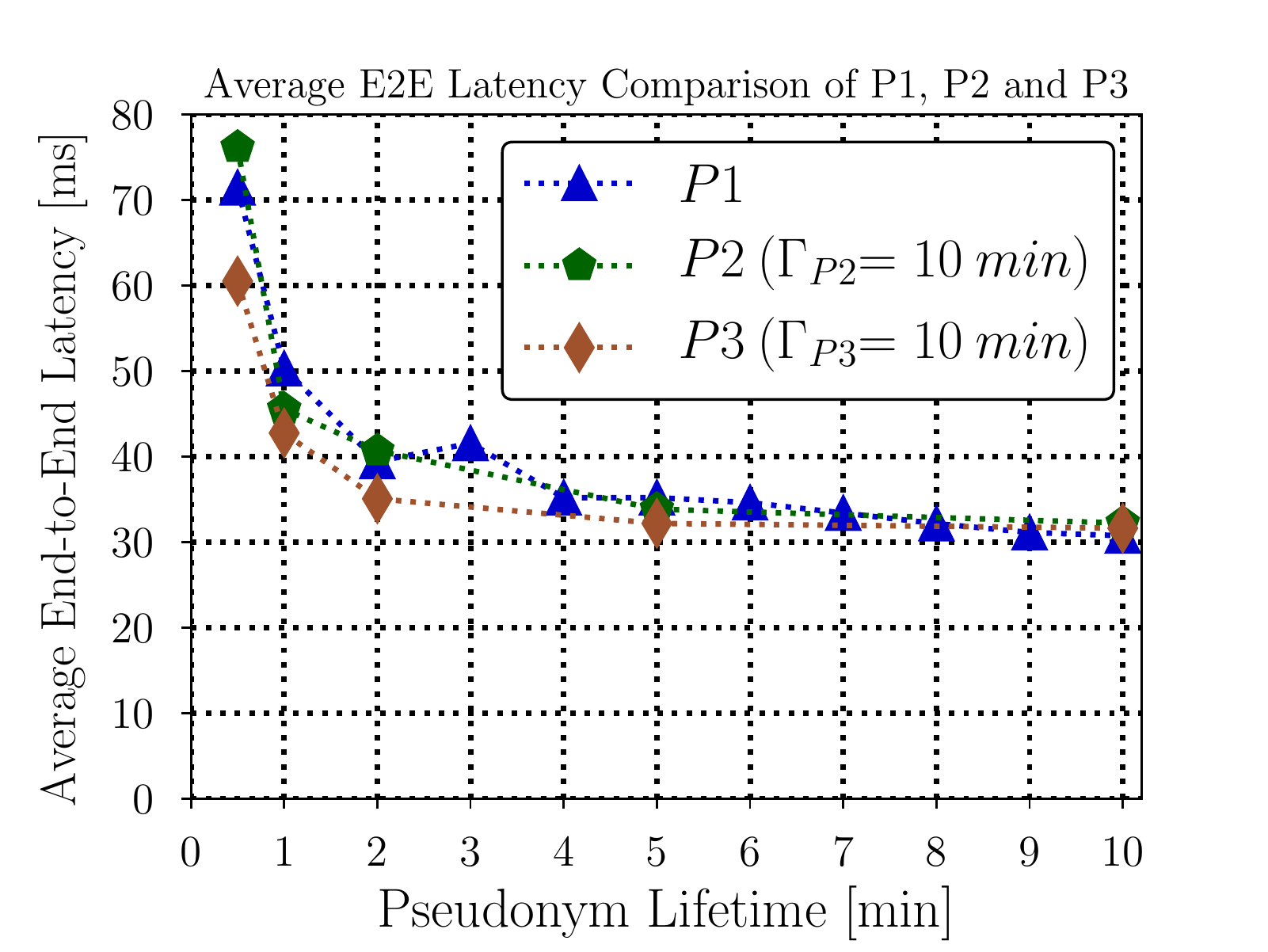}
		\caption{End-to-end latency comparison for different policies (Tapas dataset).}
		\label{fig:secmace-average-e2e-latency-for-tapas-datasets}
	\end{center}
	\vspace{-0.5em}
\end{figure}

With P2 (Figs.~\ref{fig:secmace-tapas_lust_average_latencies}.c and~\ref{fig:secmace-tapas_lust_average_latencies}.d), vehicles request a fixed amount of pseudonyms every time (for a duration of $\Gamma_{P2}$=10 min), thus never overloading the \ac{PCA} server with a large amount of pseudonyms in a single request; this results in low standard deviation and variance, and a smoother average delay in comparison to P1. The average end-to-end latency for Tapas and \ac{LuST} datasets ($\tau_{P}$=1 min) is 46 ms and 48 ms respectively; accordingly, 99\% of vehicles are served within less than 83 ms and 80 ms respectively. 

With P3 (Fig.~\ref{fig:secmace-tapas_lust_average_latencies}.e and~\ref{fig:secmace-tapas_lust_average_latencies}.f), the system enforces synchronized batch arrivals to obtain pseudonyms: each vehicle requests pseudonyms for the entire $\Gamma_{P3}$, timely aligned with the rest. 99\% of the vehicles for the two datasets are served within less than 69 ms and 70 ms respectively. This confirms that the most promising policy in terms of privacy protection incurs even lower overhead in compare to other policies. All in all, our secure and privacy preserving scheme efficiently issues pseudonyms for the requesters and an \ac{OBU} can initiate a request for pseudonyms within the lifetime of the last single valid pseudonym. 

Fig.~\ref{fig:secmace-average-e2e-latency-for-tapas-datasets} and Table~\ref{table:secmace-tapas-latency-statistics-for-each-policy} show a comparison of the average end-to-end latency for different pseudonym acquisition policies. P3 incurs the lowest delay among the three policies: for example, the average end-to-end latency for P1, P2, and P3, with $\tau_{P}$=60 sec is 50, 46, and 43 ms respectively. With P1, each vehicle requests all the required pseudonyms at once, which results in a higher workload on the \ac{PCA}, thus higher latency. In other words, for P2 and P3, a request with large number of pseudonyms is split into multiple requests, each with fewer pseudonyms, thus achieving better performance due to the parallelization in multi-core processors.

\begin{table}[!t]
	\caption{End-to-end Latency Statistics \protect\linebreak with P3 ($\Gamma_{P3}$=1 min, $\tau_{P}$=1 min)}
	\centering
	\resizebox{0.5\textwidth}{!}
	{
		\begin{tabular}{| l | *{1}{c} |*{1}{c} |*{1}{c} |*{1}{c} |*{1}{c} |*{1}{c} |}
			\hline
			& \textbf{Maximum (ms)} & \textbf{Minimum (ms)} & \textbf{Average (ms)} & \textbf{Std. Deviation} & \textbf{Variance} & $\mathbf{Pr\{t\leq x \: (ms) \}=0.99}$ \\\hline 
			\textbf{Tapas}      & 6,462 & 17 & 34.72 & 24.69 & 609.39 & 86 \\\hline 
			\textbf{\acs{LuST}} & 5,043 & 19 & 34.72 & 21.52 & 462.92 & 54 \\\hline
		\end{tabular}
		\label{table:secmace-latency-statistics-for-fully-unlinkable-pseudonyms}
	}
	\vspace{-1em}
\end{table}

Furthermore, the average end-to-end latency with P3 is lower than that with P2: the reason is that, with P3, each vehicle requests pseudonyms only for the \emph{``current''} $\Gamma_{P3}$; this results in the acquisition of only non-expired pseudonyms for the residual trip duration; while, with P2, each vehicle requests pseudonyms for an entire $\Gamma_{P2}$, out of which all pseudonyms are actually obtained. This is why the average end-to-end latency with P3 is lower than that with P2 (assuming $\Gamma_{P2} = \Gamma_{P3}$).

\subsubsection{Fully-unlinkable Pseudonym Provisioning}
\label{subsubsec:secmace-unlinkable-pseudonyms-evaluation}

Table~\ref{table:secmace-latency-statistics-for-fully-unlinkable-pseudonyms} details the latency statistics to obtain pseudonyms with P3 ($\Gamma_{P3}=\tau_{P}=1$ min). The cumulative probability of end-to-end latency for Tapas and \ac{LuST} datasets is: $Pr\{t\leq86 \: ms\} = 0.99$, and $Pr\{t\leq54 \: ms\}=0.99$ respectively. With this probability, one can be fairly assured that even under this seemingly extreme configuration, the system is workable, i.e., the servers can issue fully unlinkable pseudonyms for the requesters. Nonetheless, there are rare events where the latency jumps; this indicates that either we need to enhance servers processing power, or trade it off by requesting small sets of linkable pseudonyms.

\begin{figure} [!t] 
	\begin{center}
		\centering
		\subfloat[P2]{
			\includegraphics[trim=0.4cm 0.1cm 1cm 0.7cm, clip=true,totalheight=0.25\textheight, width=0.25\textwidth, angle=0, keepaspectratio] {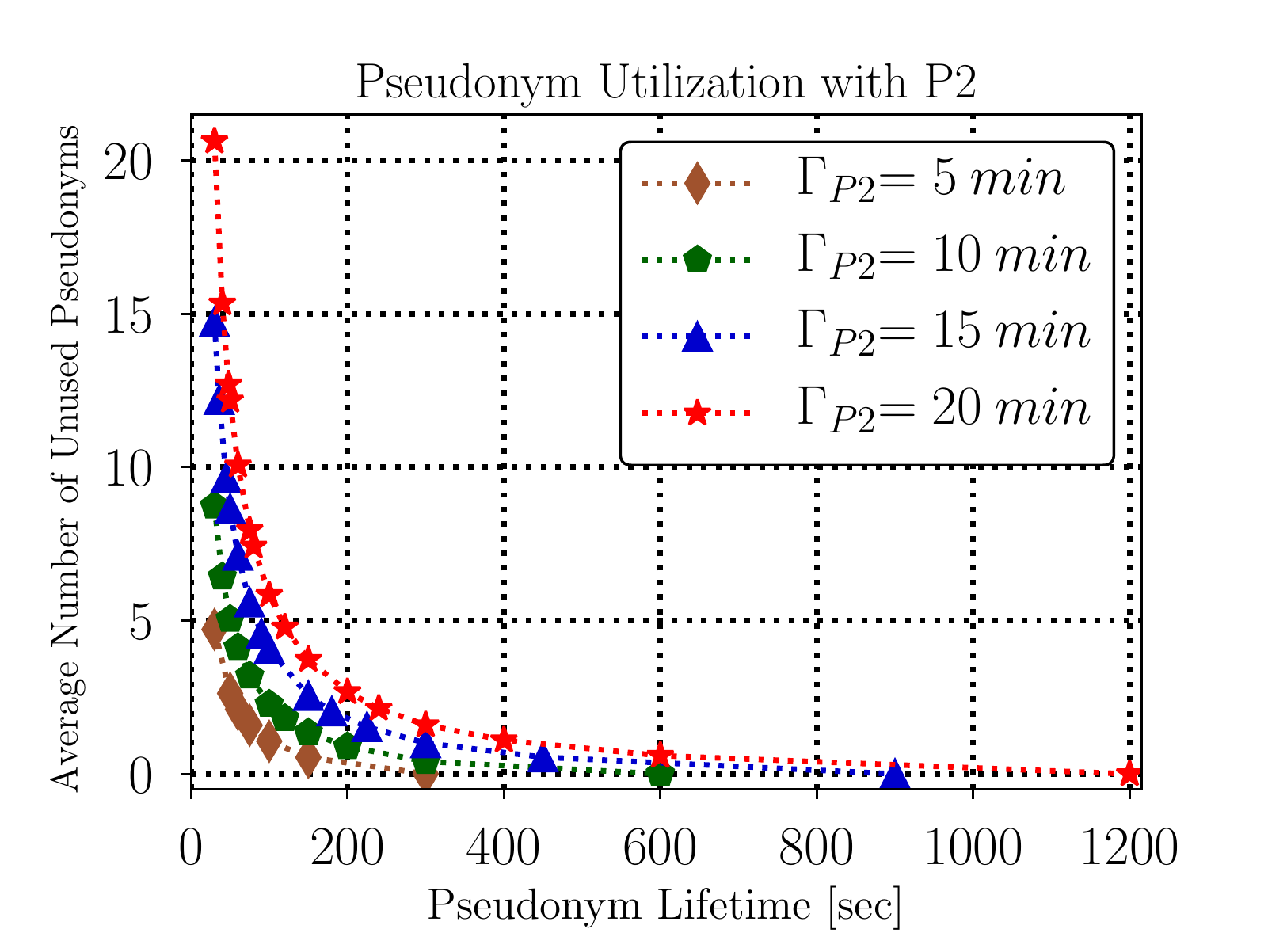}}
		\subfloat[P3]
		{\includegraphics[trim=0.4cm 0.1cm 1cm 0.7cm, clip=true,totalheight=0.25\textheight, width=0.25\textwidth, angle=0, keepaspectratio] {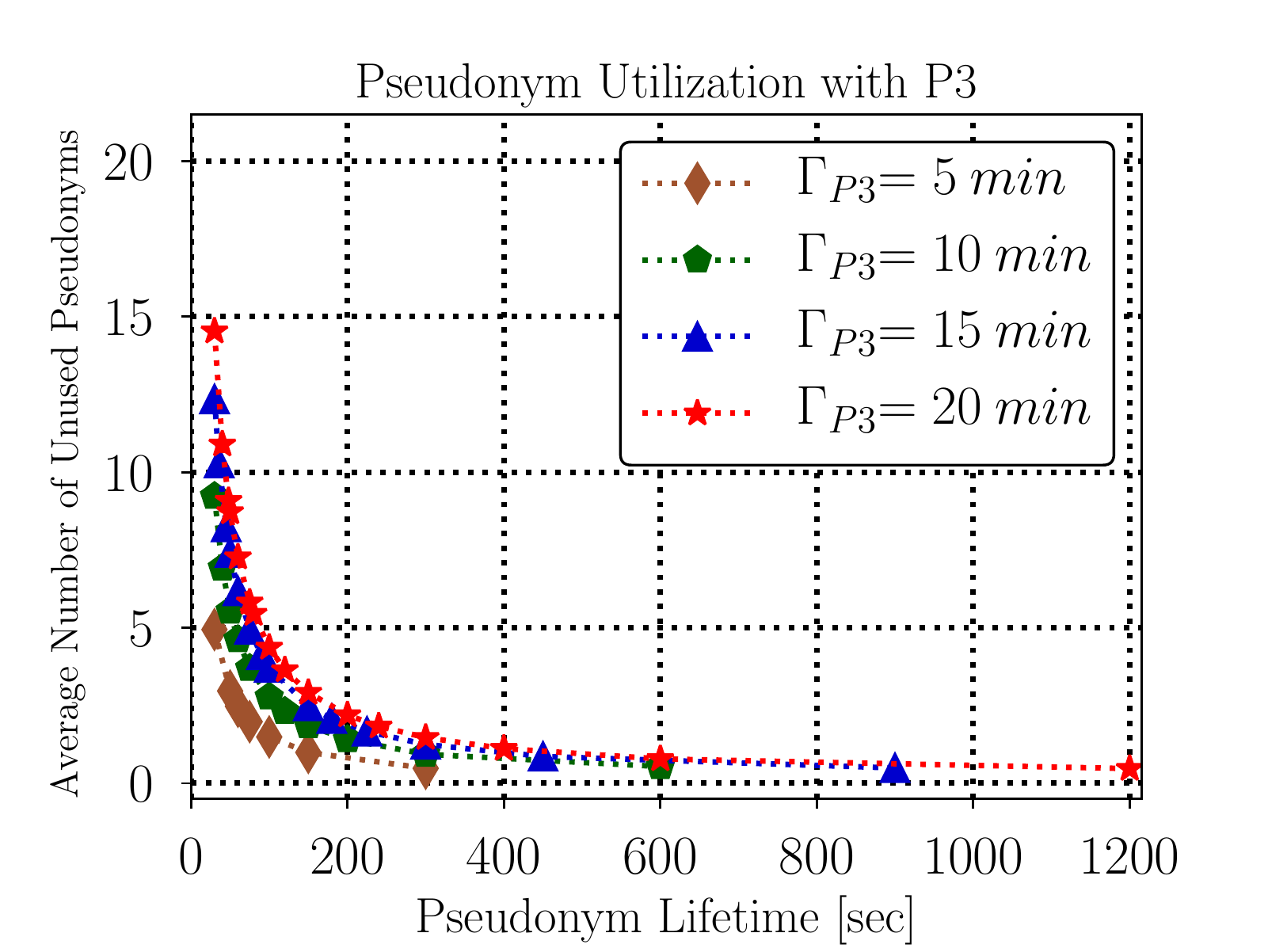}}
		\caption{Pseudonym utilization, \ac{LuST} dataset for P2 and P3.}
		\label{fig:secmace-average-psnym-utilization-for-lust-datasets-P2-P3}
	\end{center}
\end{figure}

\subsubsection{Optimal Pseudonym Utilization}
\label{subsubsec:secmace-optimal-pseudonym-utilization}

Using P1, each vehicle interacts with the \ac{VPKI} servers once to obtain the necessary pseudonyms for the entire trip duration (ideally without over-provisioning). However, according to P2 and P3, vehicles could be potentially equipped with more pseudonyms than necessary, i.e., the \ac{PCA} might issue pseudonyms that the vehicle will not use them. Fig.~\ref{fig:secmace-average-psnym-utilization-for-lust-datasets-P2-P3} shows the average number of unused pseudonyms for \ac{LuST} datasets with P2 and P3. In general, the longer the pseudonym refill interval $\Gamma_{}$, i.e., $\Gamma_{P2/P3}$, and the shorter pseudonym lifetime ($\tau_{P}$) are, the less frequent the vehicle-\ac{VPKI} interactions but the higher the chance to overprovision a vehicle. In other words, the longer the $\Gamma_{}$ intervals and the $\tau_{P}$ are, the less the average number of unused pseudonyms is, thus the higher pseudonym utilization. For example, the average number of unused pseudonyms with P2 and P3, when $\Gamma_{}$ is 5 min and $\tau_{P}$ is 30 sec, is 4.7 and 4.9 respectively; this implies that under these configurations, each vehicle on average is issued approximately 5 unused pseudonyms. The flip side is that this would allow the \ac{PCA} to have each set of pseudonyms (as a result of each request) trivially linked.

\begin{figure} [!t]
	\centering
	{\includegraphics[trim=0.1cm 0.1cm 1cm 0.5cm, clip=true, totalheight=0.35\textheight, width=0.35\textwidth, angle=0, keepaspectratio]{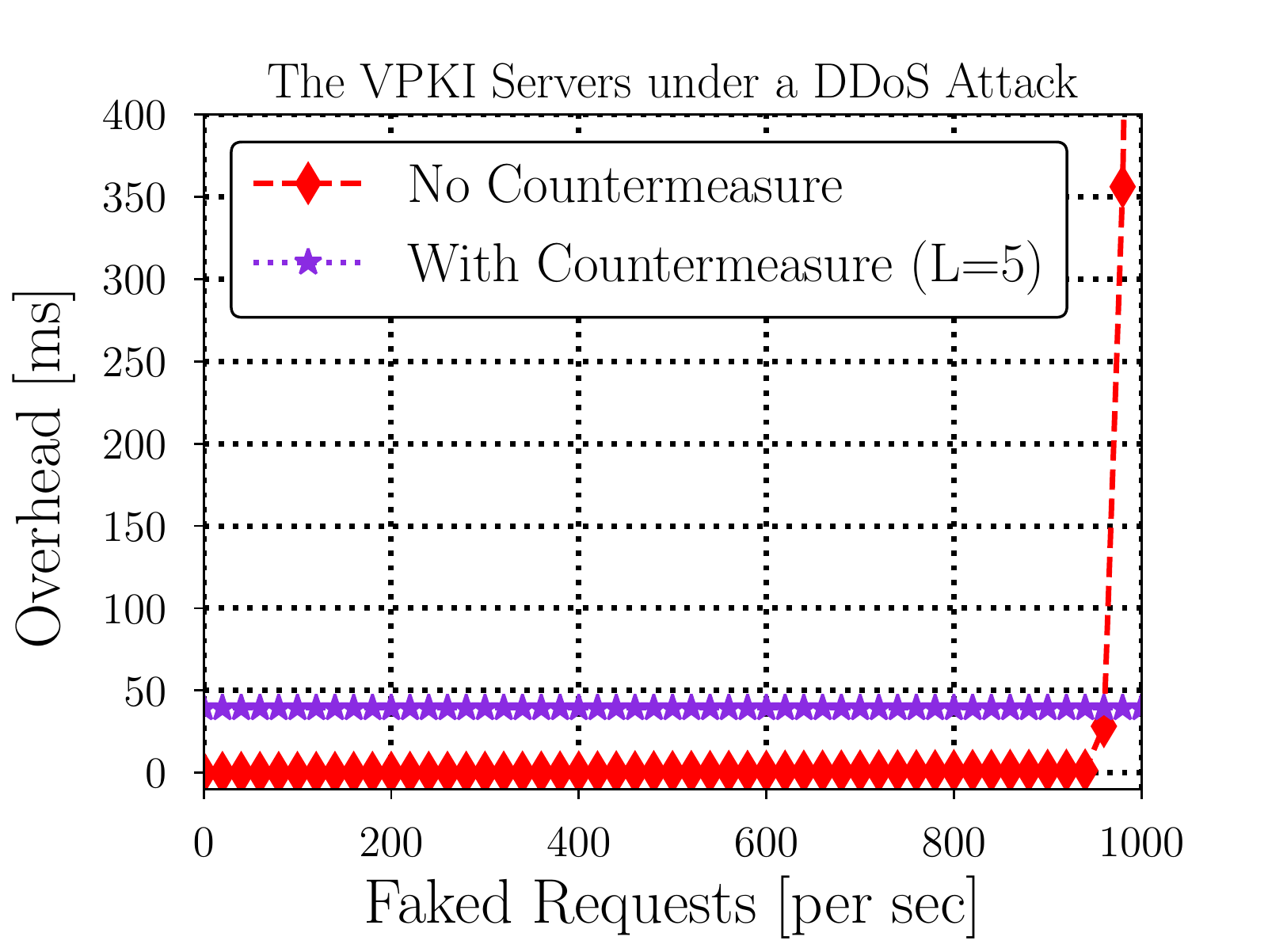}}
	\caption{The \ac{VPKI} servers are under a \acs{DDoS} attack: overhead to obtain pseudonyms for \ac{LuST} dataset with P1 (difficulty level: L=5, $\tau_P\:$=$\:$5 min).}
	\label{fig:secmace-lust-vpki-under-ddos-p1-and-countermeasure-in-place}
	\vspace{-0.5em}
\end{figure}

\subsubsection{\acs{DDoS} Attack}
\label{subsubsec:secmace-ddos-attack}

Internal adversaries could mount a clogging \ac{DoS} attack. A rate limiting mechanism prevents internal adversaries from affecting the system performance; moreover, the system flags the legitimate but misbehaving users, thus evicting them from the system. External adversaries could launch a \ac{DDoS} attack by clogging the \ac{LTCA} with faked certificates, or the \ac{PCA} with bogus tickets. 

To gauge the availability of the system, we evaluate the average system latency to issue pseudonyms under a \ac{DDoS} attack. We performed the experiments for different policies with various pseudonym lifetimes for the two datasets; we realized that the choice of policy, pseudonym lifetime, or dataset do not have a direct effect on the results; thus, we show the results for the \ac{LuST} dataset, as it represents a full-day scenario, with P1 and $\tau$=5 min. We increase the rate of adversarial requests up to 1,000 req/sec. As illustrated in Fig.~\ref{fig:secmace-lust-vpki-under-ddos-p1-and-countermeasure-in-place}, the average latency rapidly increases when the faked requests reach 1,000 req/sec. We use the guided tour puzzle~\cite{abliz2009guided} with difficulty level (L) 5 as a \ac{DDoS} mitigation technique to prevent the external adversaries from overflowing the servers with spurious requests. Using this mechanism, the power of an attacker is degraded to the power of a legitimate client; thus, an attacker cannot send high-rate spurious requests to the servers. Therefore, the attack is mitigated while the overhead to obtain pseudonyms for the legitimate vehicles increases only by approximately 50 ms.

\subsection{Revocation Update}
\label{subsec:secmace-revocation-performance}

\begin{figure} [!t]
	\begin{center}
		\centering
		\subfloat[\ac{CRL} acquisition] 
		{\includegraphics[trim=0.1cm 0.1cm 1cm 0.7cm, clip=true,totalheight=0.255\textheight, width=0.255\textwidth, angle=0, keepaspectratio]{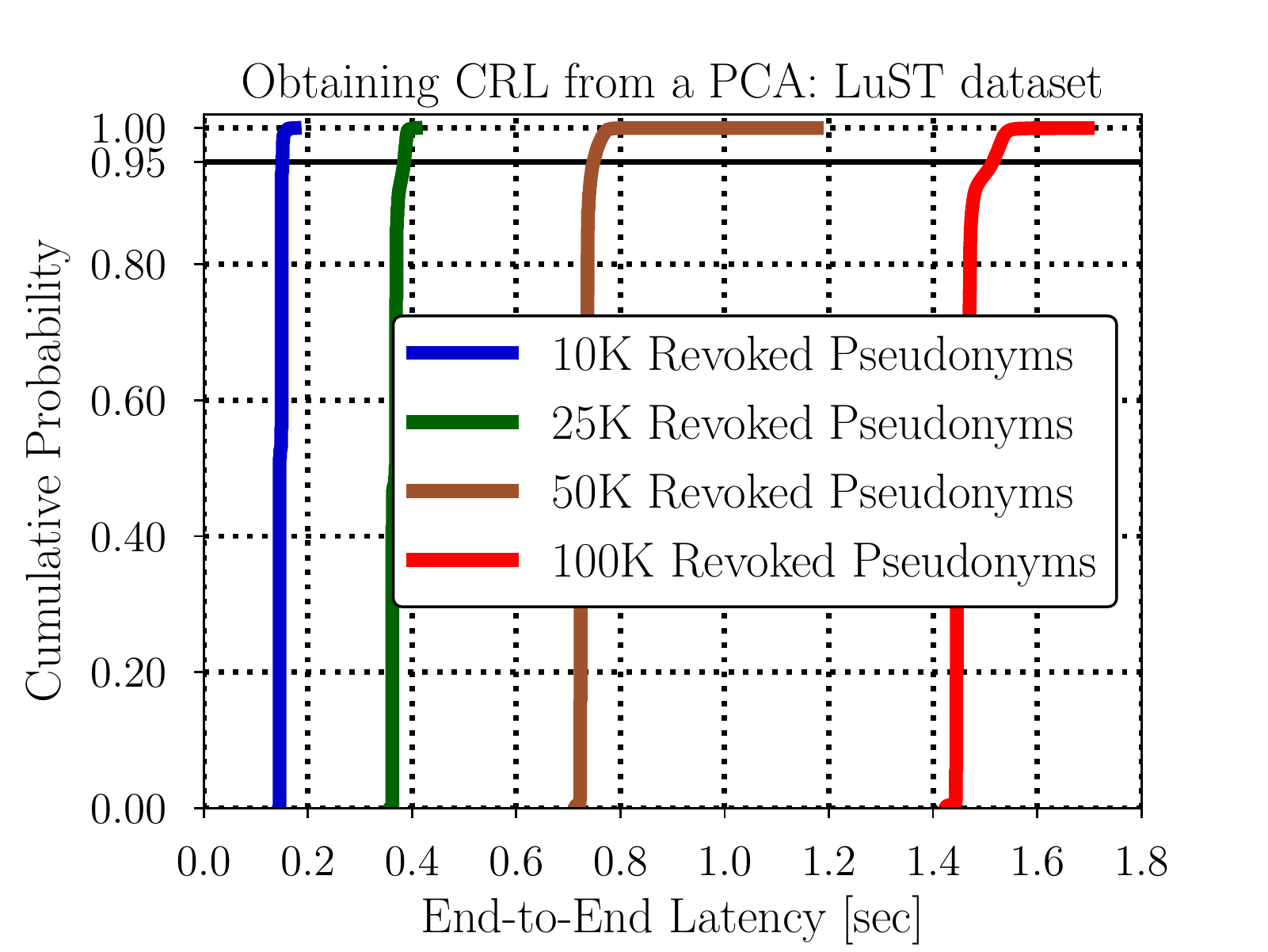}}
		\subfloat[\ac{OCSP} validation] 
		{\includegraphics[trim=0.1cm 0.1cm 1cm 0.7cm, clip=true,totalheight=0.255\textheight, width=0.255\textwidth, angle=0, keepaspectratio]{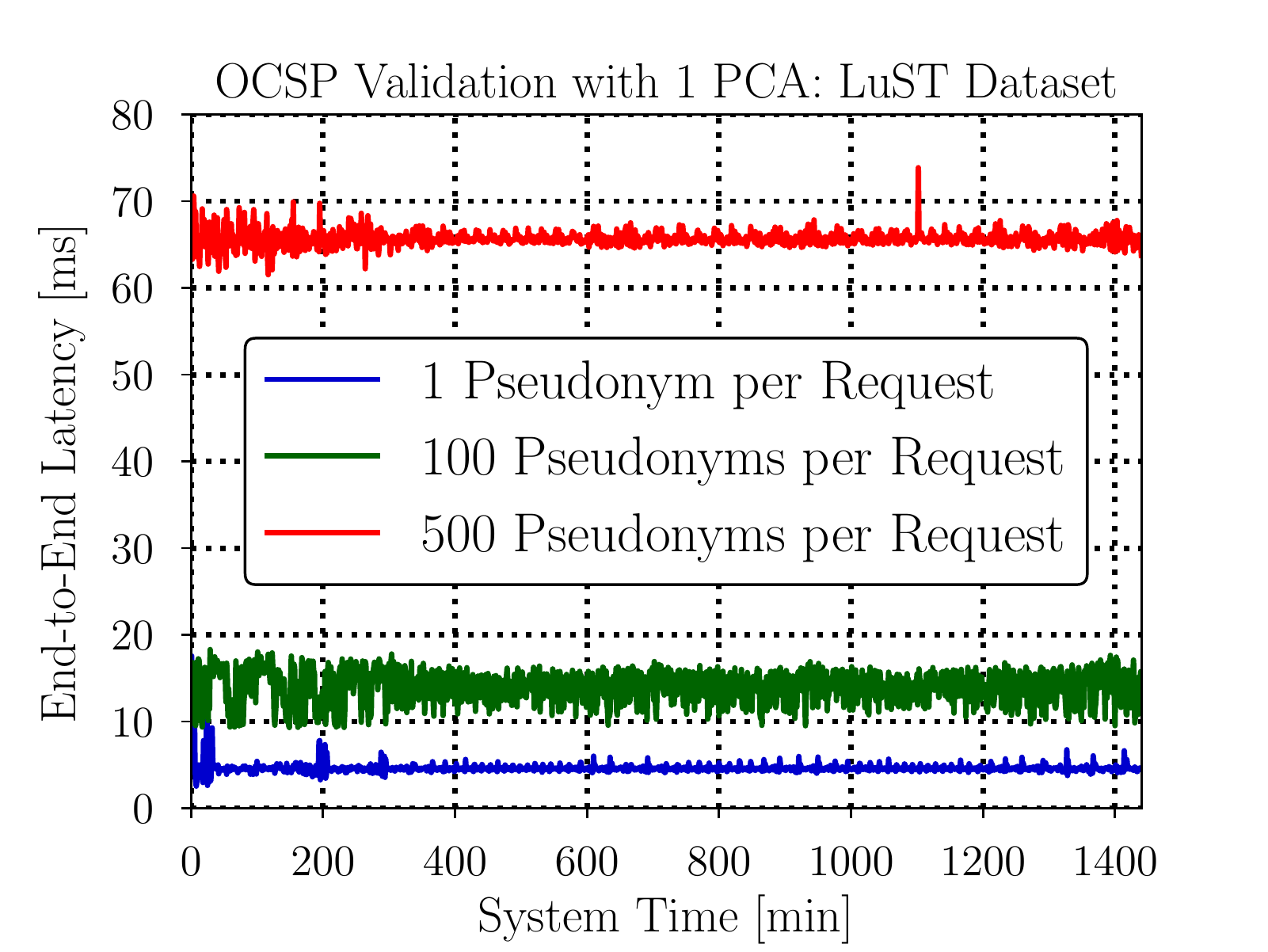}}
		\caption{End-to-end latency for revocation update (\ac{LuST} dataset).}
		\label{fig:secmace-crl-ocsp-performance-evaluation}
	\end{center}
\end{figure}

Fig.~\ref{fig:secmace-crl-ocsp-performance-evaluation} shows the CDF of latencies for obtaining a \ac{CRL} and the average latency to perform \ac{OCSP} validation for the \ac{LuST} dataset. With a modest \ac{VM} dedicated for the \ac{PCA} server, the results confirm the scalability of our system: 95\% of the requesters to fetch a \ac{CRL} with 100,000 revoked pseudonym are served within less than 1,500 ms, and the latencies for \ac{OCSP} validation with 500 pseudonyms never exceeds 75 ms.

\begin{figure} [!t]
	\begin{center}
		\centering
		\subfloat[Single domain]{
			\includegraphics[trim=0.1cm 0cm 0.5cm 0.7cm, clip=true,totalheight=0.255\textheight, width=0.255\textwidth, angle=0, keepaspectratio]{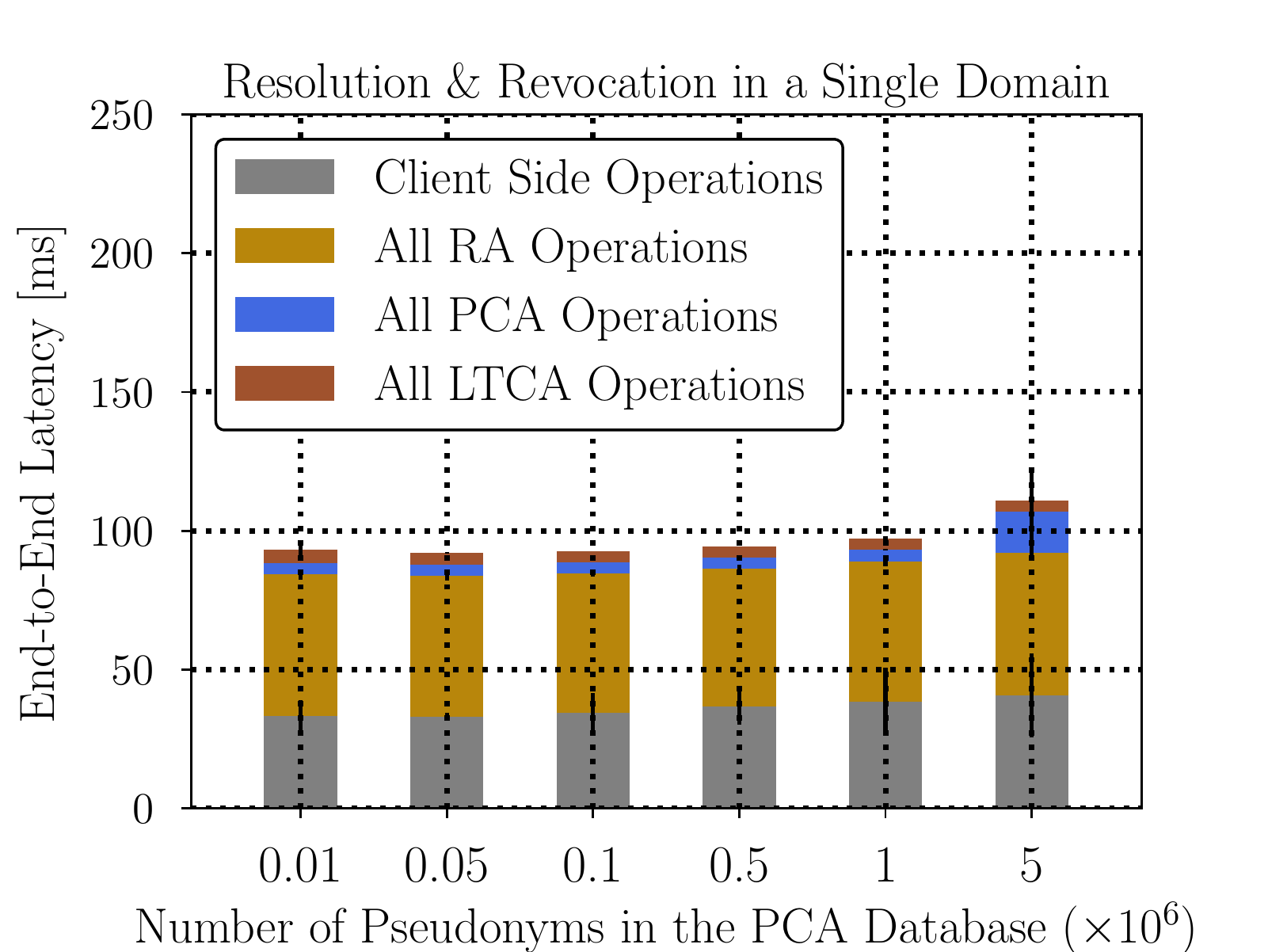}}
		\subfloat[Across domains]
			{\includegraphics[trim=0.1cm 0cm 0.5cm 0.7cm, clip=true,totalheight=0.255\textheight, width=0.255\textwidth, angle=0, keepaspectratio]{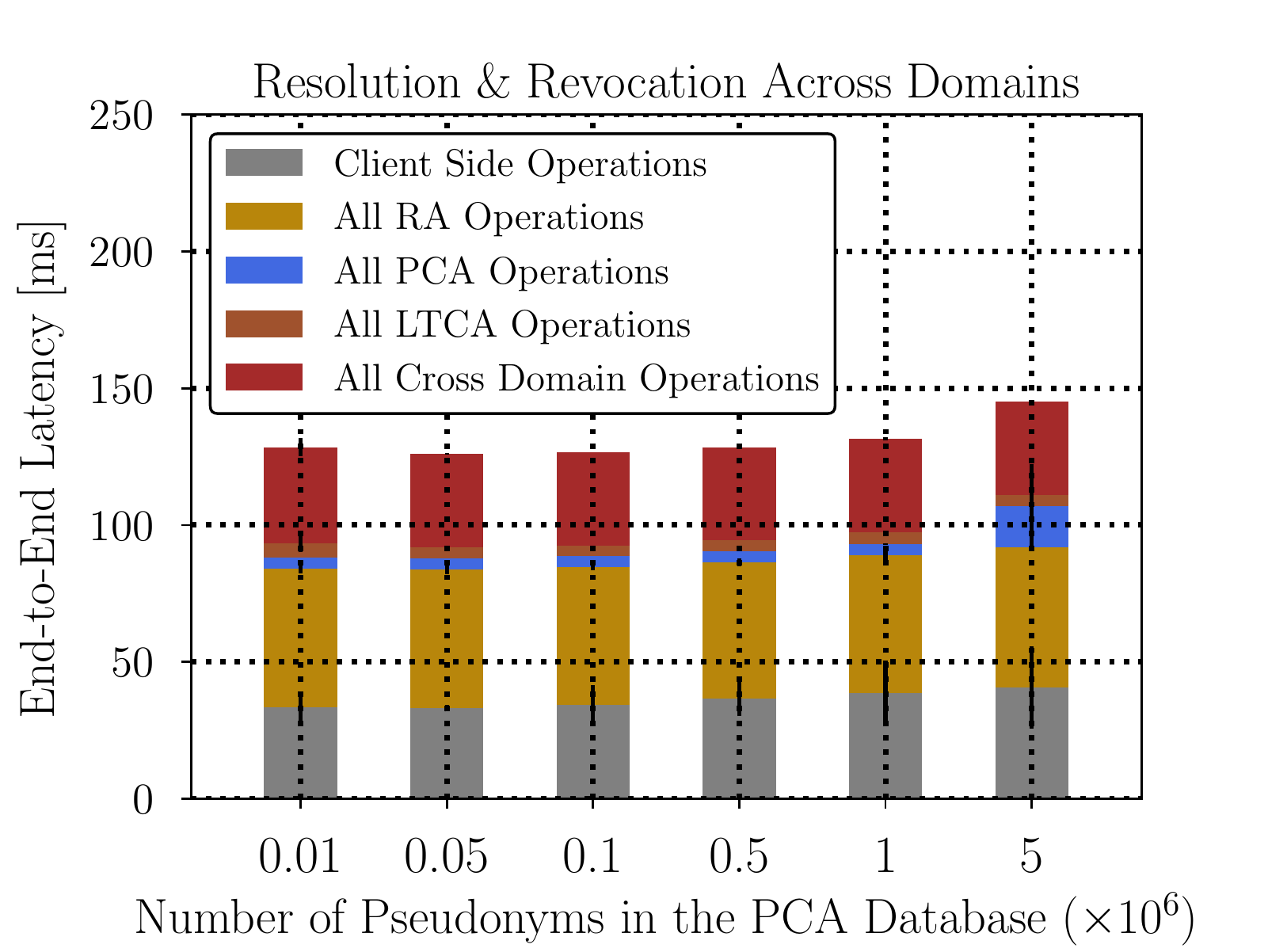}}
		\caption{End-to-end latency to resolve and revoke a pseudonym.}
		\label{fig:secmace-psnym-resolution-performance-evaluation}
	\end{center}
	\vspace{-0.8em}
\end{figure}

\subsection{Pseudonym Revocation and Resolution}
\label{subsec:secmace-pseudonym-resolution-performance}

Fig.~\ref{fig:secmace-psnym-resolution-performance-evaluation} shows the latency for each system component to resolve and revoke a pseudonym within a single domain (Fig.~\ref{fig:secmace-psnym-resolution-performance-evaluation}.a) and across domains (Fig.~\ref{fig:secmace-psnym-resolution-performance-evaluation}.b). We evaluate resolution with different number of pseudonyms in the \ac{PCA} database (from 10,000 to 5,000,000 pseudonyms). Unlike another scheme~\cite{bibmeyer2013copra} in which the performance is affected by increasing the number of revoked pseudonyms, our implementation is not. Our scheme outperforms other schemes, e.g., resolving and revoking a pseudonym in~\cite{gisdakis2013serosa} takes more than 2 sec while with the same system configuration, it takes approximately 100 ms in our implementation. 

\begin{table}[!h]
	\caption{Latency for issuing 100 pseudonyms \protect\linebreak (without communication delay)}
	\centering
	  \resizebox{0.5\textwidth}{!}
	  {
	    \begin{tabular}{l | *{6}{c} r}
		       & $Delay_{pca}$ & $CPU_{pca}$ \\ 
		      \hline
		      VeSPA~\cite{vespa2013} & 817 ms & 3.4 GHz (dual-core) \\
		      SEROSA~\cite{gisdakis2013serosa} & 650 ms & 2.0 GHz (dual-core) \\ 
		      PUCA~\cite{puca2014} & 1,000 ms & 2.53 GHz (dual-core) \\ 
		      \acs{PRESERVE} \acs{PKI} (Fraunhofer SIT)~\cite{preserve-url} & $\approx$ 4,000 ms & N/A \\
		      \acs{C2C-CC} \acs{PKI} (ESCRYPT)~\cite{c2c} & 393 ms & N/A \\ 
		      SECMACE & 260 ms & 2.0 GHz (dual-core) \\ 
	    \end{tabular}
	    \label{table:secmace-vpki-operations-efficiency-comparison}
      }
\end{table}

\subsection{Comparison with Other Implementations}
\label{subsubsec:secmace-results-comparison}

There are a few schemes with performance evaluation of their implementations. A direct comparison among these schemes based on the available information is not straightforward. However, to highlight the essential need to the experimental validation and to ensure the viability as the system scales up, Table~\ref{table:secmace-vpki-operations-efficiency-comparison} demonstrates the latency for issuing 100 pseudonyms in different schemes\footnote{We emphasize that the results of the \acs{PRESERVE} \acs{PKI} are for \acp{VM} with shared resources. Thus, the latency should not be directly compared to that for other systems. We include it here only for completeness.}. The results confirm a significant performance improvement of our scheme over prior works: a 3-fold improvement over VeSPA~\cite{vespa2013}, a 2.5-fold improvement over SEROSA~\cite{gisdakis2013serosa} and a 4-fold improvement over PUCA~\cite{puca2014}.

\section{Conclusion}
\label{sec:secmace-conclusions}

Paving the way for the deployment of a secure and privacy-preserving \ac{VC} system has been started; standardization bodies and harmonization efforts have consensus towards deploying a special-purpose identity and credential management system. However, its success requires effective security and privacy-preserving protocols to guarantee the operations of the \ac{VC} systems. To address the existing challenges, we proposed SECMACE, a novel \ac{VPKI} that improves upon prior art in terms of security and privacy protection, and efficiency, and it provides solid evidence through a detailed implementation; we proposed three pseudonym acquisition policies, one of which protects user privacy to a greater extent while the timing information cannot harm user privacy. We further provide a full-blown implementation of our system and we evaluated our scheme with real mobility traces to confirm its efficiency, scalability, and robustness. Through extensive experimental evaluation, we demonstrated that modest \acp{VM} dedicated for the servers can serve on-demand requests with very low delay, and the most promising policy in terms of privacy protection incurs moderate overhead. This supports that the deployment of \ac{VPKI} facilities can be cost-effective.

\section{Acknowledgement}
\label{sec:secmace-acknowledgement}

We would like to show our gratitude to Sebastian Mauthofer (Fraunhofer SIT) and Daniel Estor (Escrypt) to provide us the processing delays of their systems. We would like to thank Assoc. Prof. Tomas Olovsson (Chalmers University of Technology) for his helpful comments on an earlier version of the manuscript.

\bibliographystyle{IEEEtran}
\bibliography{IEEEabrv,references}

\begin{thebibliography}{10}
\providecommand{\url}[1]{#1}
\csname url@samestyle\endcsname
\providecommand{\newblock}{\relax}
\providecommand{\bibinfo}[2]{#2}
\providecommand{\BIBentrySTDinterwordspacing}{\spaceskip=0pt\relax}
\providecommand{\BIBentryALTinterwordstretchfactor}{4}
\providecommand{\BIBentryALTinterwordspacing}{\spaceskip=\fontdimen2\font plus
\BIBentryALTinterwordstretchfactor\fontdimen3\font minus
  \fontdimen4\font\relax}
\providecommand{\BIBforeignlanguage}[2]{{%
\expandafter\ifx\csname l@#1\endcsname\relax
\typeout{** WARNING: IEEEtran.bst: No hyphenation pattern has been}%
\typeout{** loaded for the language `#1'. Using the pattern for}%
\typeout{** the default language instead.}%
\else
\language=\csname l@#1\endcsname
\fi
#2}}
\providecommand{\BIBdecl}{\relax}
\BIBdecl

\bibitem{ETSI-102-638}
{European Telecommunications Standards Institute}, ``{I}ntelligent {T}ransport
  {S}ystems ({ITS}); {V}ehicular {C}ommunications; {B}asic {S}et of
  {A}pplications; {D}efinitions,'' ETSI Tech. TR-102-638, Jun. 2009.

\bibitem{papadimitratos2009vehicular}
P.~Papadimitratos, A.~La~Fortelle, K.~Evenssen, R.~Brignolo, and S.~Cosenza,
  ``{V}ehicular {C}ommunication {S}ystems: {E}nabling {T}echnologies,
  {A}pplications, and {F}uture {O}utlook on {I}ntelligent {T}ransportation,''
  \emph{IEEE Communications Magazine}, vol.~47, no.~11, pp. 84\textendash95,
  Nov. 2009.

\bibitem{jin2016security}
H.~Jin, M.~Khodaei, and P.~Papadimitratos, ``{S}ecurity and {P}rivacy in
  {V}ehicular {S}ocial {N}etworks,'' in \emph{Vehicular Social Networks}.\hskip
  1em plus 0.5em minus 0.4em\relax Taylor \& Francis Group, 2016.

\bibitem{papadimitratos2006securing}
P.~Papadimitratos, V.~Gligor, and J.-P. Hubaux, ``{S}ecuring {V}ehicular
  {C}ommunications-{A}ssumptions, {R}equirements, and {P}rinciples,'' in
  \emph{ESCAR}, Berlin, Germany, pp. 5\textendash14, Nov. 2006.

\bibitem{1609-2016}
``{IEEE} {S}tandard for {W}ireless {A}ccess in {V}ehicular {E}nvironments -
  {S}ecurity {S}ervices for {A}pplications and {M}anagement {M}essages,''
  \emph{{IEEE} {S}td 1609.2-2016 ({R}evision of {IEEE} {S}td 1609.2-2013)},
  Mar. 2016.

\bibitem{c2c}
\BIBentryALTinterwordspacing
{C}ar-to-{C}ar {C}ommunication~{C}onsortium {(C2C-CC)}, {A}ccessed {D}ate:
  9-May-2016. [Online]. Available: \url{http://www.car-2-car.org/}
\BIBentrySTDinterwordspacing

\bibitem{papadimitratos2008secure}
P.~Papadimitratos, L.~Buttyan, T.~Holczer, E.~Schoch, J.~Freudiger, M.~Raya,
  Z.~Ma, F.~Kargl, A.~Kung, and J.-P. Hubaux, ``{S}ecure {V}ehicular
  {C}ommunication {S}ystems: {D}esign and {A}rchitecture,'' \emph{IEEE
  Communications Magazine}, vol.~46, no.~11, pp. 100\textendash109, Nov.
  2008.

\bibitem{leinmuller2006sevecom}
T.~Leinm{\"u}ller, L.~Buttyan, J.-P. Hubaux, F.~Kargl, R.~Kroh,
  P.~Papadimitratos, M.~Raya, and E.~Schoch, ``{SEVECOM}-{S}ecure {V}ehicle
  {C}ommunication,'' in \emph{Proceedings of IST Mobile Summit}, Myconos,
  Greece, June 2006.

\bibitem{whyte2013security}
W.~Whyte, A.~Weimerskirch, V.~Kumar, and T.~Hehn, ``{A} {S}ecurity {C}redential
  {M}anagement {S}ystem for {V2V} {C}ommunications,'' in \emph{IEEE VNC},
  Boston, MA, USA, pp. 1\textendash8, Dec. 2013.

\bibitem{preserve-url}
\BIBentryALTinterwordspacing
``{P}reparing {S}ecure {V}ehicle-to-{X} {C}ommunication {S}ystems -
  {PRESERVE},'' {A}ccessed {D}ate: 9-May-2016. [Online]. Available:
  \url{http://www.preserve-project.eu/}
\BIBentrySTDinterwordspacing

\bibitem{douceur2002sybil}
J.~R. Douceur, ``{T}he {S}ybil {A}ttack,'' in \emph{ACM Peer-to-peer Systems},
  London, UK, Mar. 2002.

\bibitem{papadimitratos2007architecture}
P.~Papadimitratos, L.~Buttyan, J.-P. Hubaux, F.~Kargl, A.~Kung, and M.~Raya,
  ``{A}rchitecture for {S}ecure and {P}rivate {V}ehicular {C}ommunications,''
  in \emph{IEEE ITST}, Sophia Antipolis, Jun. 2007.

\bibitem{khodaei2014ScalableRobustVPKI}
M.~Khodaei, H.~Jin, and P.~Papadimitratos, ``{T}owards {D}eploying a {S}calable
  \& {R}obust {V}ehicular {I}dentity and {C}redential {M}anagement
  {I}nfrastructure,'' in \emph{IEEE VNC}, Paderborn, Germany, Dec. 2014.

\bibitem{khodaei2016evaluating}
M.~Khodaei and P.~Papadimitratos, ``{E}valuating {O}n-demand {P}seudonym
  {A}cquisition {P}olicies in {V}ehicular {C}ommunication {S}ystems,'' in
  \emph{Proceedings of the First International Workshop on Internet of Vehicles
  and Vehicles of Internet}, Paderborn, Germany, pp. 7\textendash12, July
  2016.

\bibitem{fischer2006secure}
L.~Fischer, A.~Aijaz, C.~Eckert, and D.~Vogt, ``{S}ecure {R}evocable
  {A}nonymous {A}uthenticated {I}nter-vehicle {C}ommunication ({SRAAC}),'' in
  \emph{ESCAR}, Berlin, Germany, Nov. 2006.

\bibitem{sha2006adaptive}
K.~Sha, Y.~Xi, W.~Shi, L.~Schwiebert, and T.~Zhang, ``{A}daptive
  {P}rivacy-{P}reserving {A}uthentication in {V}ehicular {N}etworks,'' in
  \emph{ChinaCom.}, Beijing, China, pp. 1\textendash8, Oct. 2006.

\bibitem{kargl2008secure}
F.~Kargl, P.~Papadimitratos, L.~Buttyan, M.~Muter, E.~Schoch, B.~Wiedersheim,
  T.-V. Thong, G.~Calandriello, A.~Held, and A.~Kung, ``{S}ecure {V}ehicular
  {C}ommunication {S}ystems: {I}mplementation, {P}erformance, and {R}esearch
  {C}hallenges,'' \emph{IEEE Communications Magazine}, vol.~46, no.~11, pp.
  110\textendash118, Nov. 2008.

\bibitem{studer2009tacking}
A.~Studer, E.~Shi, F.~Bai, and A.~Perrig, ``{TACK}ing {T}ogether {E}fficient
  {A}uthentication, {R}evocation, and {P}rivacy in {VANET}s,'' in \emph{IEEE
  SECON}, Rome, Italy, pp. 1\textendash9, Jun. 2009.

\bibitem{schaub2010v}
F.~Schaub, F.~Kargl, Z.~Ma, and M.~Weber, ``V-tokens for {C}onditional
  {P}seudonymity in {VANET}s,'' in \emph{IEEE WCNC}, Sydney, Australia, pp.
  1\textendash6, Apr. 2010.

\bibitem{khodaei2012secure}
M.~Khodaei, ``{S}ecure {V}ehicular {C}ommunication {S}ystems: {D}esign and
  {I}mplementation of a {V}ehicular {PKI} ({VPKI}),'' Master's thesis, Royal
  Institute of Technology (KTH), Stockholm, Sweden, Oct. 2012.

\bibitem{vespa2013}
N.~Alexiou, M.~Lagan\`{a}, S.~Gisdakis, M.~Khodaei, and P.~Papadimitratos,
  ``{VeSPA}: {V}ehicular {S}ecurity and {P}rivacy-preserving {A}rchitecture,''
  in \emph{ACM HotWiSec}, Budapest, Hungary, Apr. 2013.

\bibitem{bibmeyer2013copra}
N.~Bi{\ss}meyer, J.~Petit, and K.~M. Bayarou, ``{C}o{PRA}: {C}onditional
  {P}seudonym {R}esolution {A}lgorithm in {VANET}s,'' in \emph{IEEE WONS},
  Banff, Canada, pp. 9\textendash16, Mar. 2013.

\bibitem{gisdakis2013serosa}
S.~Gisdakis, M.~Lagan{\`a}, T.~Giannetsos, and P.~Papadimitratos, ``{SEROSA}:
  {SER}vice {O}riented {S}ecurity {A}rchitecture for {V}ehicular
  {C}ommunications,'' in \emph{IEEE VNC}, Boston, MA, USA, Dec. 2013.

\bibitem{US-VPKI}
\BIBentryALTinterwordspacing
``{V}ehicle {S}afety {C}ommunications {S}ecurity {S}tudies: {T}echnical
  {D}esign of the {S}ecurity {C}redential {M}anagement {S}ystem,'' July 2016.
  [Online]. Available:
  \url{https://www.regulations.gov/document?D=NHTSA-2015-0060-0004}
\BIBentrySTDinterwordspacing

\bibitem{puca2014}
D.~F{\"o}rster, H.~L{\"o}hr, and F.~Kargl, ``{PUCA}: {A} {P}seudonym {S}cheme
  with {U}ser-{C}ontrolled {A}nonymity for {V}ehicular {A}d-{H}oc {N}etworks
  ({VANET}),'' in \emph{IEEE VNC}, Paderborn, Germany, Dec. 2014.

\bibitem{sun2007secure}
X.~Sun, X.~Lin, and P.-H. Ho, ``{S}ecure {V}ehicular {C}ommunications based on
  {G}roup {S}ignature and {ID}-based {S}ignature {S}cheme,'' in \emph{IEEE
  ICC}, Glasgow, Scotland, Jun. 2007.

\bibitem{guo2007group}
J.~Guo, J.~P. Baugh, and S.~Wang, ``{A} {G}roup {S}ignature {B}ased {S}ecure
  and {P}rivacy-{P}reserving {V}ehicular {C}ommunication {F}ramework,''
  \emph{Mobile Networking for Vehicular Environments}, pp. 103\textendash108,
  May, 2007.

\bibitem{lin2007gsis}
X.~Lin, X.~Sun, P.-H. Ho, and X.~Shen, ``{GSIS}: {A} {S}ecure and
  {P}rivacy-{P}reserving {P}rotocol for {V}ehicular {C}ommunications,''
  \emph{IEEE Transactions on Vehicular Technology}, Nov. 2007.

\bibitem{wasef2008ecmv}
A.~Wasef, Y.~Jiang, and X.~S. Shen, ``{ECMV}: {E}fficient {C}ertificate
  {M}anagement {S}cheme for {V}ehicular {N}etworks,'' in \emph{IEEE GLOBECOM},
  New Orleans, LO, USA, Dec. 2008.

\bibitem{wasef2008ppgcv}
A.~Wasef and X.~S. Shen, ``{PPGCV}: {P}rivacy {P}reserving {G}roup
  {C}ommunications {P}rotocol for {V}ehicular {A}d {H}oc {N}etworks,'' in
  \emph{IEEE ICC.}, Beijing, China, May 2008.

\bibitem{lu2008ecpp}
R.~Lu, X.~Lin, H.~Zhu, P.-H. Ho, and X.~Shen, ``{ECPP}: {E}fficient
  {C}onditional {P}rivacy {P}reservation {P}rotocol for {S}ecure {V}ehicular
  {C}ommunications,'' in \emph{IEEE INFOCOM}, Phoenix, AZ, USA, pp.
  1903\textendash1911, Apr. 2008.

\bibitem{calandriello2011performance}
G.~{C}alandriello, P.~Papadimitratos, J.-P. Hubaux, and A.~Lioy, ``{O}n the
  {P}erformance of {S}ecure {V}ehicular {C}ommunication {S}ystems,'' \emph{IEEE
  TDSC}, vol.~8, no.~6, pp. 898\textendash912, Nov. 2011.

\bibitem{raya2007eviction}
M.~Raya, P.~Papadimitratos, I.~Aad, D.~Jungels, and J.-P. Hubaux, ``{E}viction
  of {M}isbehaving and {F}aulty {N}odes in {V}ehicular {N}etworks,'' \emph{IEEE
  Journal on Selected Areas in Communications}, pp. 1557\textendash1568, Oct.
  2007.

\bibitem{boneh2004group}
D.~Boneh and H.~Shacham, ``{G}roup {S}ignatures with {V}erifier-{L}ocal
  {R}evocation,'' in \emph{Proceedings of the 11th ACM conference on Computer
  and communications security}, Washington, DC, USA, Oct. 2004.

\bibitem{boneh2004short}
D.~Boneh, X.~Boyen, and H.~Shacham, ``{S}hort {G}roup {S}ignatures,'' in
  \emph{Proceedings of 24th Annual International Cryptology Conference}, Santa
  Barbara, California, USA, Aug. 2004.

\bibitem{PapadiCLH:C:08}
P.~Papadimitratos, G.~Calandriello, A.~Lioy, and J.-P. Hubaux, ``{I}mpact of
  {V}ehicular {C}ommunication {S}ecurity on {T}ransportation {S}afety,'' in
  \emph{IEEE INFOCOM MOVE}, Phoenix, AZ, USA, Apr. 2008.

\bibitem{jin2017resilient}
H.~Jin and P.~Papadimitratos, ``{R}esilient {P}rivacy {P}rotection for
  {L}ocation-{B}ased {S}ervices {T}hrough {D}ecentralization,'' in \emph{ACM
  WiSec}, Boston, MA, USA, July 2017.

\bibitem{Papadi:C:08}
P.~Papadimitratos, ``"{O}n the road" - {R}eflections on the {S}ecurity of
  {V}ehicular {C}ommunication {S}ystems,'' in \emph{{IEEE} {ICVES}}, Columbus,
  OH, USA, pp. 359\textendash363, Sep. 2008.

\bibitem{khodaei2015VTMagazine}
M.~Khodaei and P.~Papadimitratos, ``{T}he {K}ey to {I}ntelligent
  {T}ransportation: {I}dentity and {C}redential {M}anagement in {V}ehicular
  {C}ommunication {S}ystems,'' \emph{IEEE VT Magazine}, vol.~10, no.~4, pp.
  63\textendash69, Dec. 2015.

\bibitem{sermersheim2006lightweight}
J.~Sermersheim, ``{L}ightweight {D}irectory {A}ccess {P}rotocol ({LDAP}): {T}he
  {P}rotocol,'' Jun. 2006.

\bibitem{dierks2008transport}
T.~Dierks, ``The transport layer security ({TLS}) protocol version 1.2,'' Aug.
  2008.

\bibitem{solo2002internet}
D.~Solo, R.~Housley, and W.~Ford, ``{I}nternet {X}. 509 {P}ublic {K}ey
  {I}nfrastructure {C}ertificate and {C}ertificate {R}evocation {L}ist ({CRL})
  {P}rofile,'' Apr. 2002.

\bibitem{santesson2013x}
S.~Santesson, M.~Myers, R.~Ankney, A.~Malpani, S.~Galperin, and C.~Adams,
  ``{X}. 509 {I}nternet {P}ublic {K}ey {I}nfrastructure {O}nline {C}ertificate
  {S}tatus {P}rotocol - {OCSP},'' Tech. Rep., Jun. 2013.

\bibitem{heinlein1998fastcgi}
P.~Heinlein, ``{F}ast{CGI},'' \emph{Linux journal}, vol. 1998, no. 55es, p.~1,
  Nov. 1998.

\bibitem{xmlrpc-c}
\BIBentryALTinterwordspacing
``{XML-RPC} for {C} and {C++},'' {A}ccessed {D}ate: 9-May-2016. [Online].
  Available: \url{http://xmlrpc-c.sourceforge.net/}
\BIBentrySTDinterwordspacing

\bibitem{protocol-buffer}
\BIBentryALTinterwordspacing
``{G}oogle {P}rotocol {B}uffer,'' {A}ccessed {D}ate: 9-May-2016. [Online].
  Available: \url{https://developers.google.com/protocol-buffers/}
\BIBentrySTDinterwordspacing

\bibitem{uppoor2014generation}
S.~Uppoor, O.~Trullols-Cruces, M.~Fiore, and J.~M. Barcelo-Ordinas,
  ``{G}eneration and {A}nalysis of a {L}arge-scale {U}rban {V}ehicular
  {M}obility {D}ataset,'' \emph{IEEE Transactions on Mobile Computing},
  vol.~13, no.~5, pp. 1061\textendash1075, May 2014.

\bibitem{codeca2015lust}
L.~Codeca, R.~Frank, and T.~Engel, ``{L}uxembourg {S}umo {T}raffic ({LuST})
  {S}cenario: 24 {H}ours of {M}obility for {V}ehicular {N}etworking
  {R}esearch,'' in \emph{IEEE VNC}, Kyoto, Japan, pp. 1\textendash8, Dec.
  2015.

\bibitem{abliz2009guided}
M.~Abliz and T.~Znati, ``{A} {G}uided {T}our {P}uzzle for {D}enial of {S}ervice
  {P}revention,'' in \emph{IEEE Computer Security Applications Conference,
  ACSAC'09.}, Honolulu, HI, USA, pp. 279\textendash288, Dec. 2009.

\end{thebibliography}

%

\vfill
\begin{IEEEbiography}[{\includegraphics[width=1in,height=1.25in,clip,keepaspectratio]{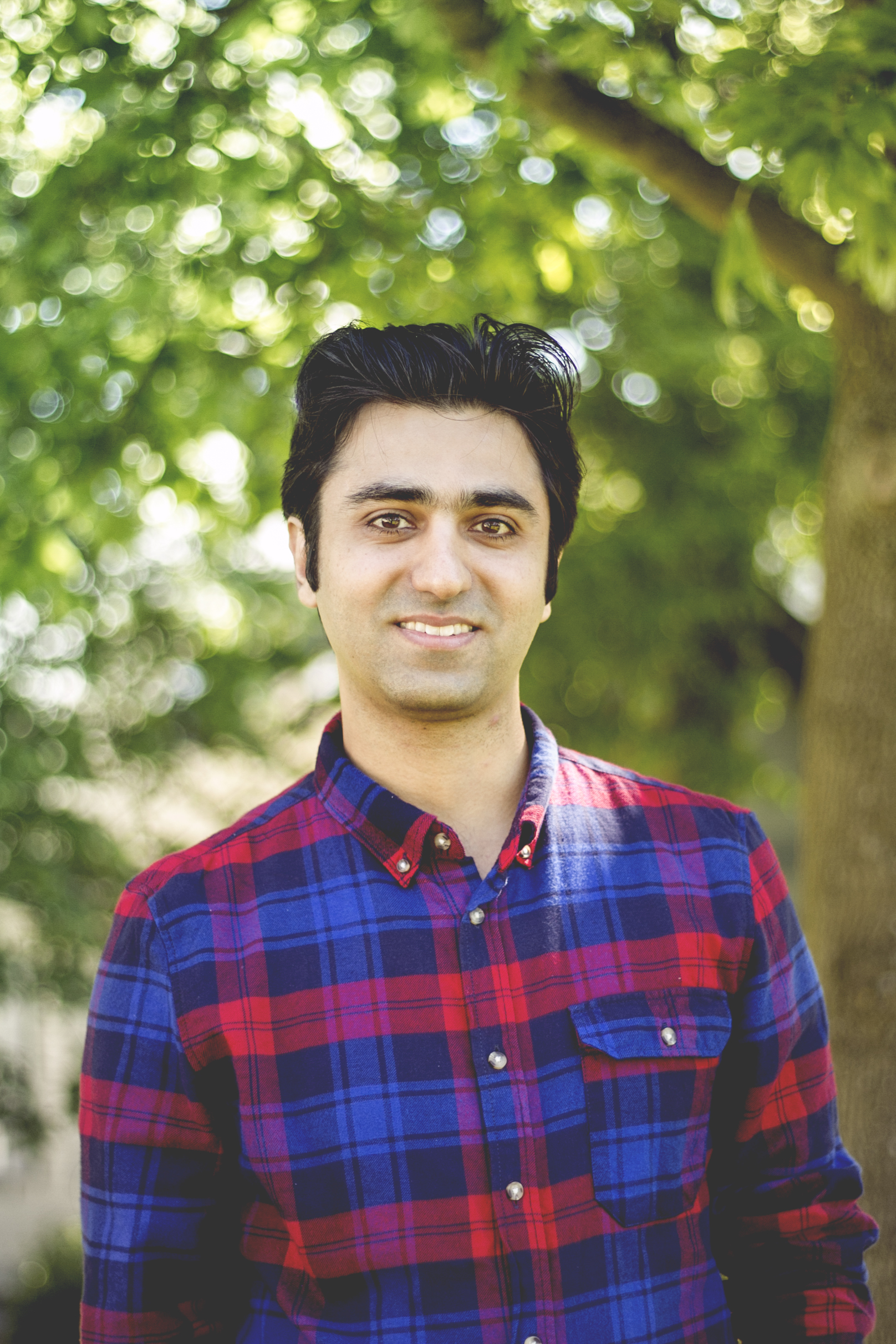}}]{Mohammad Khodaei}
earned his diploma in software engineering from Azad University of Najafabad in Isfahan, Iran, in 2006 and his M.S. degree in information and communication systems security from KTH Royal Institute of Technology, Stockholm, Sweden, in 2012. He is currently pursuing his Ph.D. degree at the Networked Systems Security Group, KTH, under the supervision of Prof. Panos Papadimitratos. His research interests include security and privacy in vehicular ad hoc networks, smart cities, and the Internet of Things. 
\end{IEEEbiography}

\vfill
\begin{IEEEbiography}[{\includegraphics[width=1in,height=1.25in,clip,keepaspectratio]{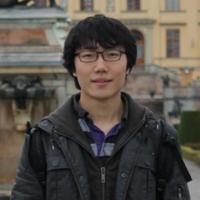}}]{Hongyu Jin}
is currently a Ph.D student at KTH Royal Institute of Technology in Networked Systems Security Group, supervised by Prof. Panos Papadimitratos. His research interests lie in security and privacy in Vehicular Communication systems and Location-based Services.  
\end{IEEEbiography}

\vfill
\begin{IEEEbiography}[{\includegraphics[width=1in,height=1.25in,clip,keepaspectratio]{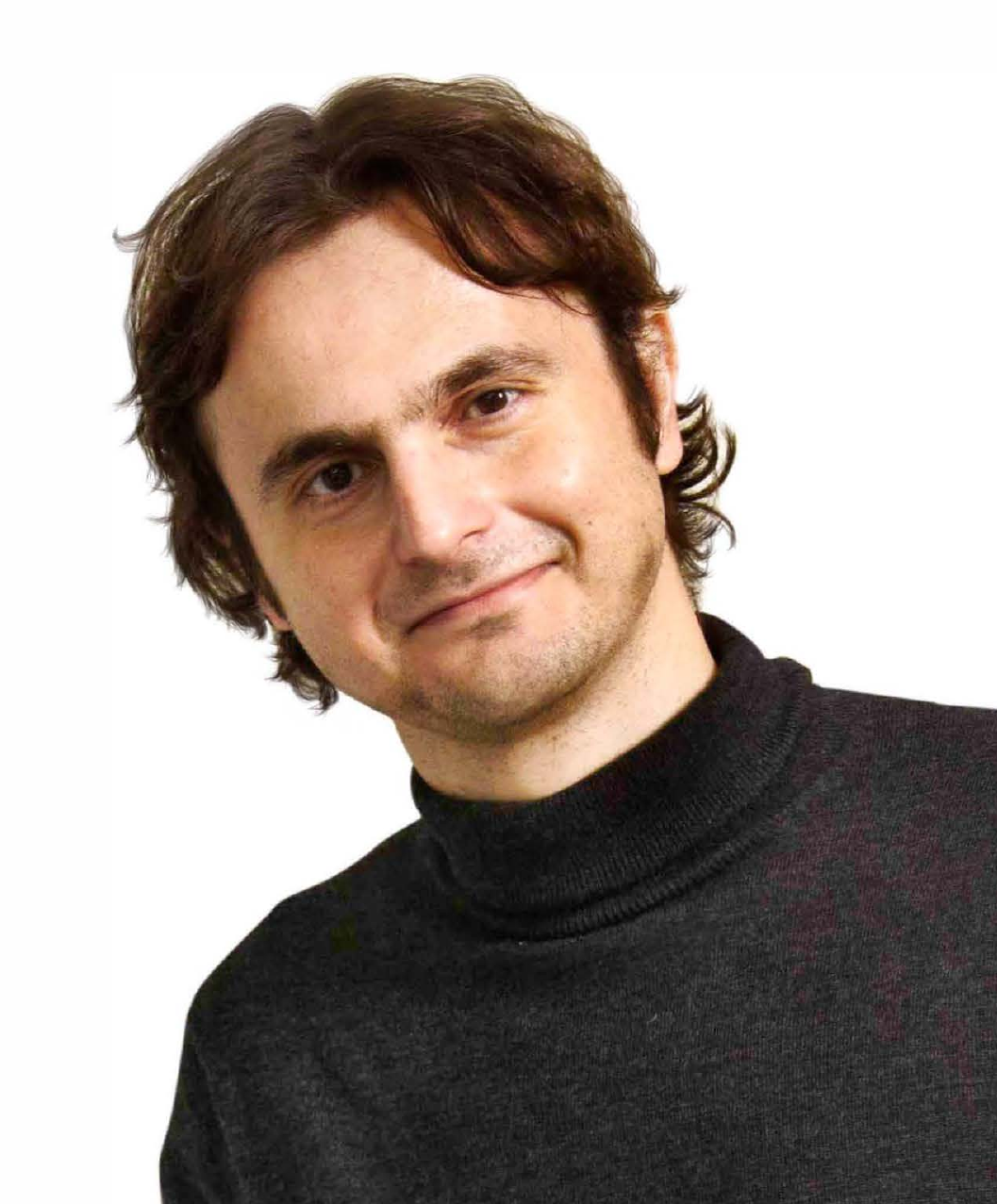}}]{Panagiotis (Panos) Papadimitratos} earned his Ph.D. degree from Cornell University, Ithaca, NY, in 2005. He then held positions at Virginia Tech, \'{E}cole Polytechnique F\'{e}d\'{e}rale de Lausanne and Politecnico of Torino. Panos is currently a tenured Professor at KTH, Stockholm, Sweden, where he leads the Networked Systems Security group. His research agenda includes a gamut of security and privacy problems, with emphasis on wireless networks. At KTH, he is affiliated with the ACCESS center, leading its Security, Privacy, and Trust thematic area, as well as the ICES center, leading its Industrial Competence Group on Security. Panos is a Knut and Alice Wallenberg Academy Fellow and he received a Swedish Science Foundation Young Researcher Award. He has delivered numerous invited talks, keynotes, and panel addresses, as well as tutorials in flagship conferences. Panos currently serves as an Associate Editor of the IEEE Transactions on Mobile Computing and the ACM/IEEE Transactions on Networking. He has served in numerous program committees, with leading roles in numerous occasions; recently, in 2016, as the program co-chair for the ACM WiSec and the TRUST conferences; he serves as the general chair of the ACM WISec (2018) and PETS (2019) conferences. Panos is a member of the Young Academy of Europe. His group webpage is: \url{www.ee.kth.se/nss}.
\end{IEEEbiography}

\vfill



\end{document}